\documentclass[journal,twocolumn, 10pt]{IEEEtran}
\usepackage{cite}
\usepackage{amsmath,amssymb,amsfonts}
\usepackage{algorithmic}
\usepackage{graphicx}
\usepackage{textcomp}
\def\BibTeX{{\rm B\kern-.05em{\sc i\kern-.025em b}\kern-.08em
    T\kern-.1667em\lower.7ex\hbox{E}\kern-.125emX}}

\usepackage{multicol}  
\usepackage{multirow}  
\usepackage{array}
\newcommand{\PreserveBackslash}[1]{\let\temp=\\#1\let\\=\temp}
\newcolumntype{C}[1]{>{\PreserveBackslash\centering}p{#1}}
\newcolumntype{R}[1]{>{\PreserveBackslash\raggedleft}p{#1}}
\newcolumntype{L}[1]{>{\PreserveBackslash\raggedright}p{#1}}
\usepackage{amssymb}

\def\be{ \begin{equation} }
\def\ee{ \end{equation} }
\usepackage{float}
\usepackage{multicol}
\usepackage{longtable}
\usepackage{xcolor}
\usepackage{soul}

\newcommand*\listnomenclature{}

\usepackage{supertabular,booktabs}

\usepackage{xparse,expl3}

\makeatletter\ExplSyntaxOn
\prop_new:N \l__ivan_nomenclature_prop
\seq_new:N  \l__ivan_nomenclature_seq

\cs_new:Npn \ivan_sort_prop:N #1
  {
    \seq_clear:N \l_tmpa_seq
    \prop_map_inline:Nn #1
      { \seq_put_right:Nn \l_tmpa_seq { ##1 } }
    \seq_sort:Nn \l_tmpa_seq
      {
        \int_compare:nTF { \pdftex_strcmp:D { ##1 } { ##2 } = -1 }
          { \sort_return_same: }
          { \sort_return_swapped: }
      }
    \prop_clear:N \l_tmpa_prop
    \seq_map_inline:Nn \l_tmpa_seq
      {
        \prop_get:NnN #1 { ##1 } \l_tmpa_tl
        \prop_put:NnV \l_tmpa_prop { ##1 } \l_tmpa_tl
      }
    \prop_set_eq:NN #1 \l_tmpa_prop
  }

\cs_new:Npn \__ivan_add_to_prop:Nw #1#2&#3 \q_stop
  { \prop_put:Nnn #1 { #2 } { #3 } }

\cs_new:Npn \ivan_list_of_nomenclature:nn #1#2
  {
    \seq_set_split:Nnn \l__ivan_nomenclature_seq { \\ } { #2 }
    \seq_map_inline:Nn \l__ivan_nomenclature_seq
      { \__ivan_add_to_prop:Nw \l__ivan_nomenclature_prop ##1 \q_stop }
    \ivan_sort_prop:N \l__ivan_nomenclature_prop
  \chapter{ \listnomenclature }
  \@mkboth
    { \MakeUppercase\listnomenclature }
    { \MakeUppercase\listnomenclature }
  \begin{supertabular}[c]{#1}
    \prop_map_inline:Nn \l__ivan_nomenclature_prop
      { ##1 & ##2 \\ }
  \end{supertabular}
}

\NewDocumentCommand \listofnomenclature { m+m }
  { \ivan_list_of_nomenclature:nn { #1 } { #2 } }

\ExplSyntaxOff\makeatother
\usepackage{array}
\newcolumntype{P}[1]{>{\centering\arraybackslash}p{#1}}
\newcolumntype{M}[1]{>{\centering\arraybackslash}m{#1}}

\DeclareMathOperator*{\argmin}{\arg\!\min}
\DeclareMathOperator*{\argmax}{\arg\!\max}

\def\rH{{\rm H}}

\def\be{ \begin{equation} }
\def\ee{ \end{equation} }
\def\bea{ \begin{eqnarray} }
\def\eea{ \end{eqnarray} }

\def\by{{\bf y}}
\def\bc{{\bf c}}

\def\bs{{\bf s}}

\def\bn{{\bf n}}

\def\bC{{\bf C}}

\def\bI{{\bf I}}

\def\b0{{\bf 0}}
\def\bPhi{{\bf \Phi}}

\def\cI{{\cal I}}

\makeatletter
\newcommand*{\rom}[1]{\expandafter\@slowromancap\romannumeral #1@}
\makeatother

\usepackage[acronym]{glossaries}
\usepackage[nolist,nohyperlinks]{acronym}

\usepackage[font=footnotesize]{caption} 
\usepackage{csquotes}

\usepackage{enumitem}

\usepackage{datatool}

\begin{document}

\title{IoT Connectivity Technologies and Applications:\\ A Survey}
\author{Jie Ding,
Mahyar Nemati, Chathurika Ranaweera, and Jinho Choi
\thanks{M. Nemati, J. Ding, C. Ranaweera, and J. Choi are with
the School of Information Technology, 
Deakin University, Victoria 3125 Australia}
\thanks{J. Ding is also with the School of Electronic information and Communications, Huazhong University of Science and Technology, 430074 China}
\thanks{Corresponding author: Jie Ding (e-mail: yxdj2010@gmail.com). }}
\date{today}
\maketitle

\begin{abstract}
The Internet of Things (IoT) is rapidly becoming an integral part of our life and also multiple industries.
We expect to see the
number of IoT connected devices explosively grows and will reach hundreds of billions during the next few years.
To support such a massive connectivity, various wireless technologies are investigated. In this survey, we provide a broad view of the existing wireless IoT connectivity technologies and discuss several new emerging technologies and solutions that can be effectively used to enable massive connectivity for IoT.
In particular, we categorize the existing wireless IoT connectivity technologies based on coverage range and review diverse types of connectivity technologies with different specifications. We also point out key technical challenges of the existing connectivity technologies for enabling massive IoT connectivity. To address the challenges, we further review and discuss
some examples of promising technologies such as compressive sensing (CS) random access, non-orthogonal multiple access (NOMA), and massive multiple input multiple output (mMIMO) based random access that could be employed in future standards for supporting IoT connectivity.
Finally, a classification of IoT applications is considered in terms of various service requirements. For each group of classified applications, we outline its suitable IoT connectivity options.
\end{abstract}

{\IEEEkeywords
IoT connectivity technologies; 5G; massive MTC; massive connectivity; compressive sensing; NOMA; massive MIMO; machine learning; IoT applications
}


\maketitle

\section{Introduction}
\label{section1}
In 1999, the MIT Auto-ID center coined the term of the Internet of Things (IoT), for the first time, where the "things" can be any physical object that sends data and communicates with a network \cite{I-1}. At the beginning, radio frequency identification (RFID) systems were the first deployed technologies for simple IoT applications that had enabled objects to communicate with other objects or a server without human interaction \cite{I-2}. Since 2003, Walmart 24, a retailer for the first time in the vertical market, has deployed RFID tags in all stores around the world \cite{I-4}. In 2009, European Commission proposed a framework, with financial support of governments, to start an extensive research on a compatible IoT network for all available and future applications \cite{I-5}. Throughout the last few years, with the introduction of the $5$th generation ($5$G) wireless technology \cite{Add_Survey1}, the IoT has drawn much attention
in particular with the emergence of machine type communications (MTC), which refers to automated data communications among devices or from devices to a central MTC server or a set of MTC servers \cite{LTEM2}.

The IoT is projected to grow significantly with a remarkable economic impact.
It is expected that there will be more devices and sensors that are to be connected to the Internet for the IoT and various new IoT applications will be emerged (e.g., smart cities and industrial IoT). According to Gartner, it is estimated that more than $8.4$ billion connected devices were in use worldwide in 2018, more than $31$\% from 2016. By 2020, it is predicted that the number will exceed $20.8$ billion and the exponential growth is expected to continue in the future \cite{I-3}.

As the number of things or devices to be connected is growing, their connectivity becomes an important issue. A number of IoT applications are used in a small coverage area and their connectivity can rely on short-range wireless  technologies such as Bluetooth, Zigbee, WiFi, and optical wireless communication (OWC) \cite{shortrange1,OWC1}. On the other hand, as there are more IoT applications that require a wide coverage area, long-range wireless connectivity technologies are required. For example, outdoor sensors for environmental monitoring and unmanned aerial vehicles (UAV) need long-range connectivity to be connected to networks. As a result, various long-range wireless technologies are developed. For example, there are Sigfox \cite{Sigfox1DJ} and LoRa \cite{Lora1} that use the unlicensed bands and have their own base stations (BS) so that things/devices can be connected to one of them, similar to conventional cellular networks. In general, Sigfox and LoRa support applications of low data rates with low power consumption so that most devices can have long life cycle (about 10 years). There are also different low-power long-range connectivity technologies that are based on cellular systems. For example, there are long-term evolution (LTE) standards, e.g., narrowband IoT (NB-IoT) and LTE MTC (LTE-M),
which are developed for MTC connectivity within LTE systems \cite{LTEM2,LTEM3}. Unlike Sigfox and LoRa, NB-IoT and LTE-M employ licensed bands and can support devices with the existing cellular infrastructure. 
In addition, $5$G is proposed to not only enhance traditional mobile broadband communications, but also expected to fulfil diverse connectivity requirements of new IoT applications like low latency and ultra-high transmission reliability.
In fact, each wireless connectivity technology has different advantages and disadvantages. In general, if IoT applications require low latency, medium to high data rates, and a wide coverage, cellular IoT connectivity technologies become suitable.  

\begin{table*}[h]
	\begin{center}
	\caption{Summary of Key Survey Papers in the Areas of IoT/MTC Connectivity. LPWAN: low power wide area networks.}\label{tablenewDJ}
	\begin{tabular}{|m{1.5cm}<{\centering}|m{3cm}<{\centering}|m{1.5cm}<{\centering}|m{1.5cm}<{\centering}|m{1cm}<{\centering}|m{1cm}<{\centering}|m{1.1cm}<{\centering}|m{1cm}<{\centering}|}
		\hline
		  \multirow{2}{*}{\textbf{Ref.}} & \multirow{2}{*} {\textbf{Main Focus}}  & \multicolumn{2}{m{3cm}<{\centering}|}{\textbf{Discussion of Existing IoT Technologies?}}& \multicolumn{4}{m{4cm}<{\centering}|}{\textbf{Discussion of Emerging Technologies for Massive Connectivity?}} \\
		 \cline{3-8}
		 &  & Short-range & Long-range & CS based & NOMA based & mMIMO based & ML based \\
		 	\hline
	     This Survey & State-of-the-art IoT connectivity technologies and their applications
       & \checkmark  &  \checkmark &  \checkmark  &  \checkmark  &  \checkmark  &  \checkmark \\
       	\hline
       \cite{Add_Survey1} & Cellular evolution challenges towards $5$G
       & $\times$  & $\times$ & $\times$  & $\times$  & $\times$  & $\times$ \\
       	\hline
       	 \cite{Survey_comp8} & IoT platforms for massive connectivity
       & $\times$  & $\times$ & $\times$  & $\times$  & $\times$  & $\times$ \\
       	\hline
       	\cite{Add_Survey2} & Spectrum sharing solutions for IoT connectivity
       &  \checkmark  &  \checkmark & $\times$  & $\times$  & $\times$  & $\times$ \\
        \hline
       	\cite{Survey_comp5} & Short-range technologies and architectures for IoT
        &  \checkmark   & $\times$  & $\times$  & $\times$  & $\times$  & $\times$ \\
       	\hline
       	\cite{Survey_comp9} & IoT communication technologies and challenges
        &  \checkmark   & \checkmark  & $\times$  & $\times$  & $\times$  & $\times$ \\
       	\hline

      	\cite{ Survey_comp6} & Comparison of Low-power technologies for IoT
        &  \checkmark   & \checkmark  & $\times$  & $\times$  & $\times$  & $\times$ \\
       	\hline
       	\cite{Survey_comp1} &  IoT enabling technologies, protocols, and applications
       &  \checkmark &  $\times$  & $\times$  & $\times$  & $\times$  & $\times$ \\
       	\hline
       	\cite{wifi4} & Different LPWAN technologies and their applications
       &  \checkmark  &  \checkmark & $\times$  & $\times$  & $\times$  & $\times$ \\
        \hline
        \cite{ LPWAN3} &  LoRa for smart city applications
       &  \checkmark &  \checkmark  & $\times$  & $\times$  & $\times$  & $\times$ \\
       	\hline
       \cite{Add_Survey3} & LoRA, NB-IoT, and semantic web
       &  $\times$  &  \checkmark & $\times$  & $\times$  & $\times$  & $\times$ \\
        \hline

        \cite{Survey_comp7} & NB-IoT and its open issues
       &  $\times$  &  \checkmark & $\times$  & \checkmark  & $\times$  & $\times$ \\
        \hline

       	\cite{LPWAN2, Survey_comp2, LoraWAN2} & Comparison of different LPWAN from various perspectives
        &  $\times$  &  \checkmark  & $\times$  & $\times$  & $\times$  & $\times$ \\
       	\hline

       	\cite{ Survey_comp4} &  CS based IoT Applications
       &  $\times$ &  $\times$  & \checkmark  & $\times$  & $\times$  & $\times$ \\
       	\hline
       	\cite{NOMA1, NOMA2,Survey_comp3} & NOMA for massive IoT connectivity
       & $\times$  & $\times$ & $\times$  & \checkmark  & $\times$  & $\times$ \\
       	\hline
     	\cite{ mMIMO12} &  mMIMO for massive IoT connectivity
       &  $\times$ &  $\times$  & $\times$  & $\times$  & \checkmark  & $\times$ \\
       	\hline
       	\cite{ML14} &  ML based solutions for massive MTC
       &  $\times$ &  \checkmark  & $\times$  & $\times$  & $\times$  & \checkmark \\
       	\hline
	
	\end{tabular}
	\end{center}
\end{table*}

In this survey, we emphasize on the state-of-the-art wireless technologies for IoT connectivity and their applications. We first provide an overview of the most dominant existing connectivity technologies that are widely debated in literature and 3rd generation partnership project (3GPP) documentation. It is noteworthy that the selected existing and conventional connectivity technologies are widely used in different industries and current applications. We outline their different specifications along with their fundamental bottleneck for enabling massive IoT connectivity. 
Then, promising emerging technologies are discussed to address the issue. Indeed, the scale of massive connectivity varies. For example, with NB-IoT, about $50,000$ devices per cell are to be connected \cite{Add_wifi}. However, in the future, the number of devices per cell will exponentially increase, which means that the existing IoT connectivity technologies may not be able to accommodate increased device connectivity without sacrificing quality of services (QoS). Therefore, new approaches are required to be developed and employed for future IoT connectivity. These new approaches should provide high spectral efficiency as spectrum resources are limited. Furthermore, it is expected that they are able to support low latency for delay-sensitive applications such as smart vehicles and collaborative IoT \cite{Mora18}. There are several survey papers that have discussed various approaches for enhancing the IoT connectivity \cite{Add_Survey3, HTCvsMTC, NOMA2, Add_Survey2}. For example,  
intelligent resource management was considered in \cite{HTCvsMTC} and non-orthogonal multiple access (NOMA) technology was reviewed in \cite{NOMA2}. In \cite{Add_Survey2}, spectrum sharing solutions for the existing IoT technologies by taking advantages of their basic features were reviewed and discussed. Different from the existing survey papers, we provide a more comprehensive overview for the cutting-edge connectivity technologies such as Compressive Sensing (CS), NOMA, massive Multiple-Input Multiple-Output (mMIMO), and Machine Learning (ML) based random access (RA). We elaborate on their abilities of enabling massive connectivity and also discuss their limitations that need to be addressed. These outlined technologies have the potential to be employed together with the existing IoT technologies to further enhance their performance.
In a nutshell, in this study, we provide the latest reviews on existing and emerging technologies along with their strengths and limitations and also new directions in terms of research topics. To further elaborate on the contribution of this survey, we summarize the features of existing key survey papers on IoT connectivity in Table \ref{tablenewDJ} while highlighting the benefits of our survey paper. As given and explained in Table \ref{tablenewDJ}, we emphasize that despite the existing key surveys, our survey mainly focuses on providing a broad overview on not just the existing IoT connectivity technologies but also diverse state-of-the-art technologies that can be used to provide connectivity for various types of IoT applications. In addition, unlike classic utilization-domain application classification, we consider a different classification approach for IoT applications with respect to their general requirements and then identify the feasible connectivity technologies for each application group.


\section{Wireless IoT Connectivity Technologies}
\label{section2}
\begin{figure}[t]
\centering 
   \includegraphics[width=8.5cm]{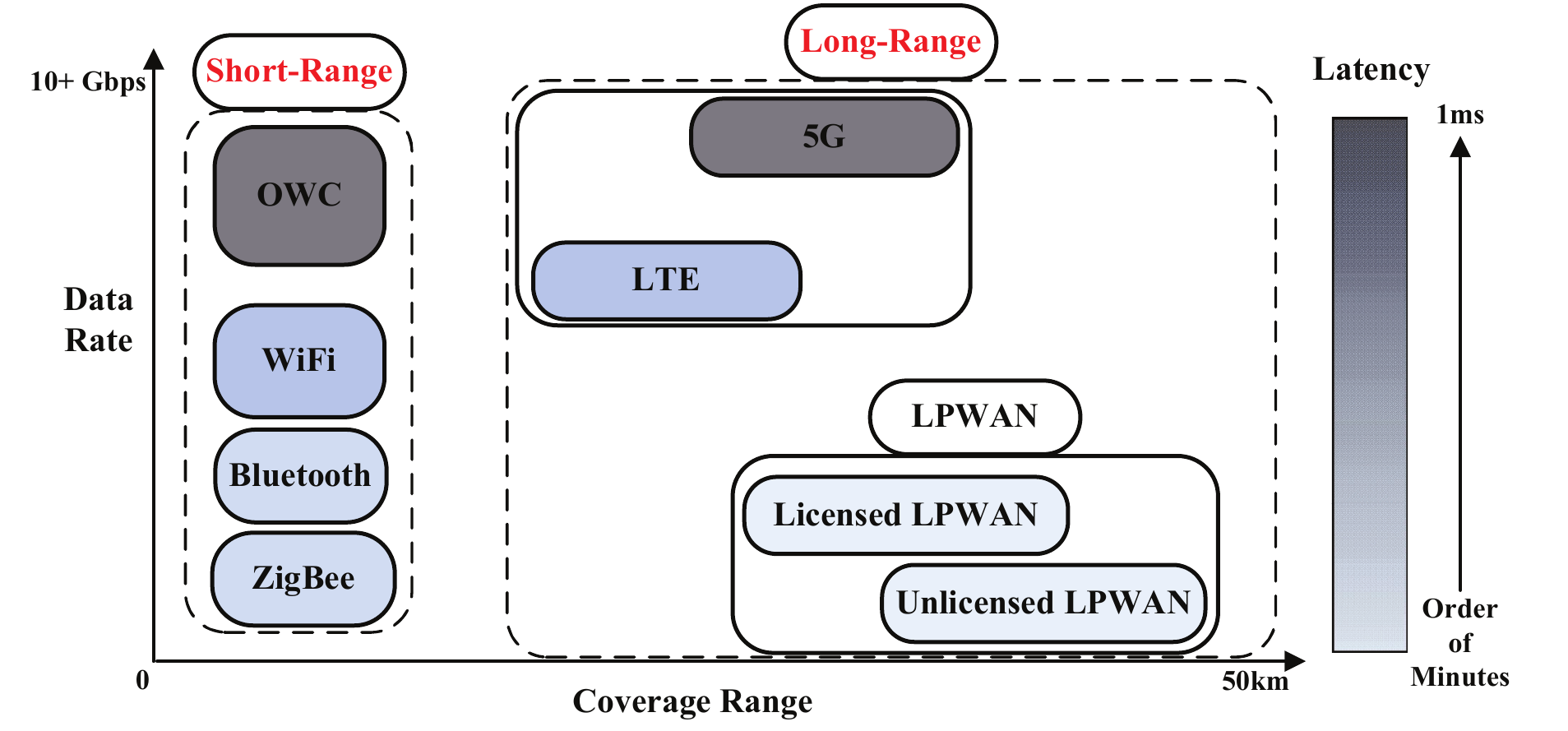}%
  \caption{Illustration of features of the IoT connectivity technologies in terms of data rate, coverage, and Latency. }%
  \label{Chap2}
\end{figure}

Since there will be billions of different kinds of connected devices in future IoT applications, it is urged to develop various technologies to support their connectivity. In this section, we discuss the existing wireless technologies for IoT connectivity and classify them into two categories in terms of coverage range, namely short-range technologies and long-range technologies.
For short-range technologies, dominant technologies like Bluetooth, ZigBee, WiFi and the emerging OWC technologies are to be discussed. For long-range technologies, depending on service features and requirements, LTE and $5$G, and LPWAN technologies including unlicensed and licensed LPWAN, are introduced. %
In Figure. \ref{Chap2}, we illustrate a diagram including the existing IoT connectivity technologies with respect to data rate, coverage range, and latency.

\subsection{Short-Range Technologies }
\label{short}
Short-range wireless technologies for IoT applications are usually used to support connectivity within a small coverage area.  
There are a number of short-range technologies with different features and performance for given application requirements. Bluetooth, ZigBee, WiFi and OWC, as the mainstream technologies of this kind, are briefly reviewed as follows.

\subsubsection{Bluetooth}
Bluetooth, standardized by the Institute of Electrical and Electronics Engineers (IEEE) $802.15.1$ \cite{bluetooth1}, is originally created by Nokia during the late $90$’s as an in-house project. However, it quickly became a popular wireless technology that is primarily used for communications between portable devices distributed in a small area (a maximum of $100$m coverage range \cite{bluetooth2}).
Technically, Bluetooth sends short data packets over several channels of bandwidth $1$MHz between $2.402$GHz to $2.480$GHz and its data rate varies from $1$Mbps to $3$Mbps \cite{bluetooth2}.
Nevertheless, the high power consumption of classic Bluetooth makes it impractical for some emerging IoT use-cases that require low-power transmissions for small and battery-limited devices \cite{bluetooth8}.
To this end, Bluetooth Low Energy (BLE) has been introduced in Bluetooth $4.0$ specifically for low-powered IoT devices \cite{Add_BLE,bluetooth4,bluetooth5}. Unlike classic Bluetooth optimized for continuous data streaming, BLE is optimized for short burst data transmissions. BLE defines $40$ usable channels. These $40$ channels are divided into $3$ primary advertisement channels and $37$ data channels.
In general, BLE employs two multiple access schemes, i.e., frequency division multiple access (FDMA) and time division multiple access (TDMA) based polling. In Bluetooth $5.0$, enhancements upon BLE’s data rates and range were presented by using increased transmit power or coded physical layer. Compared to Bluetooth $4.0$, maximum $4$x transmission range increase is expected and a maximum data rate of $2$Mbps can be achieved (as twice as fast) \cite{Add_BLE}.
In the latest Bluetooth $5.1$, direction finding feature of BLE was enhanced to better understand signal direction and achieve sub-meter location accuracy \cite{Add_BLE1}. 
To enable large-scale IoT device networks that support many-to-many device communications, BLE mesh networking has been adopted in $2017$ \cite{Add_BLE2,Add_BLE3}. BLE mesh topology operates on a managed flood routing principle for forwarding messages from one device to another. The maximum number of devices in any given Bluetooth mesh network is $32,767$, with up to $16,384$ groups. In this model, only devices that have the enabled relay feature forward received messages further into the network. In addition, a message cache is introduced to ensure that a relay device only relays a specific
message once and a time-to-live (TTL) is used to address the issues that arise with routing loops. A relay device only relays a
message if the message is not in the cache and its TTL is greater than $1$ \cite{Add_BLE5}. Each time message is received and retransmitted, TTL will be decremented by one.
If the TTL reaches zero, the message will be discarded at the relay device, eliminating endless loops. The maximum TTL supported in Bluetooth mesh is $127$ \cite{Add_BLE4}. In addition, the backwards compatibility feature and friendship feature are also defined in BLE mesh for BLE devices. In particular, the backwards compatibility feature enables the BLE devices that do not support BLE mesh to be connected to a mesh network. Furthermore, the friendship feature enables power-limited BLE devices to become part of a mesh network with the help of battery-powered devices \cite{Add_BLE5}.
Classic Bluetooth and BLE have been currently adopted by a number of use-cases including audio streaming, health and wellness monitoring, low-cost indoor positioning, and controlling and automating \cite{bluetooth3,bluetooth6,bluetooth7}.

\subsubsection{ZigBee}
ZigBee is another short-range wireless technology for wireless personal area networks (WPAN), which is built on top of IEEE $802.15.4$ \cite{zigbee1}. Currently, ZigBee has been widely considered for a variety of IoT applications including home automation, industrial monitoring, and health and aging population care \cite{zigbee2,zigbee4,zigbee5,zigbee3}.
Similar to BLE, Zigbee is also a low-power technology. Zigbee operates in the unlicensed bands, i.e., mainly at $2.4$GHz and optionally at $868$MHz or $915$MHz, and its default operation mode at $2.4$GHz uses 16 channels of $2$MHz bandwidth. ZigBee is able to connect up to $255$ devices at a time with a maximum packet size of $128$bytes. Depending on the blockage of environments, the transmission ranges between devices vary from a few meters up to $100$ meters \cite{zigbee7}. ZigBee supports star and peer-to-peer topologies for connecting devices. In ZigBee, three types of devices are defined as follows: coordinator, router, and end device. In particular, coordinator and router are normally mains-powered and end device can be battery-powered. The coordinator is the most capable device in ZigBee, which coordinates the actions of a network and might connect to another network as a bridge. The routers form a network for packet exchanges. The end devices are logically connected to a coordinator or routers. However, these end devices cannot directly communicate with each other. To enable large-scale IoT device networks, Zigbee can be extended as generic mesh where devices are clustered with a local coordinator and connected via multihop to a global coordinator \cite{zigbee6,Zigbee_add1}.
Unlike BLE, ZigBee uses carrier sense multiple access with collision avoidance (CSMA/CA) to avoid packet collisions (please refer to Appendix \ref{A_Csma} for more details and explanations on CSMA). In addition, while BLE allows four different data rates varying from $125$kbps to $2$Mbps, ZigBee can only support data rates from $20$kbps to $250$kbps. According to the performance evaluation in a realistic home automation scenario in \cite{Zigbee_add1,Zigbee_add2}, BLE is superior to ZigBee in terms of service ratio thanks to its higher bit rate and dedicated data channels.
In terms of delay, both technologies have similar performance in a basic scenario, but BLE is more delay-sensitive to the traffic load than ZigBee. It was also shown that BLE devices
consume less energy to set up the same network, which indicates the BLE devices may have a comparably longer expected lifetime.

\subsubsection{WiFi}
 WiFi, standardized by IEEE $802.11$ \cite{wifi1}, is a family of technologies commonly used for wireless local area networks (WLAN). Different from Bluetooth and Zigbee that provide connectivity between devices,  
WiFi provides the last mile wireless broadband connections for devices to the Internet with a larger coverage and higher data rates \cite{shortrange1}. 
In fact, WiFi has been evolved several generations to support higher throughputs. Specifically, IEEE $802.11$a and IEEE $802.11$b were introduced in $1999$, where IEEE $802.11$a can support a data rate up to $54$Mbps in $5$GHz, and IEEE $802.11$b makes it up to $11$Mbps in $2.4$GHz. In $2003$, IEEE $802.11$g was released with a maximum data rate of $54$Mbps in $2.4$GHz. However, IEEE $802.11$a/b/g standards were not able to meet the growing demand of hypermedia applications over WLANs due to their relatively low throughputs and capacity. Therefore, new generations of WLANs, i.e., IEEE $802.11$n \cite{wifi2} and IEEE $802.11$ac \cite{wifi3} have been released in $2008$ and $2014$, respectively. These new generations can achieve much higher data rates (up to $600$Mbps in IEEE $802.11$n and $7$Gbps in IEEE $802.11$ac) with a wider coverage compared to previous ones (IEEE $802.11$a/b/g) by using dense modulations and MIMO technology.
In addition, IEEE $802.11$ah (WiFi HaLow) was introduced in $2017$ to support IoT with extended coverage and low-power consumption requirements. It operates in the unlicensed sub-$1$GHz bands (excluding the TV white-space bands) and its bandwidth occupation is usually only $1$MHz or $2$MHz, while in some countries, wider bandwidths up to $16$MHz are also allowed. Compared to high-speed WiFi generations, the IEEE $802.11$ah aims to provide connectivity to thousands of devices with coverage of up to $1$km but its maximum data rate is about $300$Mbps utilizing $16$MHz bandwidth \cite{Add_wifi,Hal-1,Hal-2}.

\subsubsection{OWC}
Another emerging short-range wireless technology developed to support the indoor IoT device connectivity is the OWC \cite{OWC1,OWC2}. OWC is a promising architecture that can be used to resolve the issues arising from high bandwidth and low latency indoor IoT applications. In OWC, visible light (VL), infrared (IR), or ultraviolet (UV) spectrum are used as propagation media in comparison to radio frequencies used in WiFi and other WLAN technologies \cite{OWC2, wifi1}. To date, different research groups from academia and industry have demonstrated low-complex optical wireless links that can operate at multi-gigabits per second data rate in an energy-efficient manner under a typical in room environment to support various applications \cite{OWC4,OWC5,OWC6}. 

The high-speed OWC links are proposed to provide connectivity for many IoT application where we have limited or poor WiFi/other wireless connectivity \cite{OWC7}. The application includes Tactile Internet, wireless body area networks that consist of body placed sensors, in airplanes, and also connecting bandwidth demand latest medical instruments at hospitals \cite{OWC8}. Furthermore, OWC links are proposed to provide connectivity for remotely operated underwater vehicles, dense urban environments, autonomous vehicle communications, and connecting sensors in chemical and power plants where usage of radio frequency is restricted \cite{OWC2}. 

Among different types of OWC technologies have been developed, there are two major categories of OWC technologies that can be identified as potential tools to provide high bandwidth and low-latency connectivity for emerging IoT applications \cite{OWC2}. These categories are visible light communication (VLC), and beam-steered infrared light communication (BS-ILC) \cite{OWC9}.

\begin{table*}[!htb]
	\begin{center}
	\caption{Comparison of Bluetooth, Zigbee, WiFi and OWC \cite{Add_BLE,shortrange1,OWC2}. 
	GFSK: Gaussian frequency shift keying; DQPSK: differential quadrature phase shift keying; DPSK: differential phase shift keying; BPSK: binary phase shift keying; OQPSK: offset quadrature phase-shift keying; QPSK: quadrature phase shift keying; CDMA: code-division multiple access.}\label{C2_table1DJ}
	\begin{tabular}{|c|c|c|c|c|}
		\hline
		   & \textbf{Bluetooth} & \textbf{Zigbee}  & \textbf{WiFi} & \textbf{OWC}\\

		\hline
		\hline
		 \begin{tabular}{c}
       RA \\
       protocol
 \end{tabular}  & \begin{tabular}{c}
		TDMA based polling\\
		FDMA
		\end{tabular} & CSMA/CA & CSMA/CA & \begin{tabular}{c}
       CSMA/CA \\
       TDMA/CDMA
 \end{tabular} \\
		\hline
		\begin{tabular}{c}
        Modulation  \\
       type
 \end{tabular} & GFSK/DQPSK/DPSK  & BPSK/OQPSK & BPSK/QPSK/QAM & OOK/OFDM  \\ 

		\hline
		\begin{tabular}{c}
		Maximum \\
		data rate
		\end{tabular}& \begin{tabular}{c}
		Classic: $3$Mbps\\
		BLE: $2$Mbps
		\end{tabular} & $250$kbps & $7$Gbps
		& \begin{tabular}{c}
       $10$Gbps using LED \\
      $100$Gbps using LD
 \end{tabular}
		   \\

		\hline
		Coverage & \begin{tabular}{c}
		Classic: $100$m \\
		BLE: $240$m 
		\end{tabular} & $100$m & \begin{tabular}{c}
		Conv.: $100$m \\
		$802.11$ah: $1$km 
		\end{tabular}  & $200$m\\

		
		\hline
	\end{tabular}
	\end{center}
\end{table*}
\begin{enumerate}
\item VLC: 
VLC uses the laser emitting diode (LED) illumination infrastructure to provide multi-gigabit wireless connectivity by employing diverse modulation scheme ranging from simple on-off keying (OOK) to quadrature amplitude modulation (QAM) orthogonal frequency division multiplexing (OFDM) \cite{OWC10,OWC11}. In 2011, VLC was initially standardized as IEEE $802.15.7$ \cite{OWC12}. This standard was further developed in two directions based on the data rate requirements of diverse applications.  For low data rate applications, IEEE $802.15.7$m \cite{OWC13} standard was developed using the optical camera communications (OCC) that support connectivity for a range of 200m. On the other hand, for high data rate application, IEEE $802.15.13$ \cite{OWC14} was developed enabling multi gigabit data rate connectivity over few tens of meters. Recently, 100 Gbps VLC links have demonstrated using laser diodes (LD) instead of using LEDs \cite{OWC15}. The popular technology, light fidelity (LiFi) is also developed based on VLC technology \cite{OWC16}. To date, there are several commercial VLC products are available in the market such as pureLiFi to support diverse IoT applications. 

\item BS-ILC: 
Infrared light communication was first standardized by IrDA and IEEE in early 90’s. In particular, infrared light communication was included in the initial WiFi standard (IEEE $802.11$) \cite{wifi1}. In comparison to VLC, in BS-ILC systems, the IR beams are turned on when needed, for example when there are applications/users to be served. In this system, multiple beams can be used to serve several users in the same room. In term of the coverage of a single beam, there are two different types of BS-ILC systems available. In the first type, a single beam is used to serve a single user application/user within a room and hence the implementation of medium access control protocols can be avoided as no shared medium is used \cite{OWC17}. In the second type, multiple users are served within a wide IR beam and hence implementation of the medium access control protocols has also been investigated \cite{OWC18,OWC19}. To date, different types of BS-ILC systems have demonstrated their ability to provide multi-gigabit connectivity for a range of $3$m using diverse modulation formats and different beam-steering techniques that use either active beam steering devices \cite{OWC5,OWC20,OWC21} or passive beam steering devices \cite{OWC17,OWC23}.

\end{enumerate}

In Table \ref{C2_table1DJ}, the technical specifications of Bluetooth, Zigbee, WiFi, and OWC are summarized. As discussed, different technologies have different advantages. For example, the IEEE $802.11$ac and OWC are focused on supporting high-speed transmissions, while BLE, Zigbee and the IEEE $802.11$ah are targeting at low-power and low-cost communication. Among these low-power consumption candidates, the IEEE $802.11$ah can provide higher data rates and wider coverage range.

Although these technologies are able to provide connectivity to various data rate use-cases, they are not suitable for the use-cases that require a wide coverage.
As a counterpart of the short-range technologies, existing paradigm of long-range technologies is introduced in the following.

\subsection{LTE and 5G}
LTE and $5$G are the essential parts of cellular IoT technologies. As the standardized technology of the $4$th generation ($4$G), LTE/LTE-Advance (LTE-A) has now been deployed successfully worldwide, which was mainly designed to support the conventional human-type communications (HTC) for high-speed transmissions.
Since $2016$, the $5$G standardization has been progressed by the international telecommunication union (ITU) and $3$GPP \cite{ITU1,ITU2,ITU3}. Technically, the main advantage of $5$G over LTE is its ability of providing $100$x higher data rate, $10$x lower latency, and supporting $100$x more connected devices \cite{3G-5G} by utilizing a new air interface that includes much higher frequencies such as millimeter wave (mmWave) and using more advanced radio technologies, e.g., massive multiple-input multiple-output (mMIMO), edge computing, full duplex, and Polar codes \cite{Polar1}. 
Compared with LTE, $5$G is expected to not only enhance HTC by handling far more traffic at much higher data rate, but also to support unprecedented mission-critical applications \cite{3G-5G,instruments_2018}. In Table \ref{tab:t2}, comparison of the specifications of LTE and $5$G is presented.
Indeed, the current LTE has a nominal latency of $15$ms and a target block error rates (BLER) of $10^{-1}$ before retransmission \cite{qorvo_2017}. 
In future, various mission-critical applications, such as haptic communication and smart transportation, will gradually merge into our daily life. These applications are normally insensitive to power consumption and have very restrictive requirements in terms of latency ($1$ms or less) and transmission reliability (BLER as low as $10^{-9}$) \cite{LowL0}. Therefore, one of the key tasks of $5$G is to address the challenges of low latency as well as ultra-high reliability transmissions. In fact, low latency and ultra-high reliability are two conflicting requirements. On one hand, it is necessary to use a short packet to guarantee low latency, which however may have a severe impact on the channel coding. On the other hand, users usually need more resources to satisfy high transmission success rate requirements, while it may simply increase the latency for other users \cite{LowL6}.
Although research works have recently investigated and proposed the potential solutions to this technical challenge from various perspectives \cite{LowL7,LowL1,LowL2,LowL3,LowL4,LowL5}, there are open issues that still need to be addressed to enable mission-critical applications and make them practical \cite{Add_5g}. For example, resource allocation
becomes particularly challenging with the introduction of haptic communication into $5$G and 
flexible resource allocation approaches need to be investigated to enable the coexistence
of haptic communication with other types of applications. Specifically, the latency of data transmission is influenced by how quickly wireless resources can be allocated when a data packet arrives at 
the radio interface. Because of stringent latency requirements of haptic communication,
wireless resources must be provided for it on a priority basis. Furthermore, since the
available wireless resources will be shared between haptic communication
and HTC or MTC and these applications have different and often conflicting
application requirements, existing resource allocation approaches only designed for typical HTC or MTC may not result in optimal resource allocation outcome to accommodate different application with various QoS requirements. Thus,
$5$G requires flexible dynamic approaches for wireless resource management 
so that the utility of various applications can be maximized by ensuring efficient and intelligent wireless resources allocation \cite{5G_add}. In addition, accurate and fast traffic prediction approaches need to be developed in urban scenarios for smart transportation \cite{LowL7}. Specifically, traffic prediction enables the early identification of traffic jams and allows the smart vehicles or traffic authorities to take prompt measures to avoid the congestion on the roads \cite{5G_add1}. Therefore, accurate real-time traffic prediction is one of the most important component to enable traffic efficiency services in smart transportation.
Nevertheless, most existing traffic prediction approaches were developed for highway networks, which may not suitable for more complicated urban networks. In urban scenarios, traffic environments and patterns are more unpredictable and complex, which makes it difficult to use simple traffic models to predict traffic in a fast and accurate manner. Thus, some advanced and complex modeling tools are required to design effective approaches for accurate, fast, scalable traffic prediction in urban scenarios so that accurate reaction to the change in traffic flows can be carried out promptly \cite{5G_add1,5G_add2}.

\begin{table}[t]
\begin{center}
\caption{Specifications of LTE and $5$G \cite{qorvo_2017,instruments_2018,LA-1}. SC-FDMA: single-carrier frequency division multiple access; CP-OFDM: cyclic-prefix orthogonal frequency division multiplexing.}
\label{tab:t2}
\begin{tabular}{|c|c|c|} \hline
  & LTE/LTE-A & 5G \\
 \hline
 \hline
Round trip latency &  15ms & 1ms \\
 \hline
Peak data rate  & 1Gbps &  20Gbps    \\
 \hline
 Available spectrum  & 3GHz &  30GHz    \\
 \hline
 Channel bandwidth & 20MHz  &   \begin{tabular}{c}
       100MHz below 6GHz \\
      400MHz above 6GHz
 \end{tabular}    \\
 \hline
 Frequency band & 600MHz to 5.925GHz & 
 600MHz to 80GHz \\
 \hline
 Uplink waveform & \begin{tabular}{c} SC-FDMA \end{tabular} & \begin{tabular}{c} Option for CP-OFDM\end{tabular} \\
\hline
\end{tabular}
\end{center}
\end{table}


\subsection{LPWAN Technologies}
\label{long}
Currently, LPWAN has been driven to fulfill the demand of emerging IoT applications to offer a set of features including wide-area communications and
large-scale connectivity for low power, low cost, and
low data rate devices with certain delay tolerance \cite{LPWAN5}. In general, LPWAN can be divided into two categories, namely unlicensed and licensed LPWAN. In the sequel, we review the most prevailing LPWAN technologies.

\subsubsection{Unlicensed LPWAN}
The unlicensed LPWAN technologies refer to the LPWAN technologies that employ unlicensed spectrum resources over the industrial, scientific, and medical (ISM) band. Thanks to the usage of the unlicensed band, the unlicensed LPWAN providers do not necessarily pay for spectrum licensing, as a result it reduces
the cost of deployments.
For the unlicensed LPWAN, LoRa and Sigfox are the two biggest competitors \cite{LPWAN1,LPWAN4}. 

\begin{enumerate}
\item	LoRa: 
LoRa, stands for Long Range. It is a physical layer LPWAN solution that modulates signals using a spread spectrum technique designed and patented by Semtech Corporation \cite{Lora1}.
Technically, LoRa employs the chirp spread spectrum (CSS) modulation that spreads
a narrow-band signal over a wider channel bandwidth, thus enabling high interference
resilience and also reducing the signal-to-noise-and-interference ratio (SINR) required at a receiver for correct data decoding \cite{Lora2DJ}. The spreading factor of the CSS can be varied from $7$ to $12$, which makes it possible to provide variable
data rates and tradeoff between throughput and coverage range, link robustness, or energy consumption \cite{LPWAN2,LPWAN3}. Specifically, a larger spreading factor allows a longer transmission range but at the expense of lower data rate, and vice versa. Depending on the spreading factor and channel bandwidth, the data rate of LoRa can vary between $50$bps and $300$kbps. In $2015$, a LoRa-based communication protocol called LoRaWAN was standardized by LoRa-Alliance \cite{LoraWAN1}. LoRaWAN is organized in a star-of-stars topology, where gateway devices relay messages between
end-devices and a central network server \cite{LoraWAN2}. In LoRaWAN, three types of devices (Class A, B, and
C) with different capabilities are defined \cite{Lora3}. In particular, Class A
is the class of LoRaWAN devices with the lowest power
consumption that only require short downlink communication, and Class A devices use pure-ALOHA RA (please refer to Appendix \ref{A_Aloha} for more details and explanations on ALOHA protocols) for the uplink. 
Class B devices are designed for applications
with extra downlink transmission demands. 
In contrast, Class C devices have continuously receive slots, thus always listening to the channel except when they need to transmit. Among the three classes, all the devices must be compatible with Class A \cite{LoraWAN2}. 

\begin{table*}[t]
	\begin{center}
	\caption{Comparison of Sigfox and LoRa \cite{LPWAN2,LoraWAN2,Mobility_Lora}. DBPSK: differential binary phase shift keying}.\label{table2DJ} \centering
	\begin{tabular}{|c|c|c|}\hline
		
		 & \textbf{Sigfox} & \textbf{LoRa} \\
	
		\hline
		\hline
		 RA protocol  & ALOHA & ALOHA/Slotted-ALOHA\\

		\hline
		Modulation type & GFSK/DBPSK  & CSS \\

         \hline
		Frequency & Unlicensed ISM bands  & Unlicensed ISM bands\\

        \hline
		Bandwidth & $100$Hz  & $125$kHz and $250$kHz\\

        \hline
        Bidirectional & Limited/Half-duplex & Half-duplex\\

		\hline
		Link budget & $156$dB & $164$dB\\
		\hline
		Maximum data rate & $100$bps & $50$kbps\\

        \hline
        Maximum payload length & $12$bytes & $243$bytes\\

		\hline
		Coverage & $10$km (urban), $50$km (rural) & $5$km (urban), $20$km (rural)\\
        \hline
        Interference immunity & Very high & high \\
        \hline
		Battery life & $10$ years& $10$ years\\
		\hline
		Localization  & Yes  & Yes \\
			\hline
        Mobility &  No &  Yes \\

	

		\hline
	\end{tabular}
	\end{center}
\end{table*}

\item	Sigfox:
SigFox is another dominant unlicensed LPWAN solution on the market \cite{Sigfox1DJ}. 
SigFox proposes to use an ultra narrow-band (UNB) technology with only $100$Hz bandwidth for very short-payload transmission.
Thanks to the UNB technology, Sigfox enables less power consumption for devices and supports a wider coverage compared with LoRA at the cost of a lower data rate \cite{Sigfox2}.
Sigfox was initially introduced to
support only uplink communication, but later it evolved to
a bidirectional technology with a significant link asymmetry \cite{Sigfox3}. However, the
downlink transmission can only be triggered following an uplink transmission. In addition,
the uplink message number is constrained to $140$ per day and the maximum payload length for each uplink message is limited to $12$bytes \cite{LPWAN2}.
Due to these inflexible restrictions, together with its unopened business network model \cite{LPWAN3}, Sigfox has unfortunately 
shifted the interest of academia and industry to its competitor LoRaWAN, which is considered more flexible and open. In Table \ref{table2DJ}, the characteristics of Sigfox and LoRa are summarized.
\end{enumerate}

\subsubsection{Licensed LPWAN}
As a counterpart of the unlicensed LPWAN above mentioned, we briefly review the licensed LPWAN technologies in this subsection.
The licensed LPWAN refers to the LPWAN technologies using the licensed spectrum resources. They are standardized by the $3$GPP.
For the licensed LPWAN, LTE-M and NB-IoT are the two most promising standards that are introduced in $3$GPP Rel-$13$ in $2016$ \cite{LTEM2,LTEM3}. 

Since both standards are developed based on LTE, their RA procedures are compatible with that in LTE. Generally speaking, the RA procedure refers to all the procedures when a device needs to set up a radio link with the BS for data transmission and reception. In LTE, a contention-based RA procedure used on physical random access channel (PRACH) is specified for initial access \cite{PRACH1}. The PRACH consists of four-handshaking steps. In step $1$, each accessing device randomly selects a preamble from a predetermined preamble pool of size $54$. Preamble collision may occur since multiple devices may select the same preamble. However, the BS can only detect if a specific preamble is active or not in this step. In step $2$, the BS sends a RA response corresponding to each detected preamble. After receiving the RA response in step $3$,
each device sends a radio resource control (RRC) request for its data transmission. In the case of preamble collision, all the collided devices use the same resource to send their RRC request and this collision will be detected by the BS. In step $4$, contention resolution procedure is employed to resolve the collision, where all collided devices need to make a new access attempt with backoff. Since the PRACH operation is based on ALOHA-type access, its capacity is very limited \cite{PRACH1,PRACH2}.

In the following, we briefly review the two licensed LPWAN technologies for long-range connectivity.
\begin{enumerate}
 
\item	LTE-M:
LTE-M is fully compatible with existing cellular networks \cite{LTEM}. It can be considered a simplified version of LTE intending for low device cost and low power consumption IoT applications \cite{LTEM1}. The key features of LTE-M are the support of mobile MTC use-cases and voice over networks \cite{LTEM6}.
LTE-M uses orthogonal frequency division multiple access (OFDMA) in the downlink and multi-tone SC-FDMA in the uplink. To reduce hardware cost and complexity, LTE-M has a bandwidth of $1.4$MHz and typically supports one receive-antenna chain and half-duplex operations (full-duplex operations are also allowed).
In $3$GPP Rel-$14$ and Rel-$15$, new features have been proposed to enhance the performance of LTE-M in terms of data rate, latency, positioning, and voice coverage \cite{LTEM4,LTEM5}. For example, in $3$GPP Rel-$15$, coverage enhancement for higher device velocity (e.g. $200$km/h) was proposed and techniques such as wake-up signal/channel and relaxed monitoring for cell reselection during RA were used to reduce latency and power consumption. 
\begin{table*}[t]
	\begin{center}
	\caption{Comparison of LTE-M and NB-IoT \cite{LTEM6,LPWAN2,NB-IOT,NBIOT3}.}\label{table3DJ} \centering
	\begin{tabular}{|c|c|c|}\hline
		
		 & \textbf{LTE-M} & \textbf{NB-IoT} \\
	
		\hline
		\hline
		\begin{tabular}{c}
       RA protocol \\
      (based on PRACH)
 \end{tabular}  & Slotted-ALOHA & Slotted-ALOHA\\

		\hline
		Modulation type &  QPSK/QAM & BPSK/QPSK\\

         \hline
		Frequency & Licensed LTE bands  & Licensed LTE bands\\

        \hline
		Bandwidth & $1.4$MHz & $200$kHz\\

        \hline
        Bidirectional & Full/Half-duplex & Half-duplex\\
        \hline
        Link budget & $153$dB & $164$dB\\
		\hline
		Maximum data rate & $1$Mbps & $250$kbps\\

        \hline
        Maximum payload length & $1000$bits & $1000$bits\\

		\hline
		Coverage & Few kilometers & $1$km (urban), $10$km (rural)\\
        \hline
        Interference immunity & Low & Low \\
        \hline
		Battery life & $10$ years& $10$ years\\
		\hline
		Localization  &  Yes &  Yes \\

		\hline
        Mobility &  Yes &  Yes \\
		\hline
	\end{tabular}
	\end{center}
\end{table*}

\item	NB-IoT:
Compared with LTE-M, NB-IoT is a system built on the existing LTE functionality with a single narrow-band of $200$kHz with low baseband complexity, which aims at supporting wider coverage, lower device cost, longer battery life, and higher connection density \cite{NBIOT1}\cite{NBIOT3}. To be more specific, we compare the characteristics of LTE-M and NB-IoT in Table \ref{table3DJ}.
Like LTE-M, NB-IoT can coexist with the existing LTE networks, which can utilize the existing network hardware and reduce the deployment cost therefore \cite{rico2016overview,hoglund2017overview}. NB-IoT also uses OFDMA with $15$kHz subcarrier spacing in the downlink and SC-FDMA with both $15$kHz and $3.75$kHz subcarrier spacings in the uplink \cite{lee2017prediction}. Different from LTE-M, both single-tone and multi-tone SC-FDMA can be used for NB-IoT \cite{Add_Survey2} but only half-duplex operations are supported by NB-IoT. Compared to LTE-M and legacy LTE, NB-IoT has extended coverage and deep penetration in buildings and hard-to-reach areas, thanks to its narrow bandwidth and low date rate. Technically, the coverage target of NB-IoT has a link budget of $164$dB, whereas the LTE link budget is $142$dB \cite{NB-IOT, NB-IoT_Penetrate}. The $20$dB link budget margin can significantly increase the coverage range in an open environment and compensate the penetration losses caused by walls of a building to ensure high quality communication.
In addition, NB-IoT has three operation modes such as in-band, standalone, and guard-band, as illustrated in Figure \ref{fig:fignb}. In in-band mode,
one or more LTE physical resource blocks (PRBs) within an
LTE carrier are reserved for NB-IoT. In standalone
mode, NB-IoT can be deployed within one
or multiple global systems for mobile communications (GSM) carriers. In guard-band mode, NB-IoT can be utilized within the guard-band of an LTE carrier \cite{NBIOT2}. 
To prolong battery life, two main power-efficiency mechanisms are supported in NB-IoT and LTE-M, namely power saving mode (PSM) and expanded discontinuous reception (eDRX) \cite{LTEM, lee2017prediction}. In particular, PSM keeps a device registered with network, but allows it to turn off the functionalities of paging listening and
link quality measurements for energy saving. On the other hand, eDRX
allows a device to negotiate with a network when it can sleep, during which the device can turn off the receiving functionality for energy saving. Both mechanisms allow to repeat transmissions
for latency-tolerant devices to extend network coverage \cite{Add_Survey2}.

\begin{figure}[!ht]
\centering
  \includegraphics[width=8.5cm]{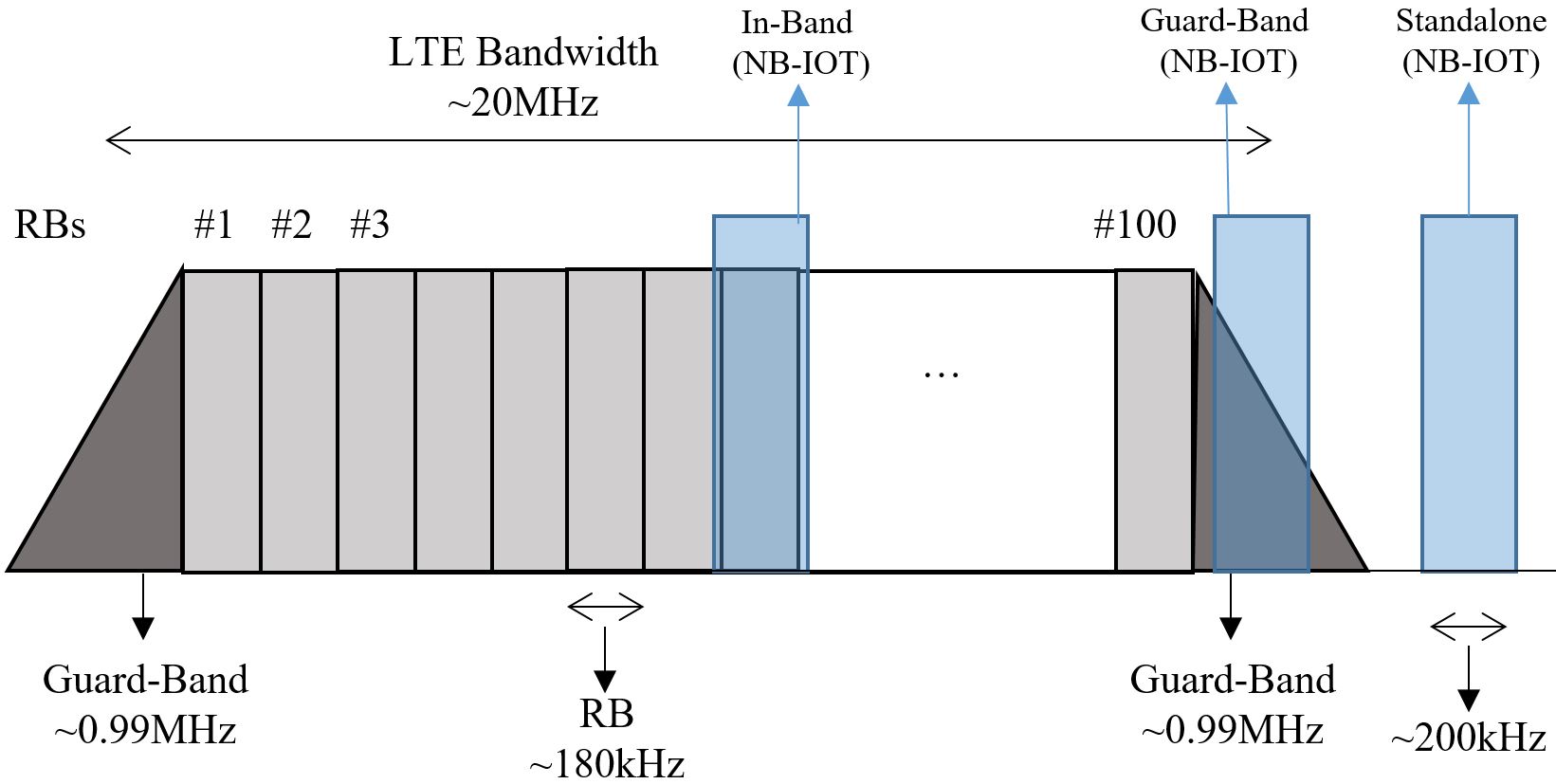}%
  \caption{In-band, standalone, and guard-band deployment of the NB-IoT in an LTE component carrier with $20$MHz bandwidth \cite{lee2017prediction}. }
  \label{fig:fignb}
\end{figure}



\end{enumerate}

The two kinds of LPWAN technologies, i.e., unlicensed and licensed LPWANs, have different features and advantages. For example, since unlicensed LPWAN uses ISM bands, this
fact favours the deployment of private BSs without the involvement of any mobile operators, but it is difficult to provide guaranteed performance due to the signals that become interferers in ISM bands. On the other hand, since licensed LPWAN is part of cellular systems, certain performance can be guaranteed using resource allocation, while its deployment and device cost are comparably higher \cite{LPWAN2}.


As mentioned earlier, both short-range and long-range technologies can be employed for various IoT applications. For example, home IoT applications can be supported using short-range technologies (e.g., WiFi), and small-scale wireless sensor networks (WSN) (e.g., specific indoor health applications) can be implemented using ZigBee.
For high data rate and low latency indoor applications such as Tactile Internet, OWC technologies can be used. 
However, to support a tremendous number of devices deployed over a large area, it is necessary to rely on long-range technologies. For multimedia and ultra reliable low latency applications, LTE and $5$G can be effectively employed to support their connectivities. For environmental monitoring and smart farming to cover a wide area (e.g., a city or a suburb), unlicensed LPWAN technologies can be used. Licensed LPWAN technologies would be required for nation-wide IoT applications that require unified supports (e.g., the connectivity for smart meters in smart grid/smart cities). 

\section{Emerging Wireless Technologies for Massive Connectivity}
\label{section4}

\begin{figure}[h]
\centering
  \includegraphics[width=8.5cm]{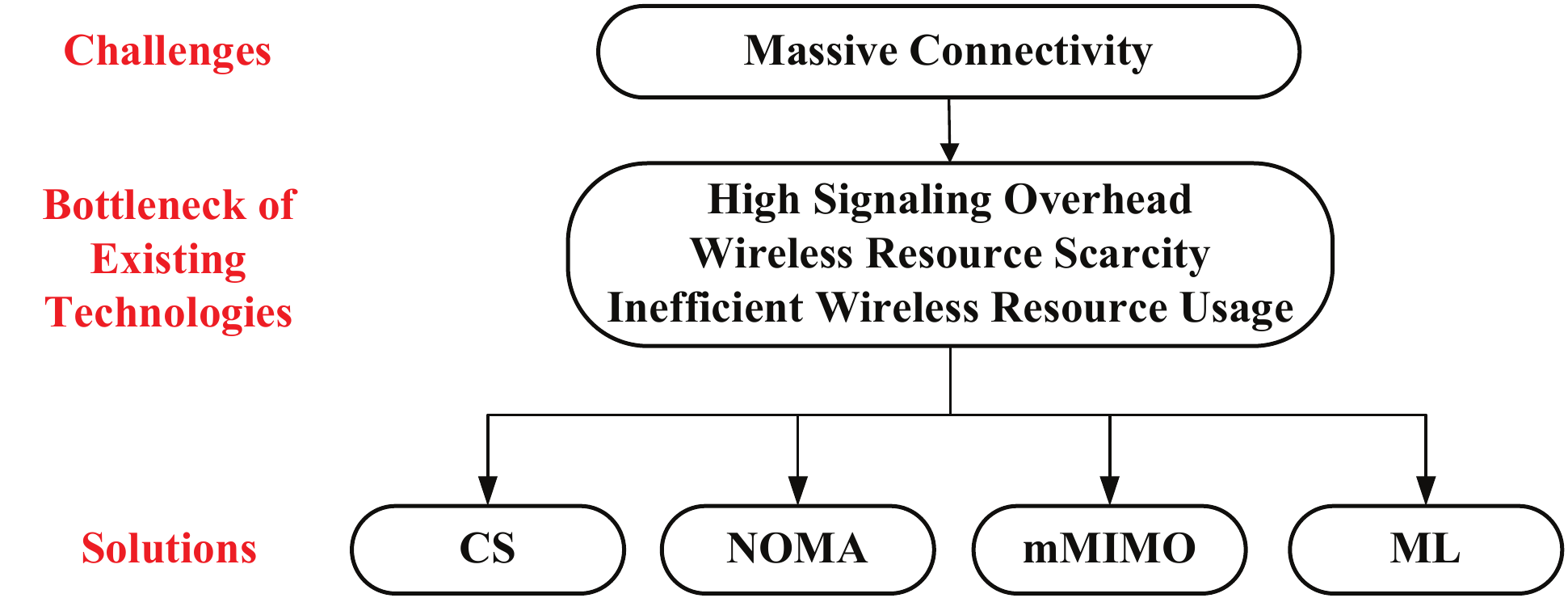}%
  \caption{Diagram of the challenge for the existing technologies and the promising solutions. }%
  \label{Challenge}
\end{figure}

Although the existing wireless IoT technologies have led to some success in supporting various IoT applications, there are still open issues and difficulties to meet the foreseeable needs of future IoT applications with hundreds of billion objects or things to be connected. One of the critical challenges is to accommodate massive connectivity from IoT devices with small-sized transmission payloads and sporadic features \cite{HTCvsMTC}\cite{Chap4_1}.
In fact, the RA protocols of the existing technologies are mainly based on ALOHA or CSMA/CA \cite{mMIMO11}, which is highly likely to cause severe access collision, increased latency, and high signalling overhead for IoT devices. Moreover, only limited wireless resources are allocated for IoT connectivity and these resources are used in an orthogonal manner, which results in wireless resource scarcity and inefficient wireless resource usage for massive connectivity. In Figure \ref{Challenge}, we summarize the main bottleneck of existing technologies for enabling future IoT connectivity.
To address these issues, ongoing efforts have been made to develop new technologies that can address the shortcomings of the existing technologies while maintaining their good characteristics. 
In this survey, an overview of four promising technologies such as CS, NOMA, mMIMO, and ML that can effectively resolve wireless resource scarcity and enhance spectrum usage efficiency is provided.

\subsection{CS based IoT Connectivity}
\begin{figure}[!ht]
\centering
  \includegraphics[width=8.5cm]{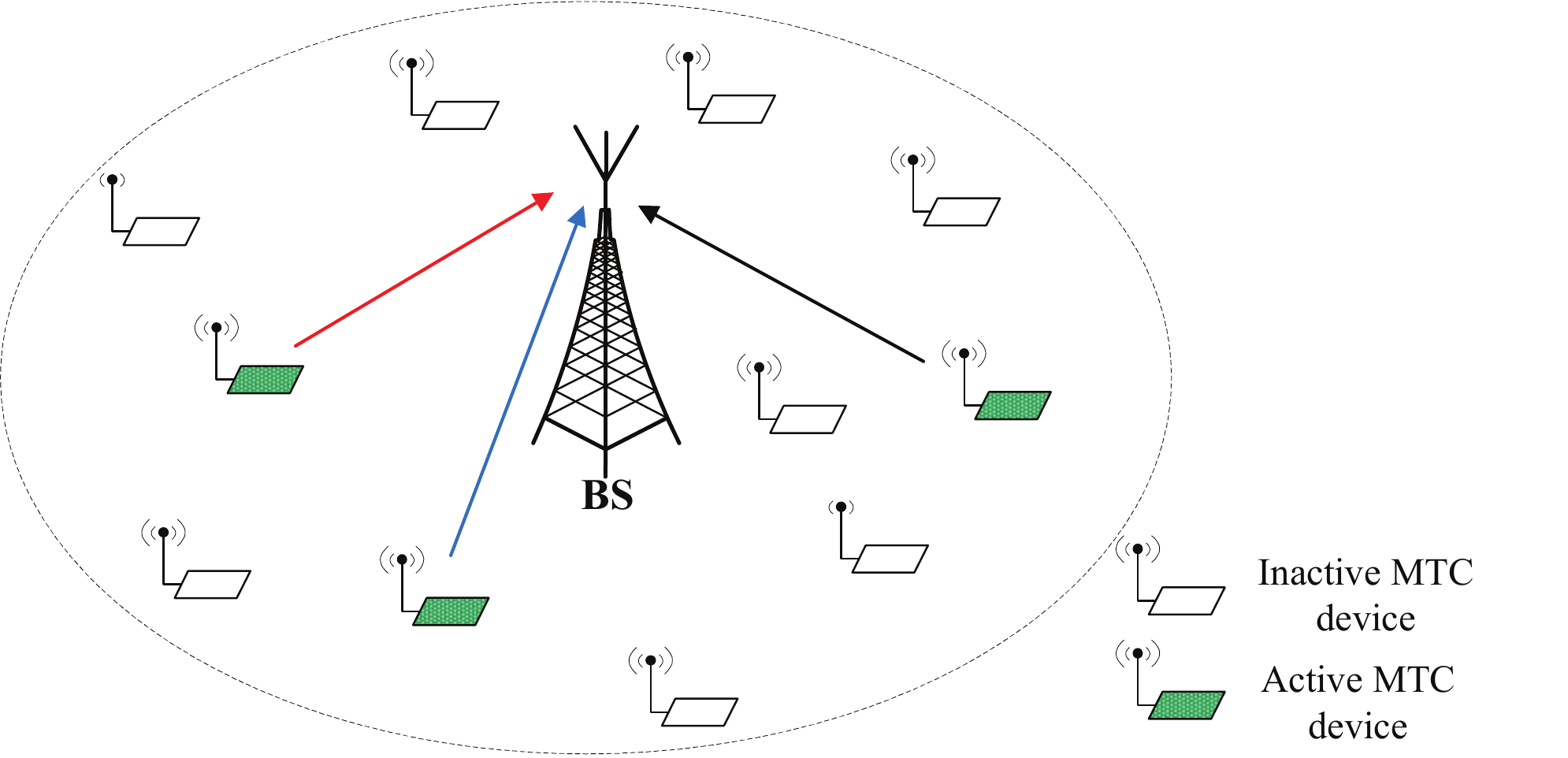}
  \caption{Illustration of sparse user activity in massive MTC. }%
  \label{CS}
\end{figure}
In order to reduce signalling overhead, grant-free RA has been proposed \cite{mMIMO17}, which does
not employ handshaking process that is employed in existing licensed LPWAN standards, i.e., LTE-M and NB-IoT (the request-grant procedure is thus omitted). In general, the grant-free RA enables IoT devices to contend with their uplink payloads directly by transmitting preamble along with data. By utilizing the natural feature of sporadic traffics in MTC, as shown in Figure \ref{CS}, various compressive grant-free RA (cGFRA) schemes have been proposed \cite{CS1,CS4,CS2,CS7,CS3}, where the sparse device activity is exploited to develop efficient multiple signal detection schemes based on CS algorithms (please refer to Appendix \ref{A_Cs} for more details and explanations on CS principle) \cite{CS5,CS6}. In particular,
cGFRA schemes have been studied where the wireless signal of each device is spread by
a unique sequence \cite{CS1,CS4}. In \cite{CS2}, sparse sequences were used instead of binary sequences for data
signal spreading in order to increase the number of MTC devices and allow device identification.
In \cite{CS7}, multiple resource blocks were used to reduce the preamble collision and improve the mGFRA throughput.
In \cite{CS3}, another cGFRA scheme was proposed where each device’s channel impulse response
is used as a unique signature to differentiate signals that are simultaneously transmitted.

Although cGFRA is well-suited to MTC with low signalling overhead to some extent, its high complexity resulted from the CS algorithm is still an issue to be addressed. In general, the complexity of cGFRA algorithm is proportional to the total number of MTC devices in a cell. In massive access with a large number of MTC devices, its complexity would be prohibitive. Thus, a low-complexity cGFRA is highly desirable for massive access.
In addition, cGFRA usually requires a bandwidth expansion to increase the number of MTC devices that can be supported simultaneously.
To efficiently utilize wireless resources and also to address the wireless resource scarcity for supporting massive access, advanced technologies such as NOMA and mMIMO have been developed, which will be introduced in the following.

\subsection{NOMA based IoT Connectivity}
\begin{figure}[!ht]
\centering
  \includegraphics[width=8.5cm]{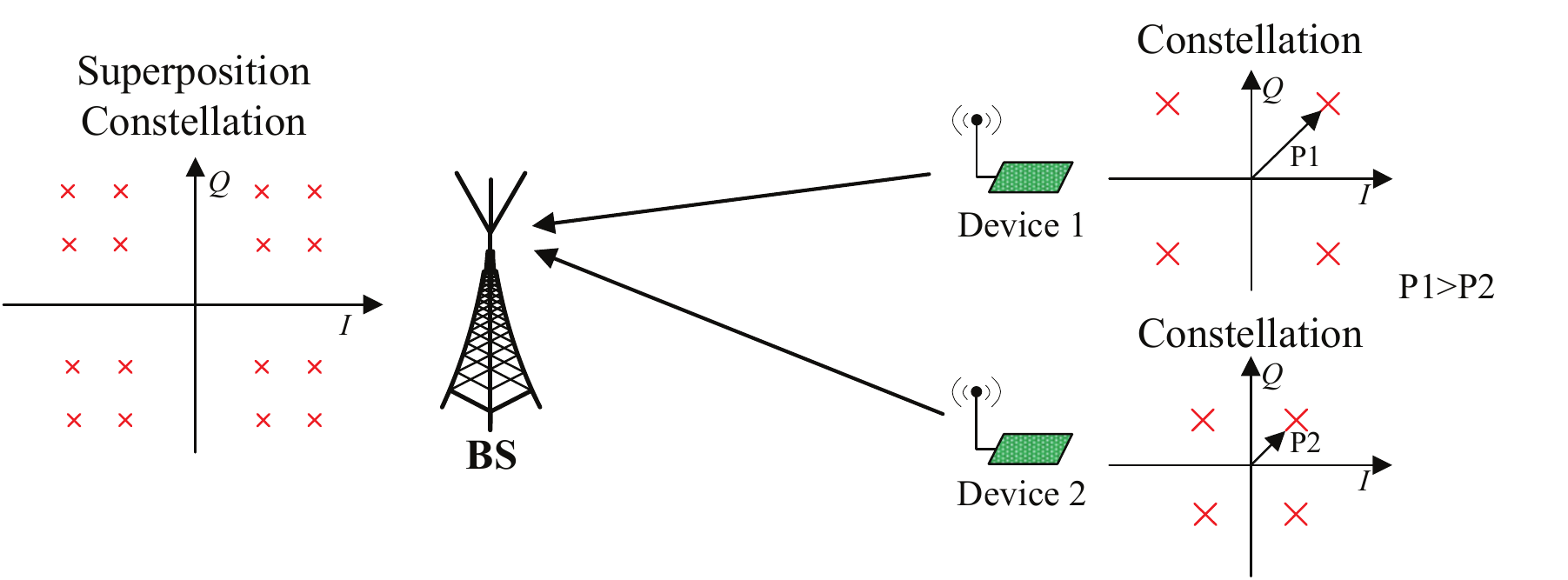}
  \caption{Simplified power-domain NOMA systems. }%
  \label{NOMA}
\end{figure}
    NOMA has recently been identified as a promising technology to make more efficient use of wireless resources \cite{NOMA4,NOMA5,NOMA6,NOMA7,choi2008h,mm-NOMA}. The key idea of NOMA for massive access is to allow overlapping among signals over the same time-frequency resource via power-domain multiplexing (PDM) or code-domain multiplexing (CDM), and to employ successive interference cancellation (SIC) at a BS to perform a separate decoding for each device \cite{NOMA1,NOMA2,NOMA3}. Figure \ref{NOMA} illustrates the basic principle of power-domain NOMA in uplink transmission. Specifically, at the BS, the strong signal from device $1$ is first decoded and removed by using SIC in the presence of the interfering signal from device $2$, which is a weak signal. Then, the weak signal, i.e, the signal from device $2$, is decoded (please refer to Appendix \ref{A_Noma} for more details and explanations on NOMA principle). The main benefit of power-domain NOMA for MTC is enabling multiple devices to perform grant-free access in the same time-frequency resource simultaneously without bandwidth spreading \cite{NOMA11,NOMA12,NOMA8,NOMA13,NOMA9,NOMA10,NOMA14}. Specifically, in \cite{NOMA11} and \cite{NOMA12}, NOMA-based RA has been investigated with multichannel ALOHA to improve the throughput for MTC. It was shown that the NOMA-based RA with multichannel ALOHA is suitable for MTC when the number of multiple access channels is limited. This is mainly due to the fact that NOMA can effectively increase the number of multiple access channels without any bandwidth expansion.
In \cite{NOMA8}, the energy efficiency of NOMA for MTC was studied, and it was shown that transmitting with minimum rate and
full time is optimal in terms of energy efficiency. In \cite{NOMA13}, a power control algorithm of NOMA was proposed to improve the energy efficiency by employing game theory. In \cite{NOMA9}, a MIMO-NOMA strategy has been designed for MTC, where two users are clustered to meet the service demands of one user while the other user is served opportunistically. In \cite{NOMA10},
a joint sub-carrier and transmission power allocation problem were considered and solved to maximize the number of MTC devices and satisfy the transmission power constraints. In \cite{NOMA14}, NOMA to cGFRA was adopted to improve the performance of cGFRA and it was revealed that the number of incorrectly detected device activity can be reduced by applying NOMA to cGFRA. In \cite{NOMA16},
a low-complexity dynamic cGFRA for NOMA was proposed to jointly realize user activities and data detections. It was shown that the proposed scheme can achieve much better performance than that of the conventional cGFRA.

Although all these works indicated that NOMA is a promising technology to enable grant-free massive access for emerging MTC standards, there are still challenges to be addressed to enable its implementation \cite{NOMA1}. For example, designing
appropriate detection algorithms and decoding strategies to increase the number of pairs of devices and suppress the error propagation is important at the stage of SIC in power-domain NOMA. On the other hand, optimizing factor graph needs to be considered for a good trade-off between overloading factor and receiver complexity in code-domain NOMA.

\subsection{mMIMO based IoT Connectivity}
\begin{figure}[!ht]
\centering
  \includegraphics[width=8.5cm]{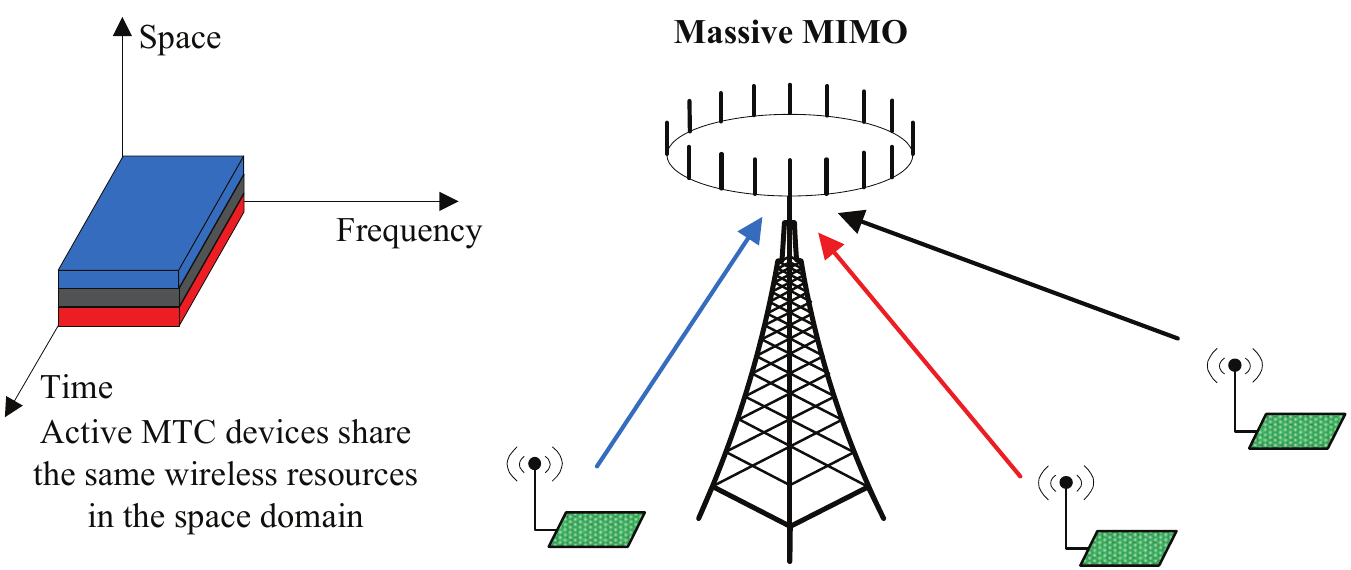}%
  \caption{Illustration of mMIMO systems. }%
  \label{mMIMO}
\end{figure}
\begin{table*}[t]
	\begin{center}
	\caption{Summary of Strengths and Limitations of the Promising Technologies for Massive Connectivity}
	\label{Comparison}
	\begin{tabular}{|m{2cm}<{\centering}|p{5cm}|p{5cm}|}\hline

		Technologies & Strengths & Limitations \\

		\hline
		\hline
		 CS  &  Efficient multiple signal detection schemes can be developed by exploiting sparse device activity in MTC. & 1) High complexity in massive access with a large number of MTC devices; 2) bandwidth expansion required to increase the number of MTC devices that can be supported simultaneously.\\

		\hline
		NOMA &  Allowing overlapping among signals over the same time-frequency resource via PDM or CDM.  & 1) Error propagation at the stage of SIC in power-domain NOMA; 2) trade-off between overloading factor and receiver complexity in code-domain NOMA needs to be optimzied\\

         \hline
		mMIMO & Thanks to favorable propagation, wireless resources in the spatial domain can be exploited to support a large number of devices simultaneously. & 1) preamble that has large preamble space but low mutual correlation needs to be designed; 2) array dimensions and hardware cost need to be considered when the number of antennas is large. \\

        \hline
		ML & Dynamic patterns of the wireless environment that are too complex to be modelled could be effectively explored.  & 1) Trade-off between the algorithm’s computational
		requirements and the learned model’s accuracy needs to be well designed; 2) it could be time consuming, which may not
		suitable for highly dynamic environments. \\

        \hline
       
	\end{tabular}
	\end{center}
\end{table*}
 Besides NOMA, mMIMO is another promising technology to mitigate wireless resource scarcity and handle the rapid growth of data traffics for $5$G and future wireless communications (please refer to Appendix \ref{A_mMIMO} for more details and explanations on mMIMO principle) \cite{mMIMO1,mMIMO2,mMIMO3,mMIMO4,mMIMO5}.
 Compared to NOMA, mMIMO exploits wireless resources in the spatial domain that can afford a large number of MTC devices, as shown in Figure \ref{mMIMO}. In a typical mMIMO, a great number of antennas are employed at the BS. Thanks to it, the channel responses between different devices tend to be orthogonal to each other. By taking advantage of this property, a large number of devices in the same time-frequency resource could be simultaneously accommodated in an efficient way. Diverse research works have shown that mMIMO can significantly improve the performance of HTC in terms of spectral efficiency \cite{mMIMO6,mMIMO7}, energy efficiency \cite{mMIMO8,mMIMO9}, and coverage \cite{mMIMO10}. For example, as shown in \cite{mMIMO2}, when the BS employs $100$ antennas to serve $40$ users, mMIMO can increase the spectrum efficiency $10$ times or more and simultaneously, improve
the radiated energy-efficiency in the order of $100$ times by using conjugate beamforming, compared to the single-antenna single-user counterpart.
Several modifications and improvements of traditional PRACH by using mMIMO have also been proposed to support MTC \cite{mMIMO12,mMIMO13,mMIMO14,mMIMO15,mMIMO16}. These works validate the effectiveness of mMIMO in resolving access collision, reducing access delay, and enhancing RA capacity in MTC.
To more efficiently accommodate massive access with low signalling overhead and access delay, mMIMO based grant-free RA (mGFRA) has been proposed as a compelling candidate for future IoT \cite{mMIMO18}.
Recently, performance analyses on mGFRA have been conducted with respect to spectral efficiency \cite{mMIMO19,mMIMO20,mMIMO21}, success probability \cite{mMIMO18}, user activity detection and channel estimation \cite{mMIMO22,mMIMO23}. Although all these research works confirmed that mMIMO is a viable and effective enabler for emerging MTC applications in IoT, they also revealed that preamble is of prime importance in mGFRA because it
not only enables RA device differentiation but also dominants the
accuracy of channel estimation, which is essential for successful data transmissions of RA devices. In general, there are two types of preambles considered in mGFRA, namely orthogonal \cite{mMIMO18,mMIMO19} and non-orthogonal preambles \cite{mMIMO22,mMIMO23}. Compared with the non-orthogonal counterpart, orthogonal
preamble detection is much more simple and effective and the
channel estimation is relatively more accurate, thanks to the
orthogonality of preambles. Nevertheless, preamble collision constraints its performance due to the limited orthogonal-preamble space. On the other hand, non-orthogonal
preamble can alleviate the preamble collision since it has larger preamble space, but its channel estimation would be affected due to non-orthogonality of preambles. Thus, designing preamble that has large preamble space but low mutual correlation is desirable in mGFRA.

On the other hand, since the number of MTC devices that could be supported by mMIMO grows with the number of antennas \cite{mMIMO18}, it is expected that hundreds or thousands of antennas are used to support massive access in various IoT applications.
However, considering the array dimensions and hardware cost, gathering massive antennas in a centralized way might become impractical. Alternatively, distributed mMIMO \cite {mMIMO24} could be a viable candidate for future IoT. Specifically, compared with the centralized scenario that a BS is essentially surrounded by devices, in distributed scenario antennas are distributed over a large geographical area so that each device is surrounded by a few antennas. A number of research works have demonstrated the performance superiority of distributed mMIMO over the traditional centralized mMIMO from different perspectives \cite {mMIMO25,mMIMO26,mMIMO27}. Nevertheless, for emerging MTC applications, only a little research has been done to discover the potential of distributed MIMO for massive access so far, for example, \cite {mMIMO28} and \cite {mMIMO29} have provided preliminary analysis on the performance of GFRA in distributed mMIMO.

\subsection{Machine Learning-assisted IoT Connectivity}
In general, machine learning (ML) algorithms can be divided into four categories, namely supervised learning, semi-supervised learning, unsupervised learning, and reinforcement learning (RL). Each category has its specific applications \cite{ML0}.
Recently, ML algorithms \cite{ML3,ML4,ML2} have drawn much attention to address various issues in wireless communications including link adaptation \cite{ML5,ML6}, traffic control \cite{ML7,ML1}, and resource allocation \cite{ML8,ML9}. 

In fact, ML is a very powerful tool that can be used to improve inefficient wireless resource usage in IoT since the resource allocation optimization related problems are usually too complex to be modelled due to the dynamic wireless environments. However, dynamic patterns of the wireless environment could be effectively explored by ML with much lower complexity than using optimization technologies. For this reason, several works have applied ML to address the challenges in the massive access for emerging MTC applications. In \cite{Chap4_1}, an RL scheme was developed to avoid access network congestion and minimize the packet delay by allocating MTC devices to appropriate BSs. In \cite{ML10}, a Q-learning algorithm (one of RL techniques) for the selection of appropriate BS for the MTC devices was proposed. With the algorithm, MTC devices are able to adapt to dynamic network traffic conditions and decide which BS is the best to be selected based on the QoS parameters. In \cite{ML11}, a Q-learning assisted PRACH scheme was proposed to control MTC traffics with the objective of reducing its impact on the mobile cellular networks. In \cite{ML13}, an online hierarchical stochastic learning algorithm was proposed to determine the access decision for MTC devices. In \cite{ML12}, the authors proposed an adaptive access control scheme by using Q-learning algorithm to solve the massive access problem. 

All these research works revealed that RL technique can be used as an efficient resource scheduler to address massive access problems \cite{ML14}. Nevertheless, there are limitations that need to be considered. For example, a trade-off between the RL algorithm's computational
requirements and the learned model's accuracy needs to be well designed, since the higher the required accuracy is, the higher the computational requirements will be, and as a result the higher energy consumption will be.
In addition,
the learning agent's observations may contain strong temporal correlations and the convergence to the steady
state can be time consuming, which may not
suitable for highly dynamic environments.

\begin{figure*}[ht]
\centering
  \includegraphics[width=11cm, height=7.5cm]{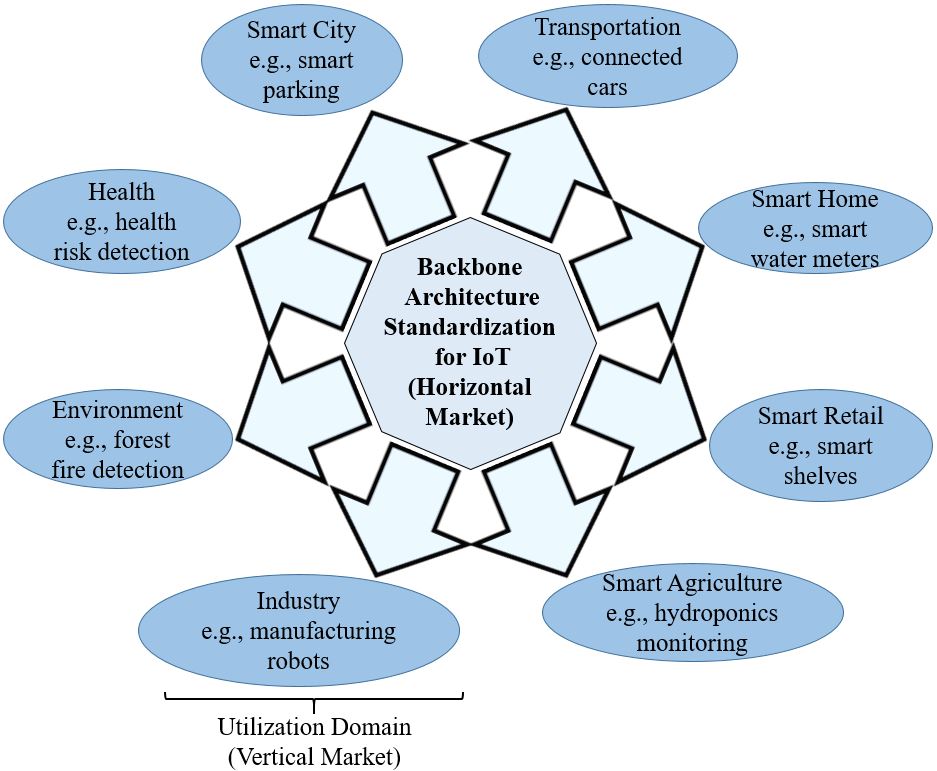}%
  \caption{Top IoT applications and their utilization domains. }%
  \label{fig:apps}
\end{figure*}

In summary, all the aforementioned technologies have the potential to be employed in future standards for IoT connectivity. Nevertheless, there are also open issues and limitations that need to be addressed for their implementation. In Table \ref{Comparison}, their strengths and limitations are highlighted. Additionally, all these emerging technologies can be not only employed to support massive connectivity, but also can be utilized to provide high reliability and low latency transmissions.
In the future, it is expected that more and more advanced technologies can be developed to address various critical challenges for IoT. In the meantime, efforts also need to be made to smartly merge the existing and emerging technologies to achieve their full potential and maximize the system performance.


\section{Classification of IoT Applications}
\label{section5}
In this section, we classify the current and future IoT applications with respect to their requirements and then identify the feasible connectivity technologies for each application category. In order to fulfil this task, first, the conventional classification of IoT applications is reviewed and then a different classification is described.

Over the last few years, in the vertical market, most of the applications are classified with respect to their utilization domains (e.g., \cite{APP-1,APP-3,APP-4,APP-5,APP-6,App-7,APP-8}). Some examples of the utilization domains in the vertical market are as follows: transportation, smart city, health-care, agriculture, environment, retail, and smart home \cite{Tr-1,Tr-2,Tr-3,Tr-5,App-2,Tr-6}. Figure \ref{fig:apps} partially illustrates the main utilization domains and their applications. However, the classification of IoT applications based on their utilization domains may result in some conflicts and overlaps \cite{OR-1}. As an example, a sensor for humidity measurement can be considered in multiple utilization domains such as industry, smart agriculture, or even smart environment. 

In order to avoid this kind of overlaps and create a straightforward pathway to identify the IoT applications categories based on their technical requirements and consequently find the nominated technologies suitable for them, we consider a different classification of IoT applications. We first focus on end-user-types of applications and then take other application requirements (i.e., data-rate, latency, coverage, power, reliability, and mobility) into account to generalize our classification. The end-user-type classification, similar to the classification used in \cite{OR-1}, is illustrated in Figure \ref{fig:fig2}. Contrary to the classic utilization-domain-classification in \cite{APP-1,APP-3,APP-4,APP-5,APP-6,App-7,APP-8}, the classification we use in our paper focuses on end-user-type for each application to classify it into one of the two main categories of human-oriented or machine-oriented applications. Human plays an essential role in human-oriented applications while machine-oriented applications automatically manage their tasks without requiring human intervention. Figure \ref{fig:fig2} shows that the aggregation of the IoT applications is mostly in machine-oriented applications. As a result, high connectivity density is required that is an important challenging topic in IoT, which comes alongside future massive machine-oriented applications/sensors. It primarily causes a high competition among smart devices to access the limited bandwidth capacity for creating a wireless data delivery link. On the other hand, sharing the available bandwidth among a massive number of devices requires more promising technologies. CS-based IoT connectivity, NOMA, and mMIMO technologies are such emerging technologies that can be deployed for such massive connectivity.

In the following two subsections, the human-oriented and machine-oriented IoT applications are reviewed.

\subsection{Human-Oriented IoT Applications}
\begin{figure*}[ht]
\centering
\captionsetup{width=1\linewidth}
  \includegraphics[width=13cm, height=7cm]{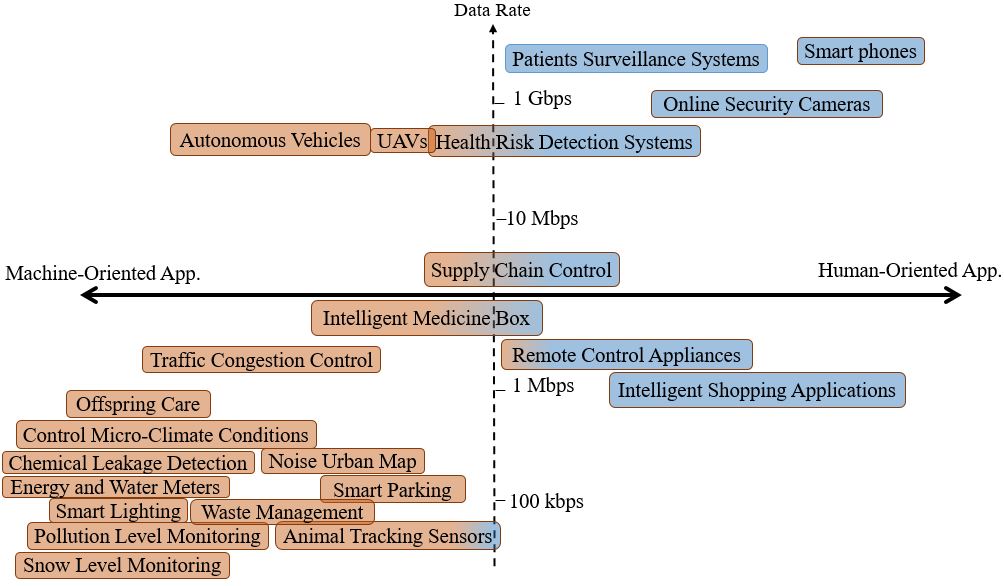}%
  \caption{Classification of IoT applications based on end-user-type and data rate. }%
  \label{fig:fig2}
\end{figure*}

Human-oriented IoT applications refer to the applications that require human interaction to communicate with a network. 
As shown in Figure \ref{fig:fig2}, conventional smartphones, security cameras, patient surveillance systems are three examples of human-oriented IoT applications \cite{PS,SC}. Authors in \cite{MR2,MR1} provided a wide range of human-oriented applications and also evaluated the role of the human in interaction with the machines. These applications usually provide a visualization to present information in an intuitive and easy-to-understand way \cite{MR2,lee15} and/or accept interaction based on natural language, e.g., through voice commands, to understand basic human orders and/or respond properly \cite{MR1}. 
Human-oriented applications are generally characterized by high data rates (i.e., from tens of Mbps up to tens of Gbps) \cite{MH,NB-IOT}.  
However, there are also a few human-oriented applications that require low data rate (e.g., form 1Mbps up to 10Mbps) like intelligent shopping applications that provide information of all items/interactions in a grocery store to the human as its end-user-type \cite{GS}. In addition, one important area in human-oriented applications is pervasive or mobile healthcare like physical activity recognition sensors \cite{rev3}. Due to the rapid increase of wearable devices and smartphones, healthcare is being evolved from conventional hub-based systems to more
personalised healthcare systems \cite{rev2}. However, enabling these kinds of human-oriented applications referring to the smart healthcare applications is significantly challenging in different issues such as cost-effective and accurate medical sensors, the multidimensionality of data, and compatibility with the current infrastructure \cite{rev2,rev3}. Data fusion techniques or ML-assisted IoT connectivities are potential technologies for classifying types of physical activity and removing the application uncertainty \cite{rev3}.

\subsection{Machine-Oriented IoT Applications}
Machine-oriented IoT applications refer to the applications that are able to automatically communicate or interact with each other or a remote server, with minimal human involvement \cite{HO-1,MO-1}.  
In the past, they were only characterized by low data rate (i.e., up to hundreds of kbps) and power consumption such as matured WSN and joint power-information transmission technologies (e.g., RFID systems) \cite{NE-1,MO-3,MR3}.
Even today, most of the applications in this class such as monitoring sensors require low data rates (as shown in Figure \ref{fig:fig2}). However, a new set of machine-oriented applications including autonomous vehicles require higher data rates (e.g., tens of Mbps) with relatively more complex designs \cite{MO-01}. 

It is worth mentioning that some applications such as health risk detection sensors can partially be either machine-oriented or human-oriented application. For instance, a health risk detection sensor can either report the risks to a human as a human-oriented application \cite{H-01} or interact with medical instruments to modify their performance automatically as a machine-oriented application \cite{H-02}. 

\subsection{Nominated Connectivity Technologies for IoT Applications}
In this subsection, first, machine-oriented and human-oriented applications are mapped into certain IoT connectivity technologies. Then, in order to be more specific, the requirements of IoT applications along with their corresponding connectivity technologies are briefly discussed. It is worthwhile to note that the mapping of technologies to the applications is not unilateral always and can be different for a specific application with respect to its unique requirements; however, in this subsection, we focus on general requirements of applications with considering the features of connectivity technologies presented in Sections \rom{2} and \rom{3}.

Most of the machine-oriented applications are suppurated by connectivity technologies such as Bluetooth/BLE, Zigbee, LoRa, and Sigfox, 
while human-oriented applications usually rely on deployment of cellular technologies such as LTE/LTE-A and 5G technologies. Recently, although cellular networks have mostly been utilized to accommodate human-oriented applications, they are being slowly overshadowed by machine-oriented applications \cite{HO-1}. Therefore, cellular technologies are also being considered the potential candidates to provide connectivity for machine-oriented applications. Today, a majority of cellular machine-oriented applications use legacy cellular technologies due to long-life-cycles of sensors \cite{Eric1-1}. However, it is expected to be replaced slowly as a broader range of use-cases evolves over time, along with the continued deployment of supporting LTE-based IoT technologies (e.g., LTE-M and NB-IoT) and future capabilities of 5G networks \cite{Eric}.

From the applications standpoint of view, the main disadvantage of LoRa or Sigfox networks deployment over cellular networks deployment is that they rely on their own IoT infrastructure, system model, and data structure, which results in interoperability issues such as difficulty in connecting different IoT applications exposing cross-platform and cross-domain, and also difficulty to use devices in different IoT platforms \cite{noura2019}. As a result, it is difficult to deploy the emergence of IoT technology at a large-scale. Exploiting cellular networks can provide an interoperable and compatible communication network for a large number of IoT applications. It can enable an IoT application to establish an association with a cellular network when the application is activated by an end-user \cite{3G-2016}. 
Consequently, instead of requiring to build a new and private network architecture to host IoT applications (e.g., LoRa and Sigfox), they can piggyback on the same cellular network as smartphones \cite{IoTALL}.

 It is worth noting that both of human-oriented and machine-oriented IoT applications demand some specific requirements including data rate, latency, coverage, power, reliability, and mobility \cite{3G-B,3G-2016}. Note that these requirements overlap with each other and may cause a trade-off for the application's performance. Therefore, in order to generalize our classification and identify the feasible technologies for more specific applications, we take them into account and describe them as follows.

\subsubsection{Data Rate}
IoT applications can have different data transmission rates from tens of kbps up to tens of Gbps. Three different application groups can be identified in terms of data rate as follows: 1) high data-rate (greater than $10$Mbps), 2) medium data-rate (less than $10$Mbps and greater than $100$kbps), and 3) low data-rate applications (less than $100$ kbps)\cite{NB-IOT}.

\begin{figure}[ht]
\centering
\captionsetup{width=1\linewidth}
  \includegraphics[width=8.5cm]{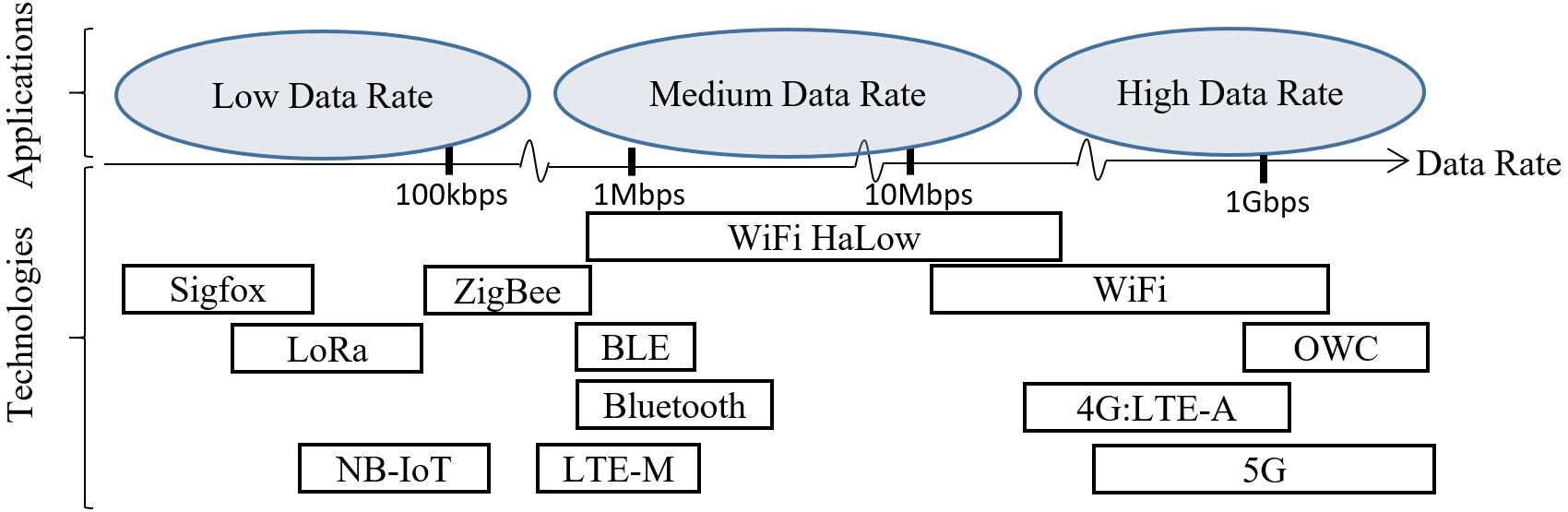}
  \caption{Nominated technologies for low, medium, and high data-rate IoT applications. }%
  \label{fig:fig_data}
\end{figure}

First, high data-rate applications such as streaming video and web applications, and smartphones are usually supported by 4G (LTE/LTE-A), 5G, OWC, and WiFi. These mentioned applications mostly transmit multimedia contents that require high data rate connectivity technologies. Moreover, mmWave wireless communications -- i.e., IEEE 802.15.3c and IEEE 802.11ad -- have recently been developed for short-range but very high data rate applications with up to tens of Gbps because of large availability of bandwidth in mmWave bands \cite{HD-1,HD-3,HD-4}. The complexity of the high data rate applications is relatively high and the market share of them is 10\% \cite{NB-IOT}. 
Second, medium data rate applications such as smart home applications are usually supported by ZigBee, Bluetooth/BLE,  and LTE-M technologies \cite{B1}. Smart home applications include a set of connected devices in homes such as connected cooking systems with medium data rate requirements \cite{KC}. Their design is less complex than high data-rate applications and their market share is estimated to be 30\% \cite{NB-IOT}.
Finally, low data-rate applications such as most of the monitoring sensors, goods tracking, smart parking and intelligent agriculture systems are mostly supported by NB-IoT, SigFox, and LoRa technologies \cite{Lora2DJ}. Low power consumption is a critical factor in these kinds of applications and consequently, their design is less complex. Moreover, the market share for this category is estimated to be 60\% \cite{NB-IOT}.  Overall, the majority of the future IoT applications require either medium or low data-rate. Therefore, ZigBee, Bluetooth/BLE, WiFi HaLow, LTE-M, NB-IoT, Sigfox, and LoRa will serve as the key connectivity technologies as shown in Figure \ref{fig:fig_data}.
\subsubsection{Latency }
\begin{figure}[ht]
\centering
\captionsetup{width=1\linewidth}
    \includegraphics[width=8.5cm]{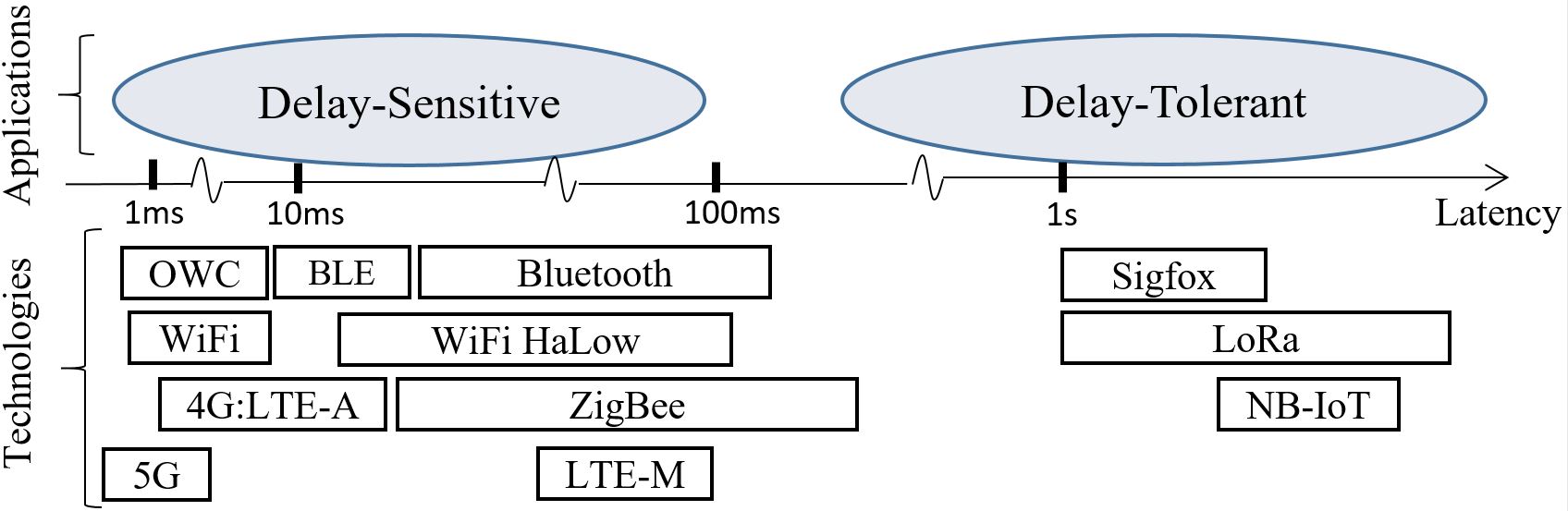}%
  \caption{Nominated technologies for delay-sensitive and delay-tolerant IoT applications. }%
  \label{fig:fig_delay}
\end{figure}

Most of IoT applications are sensitive to latency. But, the level of sensitivity varies for different applications. Due to this difference, the applications with high and low sensitivity to the latency are categorized into delay-sensitive and delay-tolerant groups, respectively \cite{L1}. Autonomous vehicles and health-care systems are two examples of delay-sensitive applications where the shortest possible latency is a critical factor that affects their performance \cite{3G-2016}. To be specific, autonomous vehicles are such driver-less cars that can move automatically and sense their environment to avoid any hazard or accident. Consequently, when the vehicles move at a high speed, latency plays a pivotal role in sensing the environment and make a decision as soon as possible. Likewise, health-care systems (e.g., cardiac telemetry) require to report the possible risks to a distant monitoring station with low latency to assist patients with early treatment. Figure \ref{fig:fig_delay} shows the current technologies such as 4G and WiFi can provide a latency up to tens of milliseconds -- e.g., the current 4G round-trip latency is about 15ms \cite{LA-1}. Although this range of latency suits most current IoT applications, it is not short enough for future applications such as autonomous vehicles that require a shorter latency. For instance, Tesla company has recently designed a connected autonomous cars system based on current 4G technologies. However, due to high latency, the cars move slowly, maintaining a large car-to-car distance, and forming platoons to cross an intersection \cite{LA-2}. Therefore, moving towards future technologies with low latency such as 5G and OWC technologies is a necessity for these kinds of latency stringent applications. 
Contrary to the delay-sensitive applications, delay-tolerant applications such as agricultural sensors, waste management systems, and smart parking applications can be supported by existing connectivity technologies as shown in Figure \ref{fig:fig_delay}. Most of these applications are low duty-cycle applications and the information transmitted by them can be received with relatively large latency (i.e., latency can be greater than 100ms). Therefore, latency, in these delay-tolerant applications, is not as important as in delay-sensitive applications.

\subsubsection{Coverage}
The maximum range of communications for IoT applications varies from couple
of meters up to tens of kilometres. The IoT applications which require a communication range of up to tens of meters 
are categorized as short-range IoT applications. For example, smart home and smart retail applications include a range of connected items/objects in the range of 100m that are considered as short-range applications. On the other hand, the applications with distant connected items/objects (i.e., up to tens of kilometres) are classified as long-range IoT applications (e.g., smart farming and UAV) \cite{NB-IOT,C001}. For instance, UAV refers to an aircraft without a human pilot onboard and can be used widely in civilian and other applications such as surveillance and product deliveries. UAV may fly long distances while they need to be connected to distant ground control stations. To support the connections in the short-range applications, Bluetooth/BLE, OWC, WiFi, and ZigBee are the nominated connectivity technologies; and for the long-range applications, Sigfox, LoRa, NB-IoT, LTE-M, WiFi HaLow, and 4G/5G are the nominated connectivity technologies as described in Subsections \ref{short} and \ref{long}. In \cite{Eric}, Ericsson forecasts that the number of long-range applications will reach $4.1$ billion by 2024 from $1$ billion in 2018; and also the number of short-range applications will increase from $7.5$ billion in 2018 to $17.8$ billion in 2024. The current technologies would not be able to support this massive connectivity. Therefore, the emerging technologies such as NOMA, mMIMO and ML-assisted cellular IoT techniques (as discussed in Section \ref{section4}) can be used in future IoT connectivity paradigms. 

\begin{table*}[ht]
\begin{center}
\centering
\caption{Summary Table of IoT applications together with their use-cases and connectivity technologies.}
\label{tab:dist}
\begin{tabular}{|c|c|m{4cm}|m{4cm}|} \hline
Requirement &  App. category &  \hspace{1cm} Use-cases (e.g.,) & Connectivity technologies\\ 
 \hline
 \hline
\multirow{2}{*}{ End-user-type} & Human-oriented &  Smart phone & 
     Legacy cellular technologies, LTE/LTE-A, 5G, WiFi/WiFi HaLow, OWC  \\
                                \cline{2-4}
                                & Machine-oriented & Monitoring sensors  & 
     Bluetooth/BLE, ZigBee, LPWAN, WiFi/WiFi HaLow, OWC \\
                                \hline
\multirow{3}{*}{Data rate} & High data-rate &  Streaming video cameras & LTE/LTE-A, 5G, OWC, WiFi\\ \cline{2-4}
& Medium data-rate &  Connected cooking systems & Bluetooth/BLE, ZigBee, LTE-M, WiFi HaLow\\ \cline{2-4}
& Low data-rate &  Energy \& water meters & NB-IoT, Sigfox, LoRa, ZigBee\\
 \hline
  \multirow{2}{*}{Latency} & Delay-sensitive & Autonomous vehicles, health-care sensors & LTE/LTE-A, 5G, OWC, WiFi/WiFi HaLow, Bluetooth/BLE, LTE-M \\ \cline{2-4}
  & Delay-tolerant & Waste management sensors & ZigBee, Sigfox, NB-IoT, LoRa\\ \hline
 \multirow{2}{*}{Coverage} & Long-range &  UAVs, smart farming sensors & LTE/LTE-A, 5G, LoRa, Sigfox, NB-IoT, LTE-M, WiFi HaLow\\ \cline{2-4}
 & Short-range & Smart home appliances & Bluetooth/BLE, ZigBee, OWC, WiFi\\ 
 \hline
  \multirow{2}{*}{Power} &   Low power & Tracking sensors, smart retail sensors  &  Bluetooth, ZigBee, LTE/LTE-A, 5G, WiFi  \\ \cline{2-4}
   & Ultra low power & Pollution monitoring sensor   &  BLE, WiFi HaLow, LPWAN: LoRa, Sigfox, LTE-M, NB-IoT \\
   \hline
  \multirow{2}{*}{Reliability} & Mission critical &  Real-time patient surveillance, autonomous vehicles & LTE/LTE-A, 5G, WiFi/WiFi HaLow, OWC \\   \cline{2-4}
  & Mission non-critical &  Smart farming sensors & LPWAN: LoRa, Sigfox, LTE-M, NB-IoT \\   \hline
  \multirow{2}{*}{Mobility} & High mobility &  Autonomous vehicles & LTE/LTE-A, 5G
  \\  \cline{2-4}
  & Low mobility & Smart traffic lights & LPWAN, Bluetooth/BLE, ZigBee \\
 \hline
\end{tabular}
\end{center}
\end{table*}

\subsubsection{Power}
Power efficiency is an important requirement that affects the cost of IoT devices. Battery production, recycling, and environmental issues are also important factors that need to be considered in designing IoT applications. For example, even though the smart electric vehicles will not be using the fossil fuel to power the vehicles, they can still cause other environmental problems if the vehicles are not recharged or recycled properly \cite{MR4,MR5}. Therefore, all the IoT applications seek the lowest possible power consumption technologies for low maintenance costs and also for achieving a lower impact on the environment. Most of the human-oriented applications (e.g., smartphones) are able to be charged regularly. However, the most challenging issues appear for ultra-low power consumption applications with LPWAN technologies where they are not able to be charged regularly. For example, applications like agricultural metering sensors normally require the terminal service life with a constant volume battery up to 10 years \cite{NB-IOT,P01,P02}. Developing such batteries requires careful engineering along with the proper low-power components selection. Besides, the key to achieving good battery life is to ensure that a sensor stays in a low-power standby mode as long as possible and also minimizing the use of wireless communications \cite{PR-1}. PSM and eDRX are two power-saving mechanisms that can be employed by NB-IoT technology to increase the battery life-time of IoT devices significantly \cite{P003}. Additionally, BLE and joint power-information transmission technologies such as backscatter communications have recently been proposed as appealing solutions to ultra-low power consumption IoT applications \cite{BC-1}.

\subsubsection{Reliability}
In terms of the reliability of the transmissions, IoT applications can be categorized into two major groups of mission critical and mission non-critical applications \cite{RE-1}. Smart grids, manufacturing robots, autonomous vehicles, and mobile health-care are some examples of mission critical applications \cite{M-01}. Ericsson forecasts that only a small portion of total IoT applications will be mission critical by 2024 \cite{Eric1-1}. On the other hand, the majority of IoT applications are mission non-critical IoT applications such as humidity sensors, smart green houses, smart parking, and energy and water meters. Overall, in order to guarantee sufficient reliability for such applications in both critical and non-critical systems, different requirements of end-to-end latency, ubiquity, availability, security, and robustness of the technologies should be assessed \cite{3G-2016}. LPWAN and current cellular technologies are the dominant technologies for mission non-critical applications. 3GPP expects that the 5G technologies with support for ultra-reliable low latency communications will enable the first series of mission critical applications such as interactive transport systems, smart grids with real-time control, and real-time control of manufacturing robots by 2020 \cite{Eric1-1}. Moreover, OWC technologies have also been proposed for short-range but mission critical applications such as real-time patient surveillance systems that report patients movement and vital signs to a monitoring station with high accuracy \cite{OWC2}.

\subsubsection{Mobility}
IoT applications can be classified into two categories in terms of mobility: low and high mobility applications. Low mobility applications can easily rely on existing connectivity technologies \cite{B010}. The challenging issues appear in high mobility applications where the speed can go up to hundreds km/h and consequently they demand for handover, redirection, and cell reselection in connected states. Additionally, high mobility increases the Doppler effect and jeopardizes the reliability of the connectivity technologies \cite{ICI}. Some examples of high mobility IoT applications are such as vehicles, trains and airplanes demanding enhanced connectivity for in-vehicle/on-board entertainment, accessing the Internet, enhanced navigation through instant and real-time information, autonomous driving, and vehicle diagnostics \cite{3G-2016}. In general, high mobility applications utilize cellular connectivity technologies. However, they require significant improvements in current cellular technologies (e.g., 4G and 5G) to overcome high mobility issues for future high mobility applications \cite{Mob-1}.

Overall, this section gives a straightforward mapping that nominates the potential connectivity technologies for each application category with respect to the application requirements and connectivity technologies specifications. It is evident that IoT applications can be mapped into multiple categories at the same time to find the best possible connectivity technology. For example, smart agricultural sensors, \cite{Tr-5}, are usually considered as machine-oriented, low data rate, delay-tolerant, long-range, low power, non-critical, and low mobility applications. Consequently, we can conclude that LPWAN (e.g., LoRa and NB-IoT) connectivity technologies suit them well. The classification which is summarized in Table \ref{tab:dist} provides a unique opportunity for all IoT applications to find their category and select the suitable connectivity technology for the deployment.


\section{Conclusions}
\label{section9}
Future IoT is expected to accommodate
an exponential growth of connected devices while
satisfying their diverse applications' requirements.
In this survey, we first reviewed the existing wireless IoT connectivity technologies with their specifications and outlined their fundamental bottleneck and challenges to support massive IoT connectivity. To shed light on addressing the bottleneck, we then reviewed the strengths and limitations of some emerging connectivity technologies, such as CS, NOMA, mMIMO, and ML based random access, that have the potential to be employed in future standards for massive IoT connectivity. We explained that although the emerging CS-based connectivity and grant-free RA protocols are proper options for signalling overhead reduction, their complexity in high-density MTC is still an open issue. We also explained that in emerging NOMA-based connectivity which has been proposed as a key idea for massive connectivity in a spectral-efficient way, the detection algorithms and interference cancellation techniques are still challenging in high-density MTC. In addition, we discussed that different from NOMA, emerging mMIMO connectivity mitigates the interference while providing resource-efficient communication. However, in high-density MTC, considering the array dimensions and hardware cost, gathering massive antennas in a centralized way might become impractical. Furthermore, we briefly discussed the limitations and strengths of the ML-assisted connectivities for massive MTC. Finally, we presented a classification of IoT applications with respect to different technical domains and also discussed the suitable IoT connectivity technology candidates for supporting various IoT applications.

\section{Acknowledgement}
This work was supported by Australian Research Council (ARC) Discovery 2020 Funding, under grant number DP200100391.

\begin{appendices}

\section{ALOHA}
ALOHA is a RA protocol \cite{Abramson} that is proposed
to share a common radio channel between multiple nodes.
In this appendix, we only focus on slotted ALOHA where 
time is divided into slots and each slot length corresponds to 
one packet duration (so that a packet can be transmitted
within a slot).
In slotted ALOHA, it is also assumed that nodes are synchronized
and there is a receiver station.

Suppose that each node can transmit a packet for a given slot
with probability $p$, which is called the access probability.
Assume that there are $K$ nodes with
the same access probability.
Then, a node that transmits a packet can successfully
transmit its packet if there are no other nodes transmitting
simultaneously, which has the following probability:
\be
P_S = \left(1 - p\right)^{K-1}.
\ee
If there are more than or equal to $2$ nodes that simultaneously
transmit packets, it is assumed that no packet is successfully
transmitted due to packet collision.
Since a node transmits a packet with probability $p$ and there
are $K$ nodes, the number of nodes that can successfully transmit packets,
which is called the throughput,
is given by
\be
\eta (K,p) = K p P_S = K p (1 - p)^{K-1}.
\ee
If $p$ is sufficiently low, we have $1 - p \approx e^{-p}$. Thus,
\be
\eta (K,p) \approx K p e^{- p(K-1)} \approx Kp e^{-Kp}.
\ee
The approximation is reasonably if $K$ is large. Letting $x = Kp$,
the throughput becomes $x e^{-x}$, which is a $\cap$-shape function of $x$
and has the maximum at $x = 1$.
In other words, if $K$ is sufficiently large, the throughput
becomes the maximum, which is $e^{-1}$, when
$x = 1$ or $p = \frac{1}{K}$.

In slotted ALOHA, as mentioned earlier,
all the nodes need to be synchronized. In addition,
it might be necessary for nodes to know whether or not transmitted
packets are successfully received at the 
receiver station. To this end, the receiver station is to periodically
broadcast a beacon signal for synchronization 
and feedback signals to let nodes know the success of packet
transmission (using a feedback signal of acknowledgment (ACK) 
or negative acknowledgment (NAK) at the end of slot.
Note that a node that transmits a packet receives 
a NAK, it can see that collision happens.
The collided packet is to be dropped or re-transmitted later.
In the case of NAK, since there are other nodes transmit packets
an immediate re-transmission
results in another collision, which should be avoided.
Thus, a random back-off time is required for re-transmissions.

Since slotted ALOHA is a distributed system, there are stability issues.
In particular, if each node has a buffer to keep
packets before transmissions, a buffer overflow can happen
due to frequent packet collisions. Thus, distributed access
control and 
re-transmission strategies are to be carefully designed to keep
buffer length (which is also proportional to 
access delay) stable.

\label{A_Aloha}

 \section{CSMA}
 Carrier-sense multiple access (CSMA)
is a RA protocol where a node attempts
to verify the absence of other traffic (by sensing the presence
of carriers or signals) in 
a common access channal before transmitting.
There are different types of CSMA protocols including
CSMA with collision detection (CSMA/CD)
and 
CSMA with collision avoidance (CSMA/CA).

In CSMA/CD, suppose that a node wants to transmit a packet. 
Then, it senses the channel and transmits a packet if the channel
is idle.
However, multiple nodes can transmit simultaneously and sense
a collision. In this case, they abort transmissions after
sending a jamming signal to notify collision.
As a result, the duration of collision can be shortened and
it can result in a better throughput.

CSMA/CD can have the following re-transmission strategies:
\begin{itemize}
\item \emph{Nonpersistent CSMA}: If the channel is idle, the node
transmits a packet. 
If the channel is not idle, the node waits
for a random time (according to a certain distribution).

\item \emph{1-persistent CSMA}: If the channel is idle, the node
transmits a packet.
If the channel is not idle, the node waits
until the channel becomes idle and transmits a packet immediately.
In this case, a collision always occurs if there are multiple nodes
(with packets) sensing at the same time.

\item \emph{$p$-persistent CSMA}: When
a node with a packet senses that the channel is idle, it
transmits a packet with probability $p$ and
delays by $\tau$ with probability $1-p$,
where $\tau$ is the duration of minislots.
A node waiting for a time duration of $\tau$ 
is to repeat the same process above. That is,
it senses the channel: if the channel is idle,
it transmits with probability $p$ and delays by $\tau$
with probability $1-p$. If the channel is busy,
it waits until the channel is idle (and repeats
the same process again). 
\end{itemize}

\begin{figure}[t]
\centering
\captionsetup{width=1\linewidth}
  \includegraphics[width=9cm, height=5.5cm]{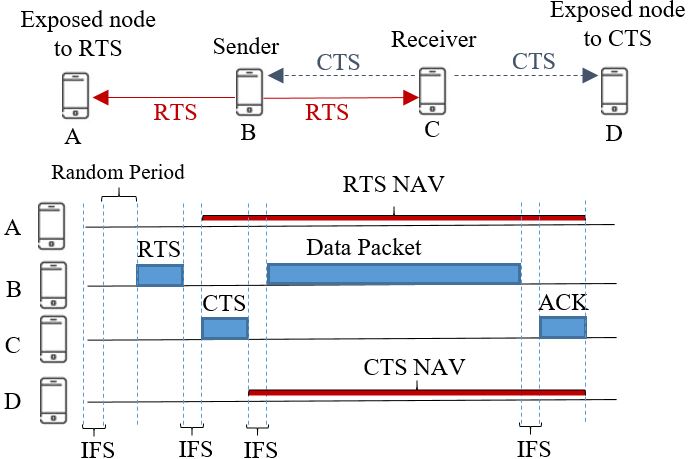}%
  \caption{Standard CSMA/CA mechanism with RTS/CTS packet transmission. }%
  \label{fig:fig_CSMA}
\end{figure}

CSMA/CD is usually used for wired networks where a node can
simultaneously sense the channel when it transmits a packet.
In wireless channels, however, a node cannot sense when it
transmits. In this case, CSMA/CA can be used,
where collision is to be avoided using a few strategies.
In CSMA/CD, inter-frame space
(IFS) is introduced to wait a certain period of time
although the channel appears idle after sensing
as another node may start transmitting, but its signal
is not yet reached at the node.
If a node is ready to transmit, a random number
is generated to wait 
and the range of the random number
is called the contention window.
The waiting time is proportional to the random number
and the length of the contention window
is varying. Initially, the length of contention window
is set to 1 and doubled if the node sees that the channel is not idle
after the IFS time.
In CSMA/CD, although collisions are to be avoided with CS at the sender,
they can happen because the collisions happen at the receiver. Thus, feedback signals (ACK or NAK) 
are given to the nodes to inform collisions. Moreover, in wireless channels, signal strength decreases proportional to the square of the distance and may cause near-far terminal problems in CSMA/CD. Therefore, CSMA/CA usually utilizes two short signaling packets to avoid collisions as follows
\begin{itemize}
    \item RTS (request to send): a node requests the right to send from a receiver with a short RTS packet before it sends a data packet.
    \item CTS (clear to send): the receiver grants the right to send as soon as it is ready to receive.
\end{itemize}
Both of RTS and CTS packets contain information such as node address, receiver address, and packet size. Fig. \ref{fig:fig_CSMA} shows the standard CSMA/CA mechanism. A sender node senses the channel and sends RTS when channel is idle. Other exposed nodes that receive the RTS, hold their requests for a RTS network allocation vector (NAV) period. Receiver receives the RTS and sends the CTS after a short IFS period. Accordingly, other exposed nodes that receive the CTS hold their transmission requests for a CTS NAV period. The sender node receives the CTS and sends its data after a short IFS period. The receiver receives the data and sends ACK to the sender \cite{ARS-1}. Consequently, no collision occurs in CSMA/CA. Different variations of this model can be found in IEEE 802.11 as distributed foundation wireless MAC \cite{Ars}.

\label{A_Csma}

 \section{CS}
 Compressive sensing (CS) is to deal with sparse signals
\cite{Candes,Donoho}.
There are a number of applications of CS
including image compression and radar systems.
In this appendix, we briefly discuss
the sparse signal recovery in the context of CS
and explain how the notion of CS is applied to RA.

The set of $k$-sparse signals is defined as
\be
\Sigma_k = \{\bs: \  ||\bs||_0 \le k\}.
\ee
A group of signals can have a sparse representation if
$\bs$ can be expressed as
\be
\bs = \bPhi \bc,
\ee
where $\bc \in \Sigma_k$ and $\bPhi$ is a (known) basis.
For convenience, we assume that the length of $\bs$ is $n$.
For a given $\bs$, suppose that the following vector is available:
\be
\by = \bC \bs,
	\label{EQ:yCs}
\ee
where $\bC$ is an $M \times L$ matrix that is called the measurement
matrix. Here, it is assumed that $M < L$ for a dimensionality reduction.
In general, it is not possible to recover $\bs$ from $\by$ 
unless $\bs$ and $\bC$ have certain conditions (as \eqref{EQ:yCs}
is an underdetermined linear system).

Suppose that the sparsity of $\bs$ is known in \eqref{EQ:yCs}. 
For convenience, let $q = ||\bs||_0$.
Consider the estimation of $\bs$ 
based on the ML 
criterion:
\begin{align}
\hat \bs 
& = \argmax_{\bs: \ ||\bs||_0 = q} f(\bs | \by) \cr
& = \argmin_{\bs: \ ||\bs||_0 = q} ||\by - \bC \bs ||^2.
	\label{EQ:ML_bs}
\end{align}
Since $\bC$ is an $M \times L$ matrix,
there are $L$ columns. For a given sparsity $q$, there can be
$L_q = \binom{L}{q}$ possible supports.
Denote by $\cI_i$ the $i$th support, where $i = 1, \ldots, L_q$.
For example, if $L = 4$ and $q = 2$, $L_q = 6$
and
$$
\cI_1 = \{1,2\},
\cI_2 = \{1,3\},
\cI_3 = \{1,4\},
\cI_4 = \{2,3\},
\cI_5 = \{2,4\},
\cI_6 = \{3,4\}.
$$
In addition, denote by $\bC_i$ and $\bs_i$ the submatrix
of $\bC$ and the subvector of $\bs$ corresponding to $\cI_i$, 
respectively.
Then, for given $\cI_i$, it can be shown that
\be
\min_{\bs: \ ||\bs||_0 = q} ||\by - \bC \bs ||^2
= \min_{i \in \{1,\ldots, L_q\}} \min_{\bs_i} ||\by - \bC_i \bs_i ||^2.
\ee
If $q \le M$, the inner minimization can be solved
by the method of least squares (LS), i.e.,
\begin{align}
\hat \bs_i & = \argmin_{\bs_i} ||\by - \bC_i \bs_i ||^2 \cr
& = \bC_i^\dagger \by,
\end{align}
where $\bC_i^\dagger$ is the pseudo-inverse of $\bC_i$.
If the rank of $\bC_i$ is $q$, 
$\bC_i^\dagger = 
(\bC_i^\rH \bC_i)^{-1} \bC_i^\rH$.
In addition, it follows that
\be
\min_{\bs_i} ||\by - \bC_i \bs_i ||^2 = ||(\bI - \bC_i 
(\bC_i^\rH \bC_i)^{-1} \bC_i^\rH) \by||^2.
\ee
As a result, the ML solution in \eqref{EQ:ML_bs}
can be found if all possible supports are considered.
However, the computational complexity becomes
proportional to $L_q = \binom{L}{q}$. Thus, for a large $L$,
this approach might be prohibitive.

\begin{figure}[thb]
\begin{center}
\includegraphics[width=8cm]{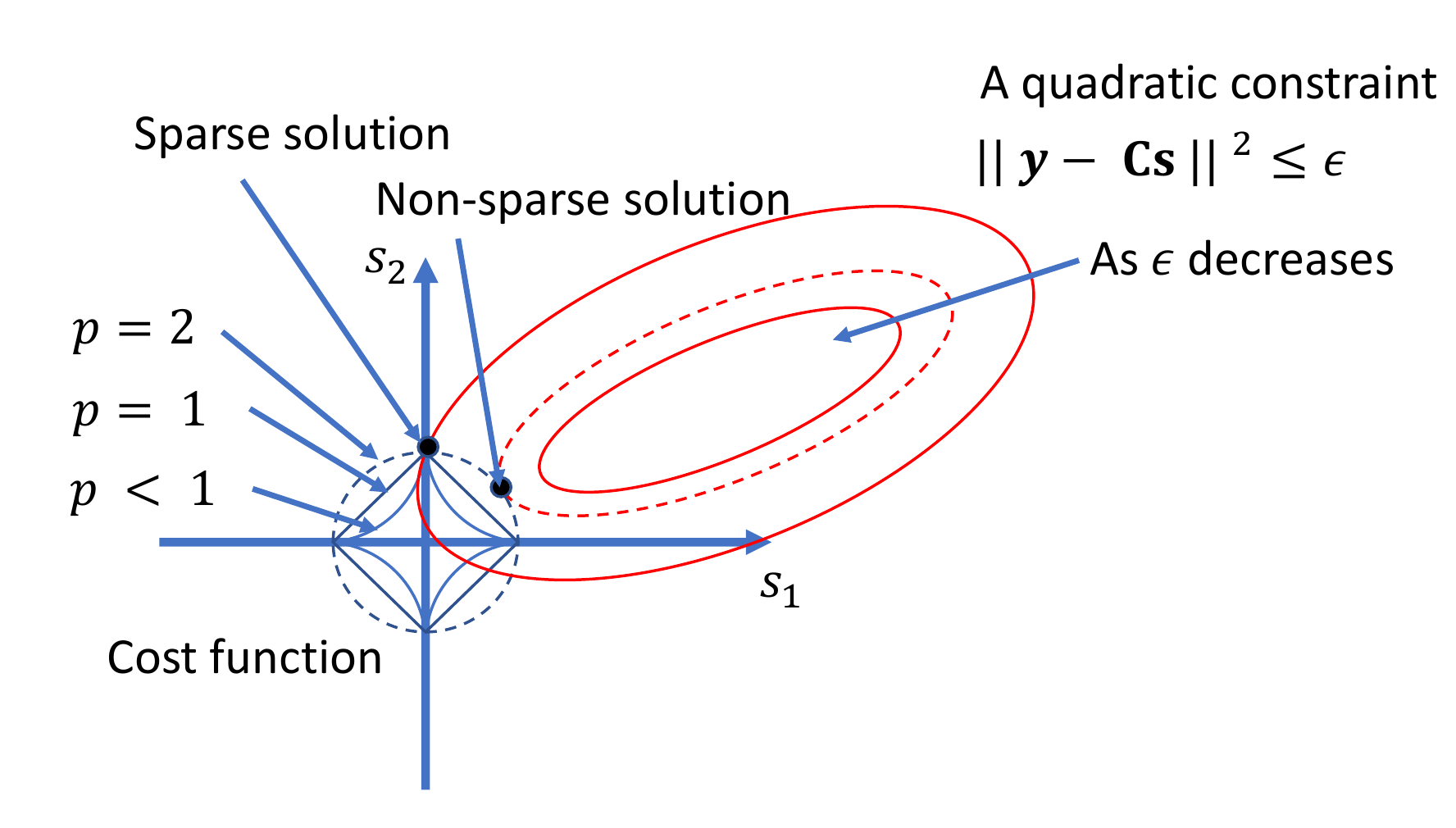}
\end{center}
\caption{Optimization problems with a quadratic constraint.}
        \label{Fig:norm_p2}
\end{figure}

From \eqref{EQ:ML_bs}, a different approach can be considered by noting
that $\bs$ is sparse (i.e., $q \ll L$).
Let us assume that $\bn = \b0$. Then, it is expected
to find a sparse solution that satisfies $\by = \bC \bs$.
Since $M < L$, the resulting system is considered underdetermined
(i.e., more unknown variables than equations). Since
an underdetermined linear system has 
either no solution or infinitely many solutions,
it is necessary to take into account the sparsity of $\bs$.
Since the sparsity of $\bs$ can be measured by the $\ell_0$-norm,
in order to find the most sparse solution, the following
optimization problem can be formulated:
\begin{eqnarray}
& \min ||\bs||_0 & \cr
& \mbox{subject to} \ \by = \bC \bs. &
	\label{EQ:l0}
\end{eqnarray}
Unfortunately, \eqref{EQ:l0} is not a convex optimization problem
since $||\bs||_0$ is not a convex function. To generalize
\eqref{EQ:l0}, the $p$-norm can be used, which results in the
following problem:
\begin{eqnarray}
& \min ||\bs||_p & \cr
& \mbox{subject to} \ \by = \bC \bs. &
	\label{EQ:lp}
\end{eqnarray}
If $p \ge 1$, the problem becomes a convex optimization problem.
Furthermore, in the presence of error or background noise,
the constraint can be relaxed and the following
convex-optimization problem can be formulated:
\begin{eqnarray}
& \min ||\bs||_p & \cr
& \mbox{subject to} \ ||\by - \bC \bs||^2 \le \epsilon, &
	\label{EQ:lpe}
\end{eqnarray}
where $\epsilon > 0$. To obtain a sparse solution, it is
desirable to have $p \le 1$ as illustrated in
Fig.~\ref{Fig:norm_p2}. That is, 
since the cost function in \eqref{EQ:lpe}
is spike with $p \le 1$, the solution of \eqref{EQ:lpe}
tends to be sparse when $p = 1$ (although $\ell_0$-norm is not used),
while the solution with $p = 2$ (which
corresponds to the least squares solution
of an underdetermined system) is not sparse.

As mentioned earlier, since the problem in \eqref{EQ:lpe}
with $p = 1$ is a convex optimization problem, its sparse solution
can be obtained by a number of convex optimization tools.

The notion of 
CS can be applied to the user activity detection in RA.
Suppose that 
there are $L$ users and each user has a unique signature sequence,
denoted by $\bc_l$. In addition, denote by $s_l$ the user activity
variable. That is, if user $l$ is to transmit a signal, it can
send its unique
signature sequence, $\bc_l$. Let the length of $\bc_l$ be $M$
(if $L > M$, the $\bc_l$'s are not orthogonal to each other).
Thus, the received
signal at a receiver is given by
\be
\by = \sum_{l=1}^L \bc_l s_l + \bn = \bC \bs + \bn,
\ee
where $\bn$ is the background noise. If a few users are active at a time,
$\bs$ becomes sparse.
In RA to support a large number of users,
it is desirable to have a large $L$ for a fixed $M$.
This shows that the receiver can employ the notion of CS
to detect active users when $L > M$ as shown above.
The resulting RA (with non-orthogonal sequences, $\{\bc_l\}$)
is referred to as compressive RA \cite{CS3,CS7}.

 \label{A_Cs}

 \section{NOMA}
 Non-orthogonal multiple access (NOMA)
refers to a set of multiple access schemes where
multiple access channels are not orthogonal
as opposed to orthogonal multiple access (OMA), e.g.,
time division multiple access (TDMA) and 
frequency division multiple access (FDMA).
While there are various ways to form NOMA schemes,
the most popular one is based on the power-domain,
which is often called power-domain NOMA \cite{Choi_ISWCS17}.

Power-domain NOMA employs the superposition coding
where multiple signals are transmitted 
through a shared channel or radio resource block with different
power levels in downlink transmissions.
In power-domain NOMA, user pairing is also an important
technique where one user is usually close to a BS
the stong user) and 
the other user is far away from the BS.
The former and latter users are referred to as
the near and far users, respectively.
Due to different distances,
the transmit signal power to the near user 
is lower than that to the far user.
Thus, at the near user, the signal to the far user is a strong interfering
signal that can be decoded and then removed using
successive interference cancellation (SIC).
For convenience, denote by $s_1$ and $s_2$
the signals to the near and far users, respectively,
and the transmit powers are accordingly denoted by $P_k$,
$k = 1,2$. 
The received signal at the near user is given by
\be
y_1 = h_1 \left(\sqrt{P_1} s_1 + \sqrt{P_2} s_2 \right) + n_1,
\ee
where $h_1$ and $n_1$ are
the channel coefficient and the background noise at the near user,
respectively.
As mentioned earlier, it is assumed that $P_1 \ll P_2$.
Taking $s_1$ as the interference, the near user
can decode $s_2$ and remove it as follows:
\be
\tilde y_1 = y_1 - h_1 \sqrt{P_2} \hat s_2,
\ee
where $\hat s_2$ is the estimate of $s_2$. If $\hat s_2 = s_2$,
$\tilde y_1 = h_1 \sqrt{P_1} \hat s_1 + n_1$.
Then, $s_1$ is to be decoded from $\tilde y_1$. The above
procedure is called SIC.

If the $s_k$'s are coded signals using a capacity-achieving code,
with power-domain NOMA, using the capacity
formula \cite{CoverBook}, the code rate for $s_k$,
denoted by $R_k$, has the following constraints:
\begin{align}
R_2 & \le \log_2 \left(1 + \frac{|h_1|^2 P_2}{N_0 + |h_1|^2 P_1}\right) \cr
R_1 & \le \log_2 \left(1 + \frac{|h_1|^2 P_1}{N_0} \right),
	\label{EQ:NOMA_C1}
\end{align}
where $N_0$ stands from the noise variance.
The first and inequalities in \eqref{EQ:NOMA_C1}
are to successfully decode $s_2$ and $s_1$
(after SIC) at the near user, respectively.

At the far user, the received signal is given by
\be
y_2 = h_2 \left(\sqrt{P_1} s_1 + \sqrt{P_2} s_2 \right) + n_2,
\ee
where $h_2$ and $n_2$ are
the channel coefficient and the background noise at the near user,
respectively.
The far user is to decode $s_2$ and requires the following
condition for successful decoding:
\be
R_2 \le \log_2 \left(1 + \frac{|h_2|^2 P_2}{N_0 + |h_2|^2 P_1}\right).
	\label{EQ:NOMA_C2}
\ee
As a result, the rate constraints from 
\eqref{EQ:NOMA_C1} and \eqref{EQ:NOMA_C2} can be combined as follows:
\begin{align}
R_2 & \le 
\min_k \log_2 \left(1 + \frac{|h_k|^2 P_2}{N_0 + |h_k|^2 P_1}\right) \cr
R_1 & \le \log_2 \left(1 + \frac{|h_1|^2 P_1}{N_0} \right),
	\label{EQ:NOMA_C}
\end{align}
which plays a key role in the power allocation for power-domain NOMA.

While power-domain NOMA 
is usually studied for downlink transmissions,
it can be naturally applied to uplink transmissions
where the received signal at the BS becomes a superposition
of transmitted signals from a number of users.
For example, with two users, the received signal at the BS is
given by
\be
y = h_1 \sqrt{P_1} s_1 + h_2 \sqrt{P_2} s_2 + n,
	\label{EQ:MAC}
\ee
where $n$ is the background noise at the BS.
Here, $s_k$ is the transmitted signal by user $k$
and $P_k$ is the transmit power at user $k$.
While the BS is able to perform joint decoding
to recover $s_1$ and $s_2$, its complexity
is usually high. However, by exploiting the notion of SIC,
the complexity can be lowered. For example, if
$|h_1|^2 P_1 \gg |h_2|^2 P_2$, $s_1$ is decoded
first (where $s_2$ is regarded as an interfering signal).
Then, $s_1$ is removed and $s_2$ is decoded
from $y - h_1 \sqrt{P_1} s_1$.
As a result, the rate constraints are given by
\begin{align}
R_1 & \le 
\log_2 \left(1 + \frac{|h_1|^2 P_1}{N_0 + |h_2|^2 P_2}\right) \cr
R_2 & \le \log_2 \left(1 + \frac{|h_2|^2 P_2}{N_0} \right).
	\label{EQ:NOMA_U}
\end{align}
We note that the sum rate becomes
\begin{align}
R_1 + R_2 & \le 
\log_2 \left(1 + \frac{|h_1|^2 P_1}{N_0 + |h_2|^2 P_2}\right) +
 \log_2 \left(1 + \frac{|h_2|^2 P_2}{N_0} \right) \cr
& = 
 \log_2 \left(1 + \frac{|h_1|^2 P_1+|h_2|^2 P_2}{N_0} \right),
	\label{EQ:NOMA_S}
\end{align}
which implies that power-domain NOMA is also optimal
in terms of the sum rate
as the sum rate in \eqref{EQ:NOMA_S}
is also the achievable rate of multiple access
channel with \eqref{EQ:MAC} \cite{CoverBook}.

 \label{A_Noma}

 \section{mMIMO}
 Massive multiple input multiple output (mMIMO) is a extended form of multi-user MIMO (MU-MIMO) systems where hundreds or thousands of BS antennas simultaneously serve tens or hundreds of users over the same wireless time-frequency resource. 
In mMIMO, time division duplex (TDD) operation is more favorable than frequency division duplex (FDD) operation since the TDD can take the advantage of reciprocity between uplink channel
and downlink channel within a given coherence interval and thus remove the need for downlink channel estimation \cite{mMIMO00}.

Since the number of antennas, $M$, at the BS is usually much larger than the number of users $K$, i.e., $M \gg K$, favorable propagation (FP)
can be approximately achieved in mMIMO systems due to the law of large numbers  \cite{mMIMO1}, which means users' channel vectors are mutually orthogonal/quasi-orthogonal. Under the property of FP, simple
linear processing (receive beamforming in the uplink and transmit beamforming in the downlink), such as conjugate beamforming (CB) and zero-forcing beamforming (ZFB), can be nearly optimal to discriminate the signal transmitted by each user from the signals of other users in mMIMO, since the effect of user interference and noise can be eliminated. Furthermore, thanks to the large number of antennas, channel hardening is another key property in mMIMO \cite{mMIMO2}, upon which the channel becomes nearly deterministic. As a result, the effect of small-scale fading is averaged out. This also simplifies the signal processing significantly in mMIMO.

Consider the downlink transmission in mMIMO (the same argument can be used for the uplink transmission), with the transmit CB, the
transmitted signal vector from the BS to all users is given by
\be
\mathbf{s}=\sqrt{\frac{P_t}{KM}}\sum_{k=1}^{K}\mathbf{h}^H_kx_k,
\ee
where $P_t$ is the total power transmitted by the BS and $\mathbf{h}_k \in \mathbb{C}^{M} \sim \mathcal{CN}(0,\mathbf{I}_{M})$ stands for the channel response vector between the $k$th user and the BS. $\mathbf{h}^H_k$ is the transmit conjugate beamformer and $(\cdot)^H$ stands for the matrix Hermitian. $x_k$ is the data symbol intended
for the $k$th user with power normalization, i.e., $\mathbb{E}[|x_{k}|^2]=1$.

Accordingly, the received signal at the $k$th user is given by
\be
y_k=\underbrace{\sqrt{\frac{P_t}{KM}}\mathbf{h}^H_k\mathbf{h}_kx_k}_{\mathrm{Desired~signal}}+\underbrace{\sqrt{\frac{P_t}{KM}}\sum_{j=1,j\neq k}^{K}\mathbf{h}^H_k\mathbf{h}_jx_j}_{\mathrm{Multiuser~interference}}+n_k,
\ee
where $n_k$ is the additive Gaussian noise with zero-mean and unit-variance.

In mMIMO, when $M \to \infty$, under the law of large numbers, we have,
\be
\frac{\mathbf{{h}}_k^{\mathrm{H}}\mathbf{h}_k}{M} \stackrel{M \to \infty}{\longrightarrow} 1,
\ee
\be
\frac{\mathbf{{h}}_k^{\mathrm{H}}\mathbf{h}_j}{M} \stackrel{M \to \infty}{\longrightarrow} 0, ~k \ne j.
\ee  
Then, we have
\be
\frac{y_k}{\sqrt{M}} \stackrel{M \to \infty}{\longrightarrow} \sqrt{\frac{P_t}{K}}x_k,
\ee
which indicates that the multiuser interference and noise can be eliminated in mMIMO when $M$ is sufficiently large.

In addition, the SINR can be written as
\be
\mathrm{SINR}_k=\frac{\frac{P_t}{KM}|\mathbf{h}^H_k\mathbf{h}_k|^2}{1+\frac{P_t}{KM}\sum_{j=1,j\neq k}^{K}|\mathbf{h}^H_k\mathbf{h}_j|^2}.
\ee
Considering $M, K \to \infty$ with a fixed ratio, under the law of large numbers, we have, 
\be
\frac{|\mathbf{h}^H_k\mathbf{h}_k|^2}{M}\stackrel{M \to \infty}{\longrightarrow} M,
\ee
\be
\frac{|\mathbf{h}^H_k\mathbf{h}_j|^2}{M} \stackrel{M \to \infty}{\longrightarrow} 1, ~k \ne j.
\ee
Thus, the asymptotic
deterministic equivalence of the SINR can be obtained as
\begin{align}
\mathrm{\overline{SINR}}_k
=\frac{M}{K}\frac{P_t}{1+P_t}.
\end{align} 
Accordingly, the asymptotic sum rate in mMIMO is given by
\be
R=K\log(1+\frac{M}{K}\frac{P_t}{1+P_t}).
\label{Sumrate}
\ee 
From \eqref{Sumrate}, it can be seen that a huge spectral efficiency and energy efficiency are obtained when $M$ and $K$ are large.
Without the need of increase in transmitted power $P_t$, by increasing $M$, we can increase the throughput per user and serve more users simultaneously. On the other hand, given a targeted throughput per user, more power can be saved as $M$ grows.

 \label{A_mMIMO}

\end{appendices}

\section*{List of Abbreviations}

\begin{table}[H]
    \centering
    \begin{tabular}{|l|p{6.2cm}|}
    \hline
 3GPP & 3rd Generation Partnership Project \\\hline
 4G & The 4th Generation\\\hline
 5G  &  The 5th Generation  \\\hline
 BLE  &  Bluetooth Low Energy  \\\hline
BLER  &  Block Error Rate  \\\hline
 BPSK  &  Binary Phase Shift Keying  \\ \hline
  BS  &  Base Station  \\\hline
 BS-ILC  & Beam-Steered Infrared Light Communication\\ \hline
CDM  &  Code-Domain Multiplexing \\\hline
 CDMA  &  Code-Division Multiple Access \\\hline
 CP-OFDM  &   Cyclic-Prefix Orthogonal Frequency Division Multiplexing    \\\hline
CS  &  Compressive Sensing  \\\hline
 CSMA  & Carrier Sense Multiple Access \\\hline
 CSMA/CA  & Carrier Sense Multiple Access with Collision Avoidance \\\hline
  CSS  &  Chirp Spread Spectrum  \\\hline
   DBPSK &   Differential Binary Phase Shift Keying\\\hline
   DPSK  &  Differential Phase Shift Keying \\ \hline
       DQPSK  &   Differential Quadrature Phase Shift Keying \\ \hline
        FDMA  &  Frequency-Division Multiple Access \\\hline
          GFSK  &  Gaussian Frequency Shift Keying \\ \hline
           GSM & Global System for Mobile Communication \\\hline
     HTC  &  Human Type Communications  \\\hline 
 IR & Infrared\\\hline
  ISM  &  Industrial, Scientific, and Medical  \\\hline
  ITU & International Telecommunication Union  \\\hline
  IoT & Internet of Things \\\hline
   LD  &  Laser Diodes \\\hline
    LED  &  Laser Emitting Diode \\\hline
 LPWAN & Low Power Wide Area Networks\\ \hline
  LTE  &  Long-Term Evolution \\\hline
 LTE-A  &  Long-Term Evolution-Advance  \\\hline
  LTE-M  &  Long-Term Evolution Machine Type Communications  \\\hline
 LiFi  &  Light Fidelity\\\hline
  LoRa  &  Long Range  \\\hline
   ML  &  Machine Learning  \\\hline
     MTC & Machine Type Communications\\\hline
    NB-IoT  &  Narrow-Band IoT  \\\hline
  NOMA  &  Non-Orthogonal Multiple Access  \\ \hline
    OCC  &  Optical Camera Communication \\\hline
   OFDM  &  Orthogonal Frequency Division Multiplexing \\\hline
        OFDMA  &  Orthogonal Frequency Division Multiplexing Access \\\hline
                      OOK  &  On-Off Keying \\\hline
   OQPSK  &  Offset Quadrature Phase-Shift Keying \\ \hline
   OWC & Optical Wireless Communication\\\hline
    PDM  &  Power-Domain Multiplexing \\\hline
     POS  &  Point of Sale \\\hline
      PRACH  &  Physical Random Access Channel  \\\hline
       PRBs  &  Physical Resource Blocks  \\\hline
        PSM  &  Power Saving Mode  \\\hline
        QAM  &  Quadrature Amplitude Modulation \\\hline
       QPSK  &  Quadrature Phase-Shift Keying \\ \hline
      QoS  &  Quality of Services  \\\hline
      RA  &  Random Access  \\\hline
      RFID  &  Radio Frequency Identification  \\\hline
       RL  &  Reinforcement Learning \\\hline
        RRC  &  Radio Resource Control \\\hline
        SC-FDMA  &  Single-carrier Frequency-Division Multiple Access \\\hline
        SIC  &  Successive Interference Cancellation  \\\hline
        SINR  &  Signal-to-Noise-and-Interference Ratio \\\hline
        TDMA  &  Time-Division Multiple Access \\\hline
     TTL & Time-To-Live\\\hline
      UAV  &  Unmanned Aerial Vehicles  \\\hline
       UNB  &  Ultra Narrow-Band  \\\hline
      UV &Ultraviolet\\\hline
      VL &Visible Light\\\hline
  VLC &Visible Light Communication\\ \hline
   WLAN  &  Wireless Local Area Networks  \\\hline
    WPAN  &  Wireless Personal Area Networks  \\\hline
     WSN  &  Wireless Sensor Networks  \\\hline
      cGFRA  &  Compressive Grant-Free Random Access  \\\hline
          \end{tabular}
\end{table}
\begin{table}[!htbp]
    \centering
    \begin{tabular}{|l|p{6cm}|}
    \hline
      eDRX  &  Expanded Discontinuous Reception  \\\hline
       mGFRA  &  Massive Multiple-Input Multiple Output based Grant-Free Random Access  \\\hline
        mMIMO  &  Massive Multiple-Input Multiple Output  \\\hline
        mmWave  &  Millimeter Waves \\ \hline
    \end{tabular}
\end{table}

\bibliographystyle{ieeetr}
\bibliography{ref}

\begin{thebibliography}{100}

\bibitem{I-1}
T.~Liu and D.~Lu, ``The application and development of iot,'' in {\em 2012
  International Symposium on Information Technologies in Medicine and
  Education}, vol.~2, pp.~991--994, IEEE, 2012.

\bibitem{I-2}
K.~Ashton {\em et~al.}, ``That ‘internet of things’ thing,'' {\em RFID
  journal}, vol.~22, no.~7, pp.~97--114, 2009.

\bibitem{I-4}
P.~Suresh, J.~V. Daniel, V.~Parthasarathy, and R.~Aswathy, ``A state of the art
  review on the internet of things (iot) history, technology and fields of
  deployment,'' in {\em 2014 International conference on science engineering
  and management research (ICSEMR)}, pp.~1--8, IEEE, 2014.

\bibitem{I-5}
V.~Reding, ``Internet of the future: Europe must be a key player,'' {\em speech
  to the Lisbon Council on}, vol.~2, 2009.

\bibitem{Add_Survey1}
M.~{Agiwal}, A.~{Roy}, and N.~{Saxena}, ``Next generation 5g wireless networks:
  A comprehensive survey,'' {\em IEEE Communications Surveys Tutorials},
  vol.~18, pp.~1617--1655, thirdquarter 2016.

\bibitem{LTEM2}
{3GPP TS 22.368}, ``Service requirements for machine-type communications
  (mtc),'' {\em V13.1.0}, 2014.

\bibitem{I-3}
D.~Hamilton, ``Best practices for iot security,'' {\em
  https://www.networkworld.com/article/3266375/internet-of-things\\/best-practices-for-iot-security.html
  (Accessed 09/07/2019)}, 2018.

\bibitem{shortrange1}
Q.~M. {Qadir}, T.~A. {Rashid}, N.~K. {Al-Salihi}, B.~{Ismael}, A.~A. {Kist},
  and Z.~{Zhang}, ``Low power wide area networks: A survey of enabling
  technologies, applications and interoperability needs,'' {\em IEEE Access},
  vol.~6, pp.~77454--77473, 2018.

\bibitem{OWC1}
D.~C.~O. {Brien}, ``Optical wireless communications: Current status and future
  prospects,'' {\em IEEE Summer Top.}, 2016.

\bibitem{Sigfox1DJ}
``Sigfox,''
\newblock http://www.sigfox.com (Accessed 09/07/2019).

\bibitem{Lora1}
{3GPP TSG GERAN 65}, ``Combined narrow-band and spread spectrum physical layer
  coverage and capacity simulations,'' {\em Semtech}, 2015.

\bibitem{LTEM3}
{3GPP TS 22.261}, ``Service requirements for next generation new services and
  markets,'' {\em V1.0.0}, 2016.

\bibitem{Survey_comp8}
H.~{Hejazi}, H.~{Rajab}, T.~{Cinkler}, and L.~{Lengyel}, ``Survey of platforms
  for massive iot,'' in {\em 2018 IEEE International Conference on Future IoT
  Technologies (Future IoT)}, pp.~1--8, Jan 2018.

\bibitem{Add_Survey2}
L.~{Zhang}, Y.~{Liang}, and M.~{Xiao}, ``Spectrum sharing for internet of
  things: A survey,'' {\em IEEE Wireless Communications}, vol.~26,
  pp.~132--139, June 2019.

\bibitem{Survey_comp5}
S.~Madakam, R.~Ramaswamy, and S.~Tripathi, ``Internet of things (iot): A
  literature review,'' {\em Journal of Computer and Communications}, vol.~3,
  pp.~164--173, 04 2015.

\bibitem{Survey_comp9}
G.~A. {Akpakwu}, B.~J. {Silva}, G.~P. {Hancke}, and A.~M. {Abu-Mahfouz}, ``A
  survey on 5g networks for the internet of things: Communication technologies
  and challenges,'' {\em IEEE Access}, vol.~6, pp.~3619--3647, 2018.

\bibitem{Survey_comp6}
M.~Elkhodr, S.~Shahrestani, and H.~Cheung, ``Emerging wireless technologies in
  the internet of things : A comparative study,'' {\em International Journal of
  Wireless \& Mobile Networks}, vol.~8, 11 2016.

\bibitem{Survey_comp1}
A.~{Al-Fuqaha}, M.~{Guizani}, M.~{Mohammadi}, M.~{Aledhari}, and M.~{Ayyash},
  ``Internet of things: A survey on enabling technologies, protocols, and
  applications,'' {\em IEEE Communications Surveys Tutorials}, vol.~17,
  pp.~2347--2376, Fourthquarter 2015.

\bibitem{wifi4}
Q.~M. {Qadir}, T.~A. {Rashid}, N.~K. {Al-Salihi}, B.~{Ismael}, A.~A. {Kist},
  and Z.~{Zhang}, ``Low power wide area networks: A survey of enabling
  technologies, applications and interoperability needs,'' {\em IEEE Access},
  vol.~6, pp.~77454--77473, 2018.

\bibitem{LPWAN3}
M.~{Centenaro}, L.~{Vangelista}, A.~{Zanella}, and M.~{Zorzi}, ``Long-range
  communications in unlicensed bands: the rising stars in the iot and smart
  city scenarios,'' {\em IEEE Wireless Communications}, vol.~23, pp.~60--67,
  October 2016.

\bibitem{Add_Survey3}
S.~K. Goudos, P.~I. Dallas, S.~Chatziefthymiou, and S.~Kyriazakos, ``A survey
  of iot key enabling and future technologies: 5g, mobile iot, sematic web and
  applications,'' {\em Wirel. Pers. Commun.}, vol.~97, pp.~1645--1675, Nov
  2017.

\bibitem{Survey_comp7}
J.~{Xu}, J.~{Yao}, L.~{Wang}, Z.~{Ming}, K.~{Wu}, and L.~{Chen}, ``Narrowband
  internet of things: Evolutions, technologies, and open issues,'' {\em IEEE
  Internet of Things Journal}, vol.~5, pp.~1449--1462, June 2018.

\bibitem{LPWAN2}
K.~Mekki, E.~Bajic, F.~Chaxel, and F.~Meyer, ``A comparative study of lpwan
  technologies for large-scale iot deployment,'' {\em ICT Express}, vol.~5,
  no.~1, pp.~1 -- 7, 2019.

\bibitem{Survey_comp2}
W.~{Yang}, M.~{Wang}, J.~{Zhang}, J.~{Zou}, M.~{Hua}, T.~{Xia}, and X.~{You},
  ``Narrowband wireless access for low-power massive internet of things: A
  bandwidth perspective,'' {\em IEEE Wireless Communications}, vol.~24,
  pp.~138--145, June 2017.

\bibitem{LoraWAN2}
A.~{Ikpehai}, B.~{Adebisi}, K.~M. {Rabie}, K.~{Anoh}, R.~E. {Ande},
  M.~{Hammoudeh}, H.~{Gacanin}, and U.~M. {Mbanaso}, ``Low-power wide area
  network technologies for internet-of-things: A comparative review,'' {\em
  IEEE Internet of Things Journal}, vol.~6, pp.~2225--2240, April 2019.

\bibitem{Survey_comp4}
H.~Djelouat, A.~Amira, and F.~Bensaali, ``Compressive sensing-based iot
  applications: A review,'' {\em Journal of Sensor and Actuator Networks},
  vol.~7, no.~4, 2018.

\bibitem{NOMA1}
L.~{Dai}, B.~{Wang}, Y.~{Yuan}, S.~{Han}, C.~{I}, and Z.~{Wang},
  ``Non-orthogonal multiple access for 5g: solutions, challenges,
  opportunities, and future research trends,'' {\em IEEE Communications
  Magazine}, vol.~53, pp.~74--81, Sep. 2015.

\bibitem{NOMA2}
M.~{Shirvanimoghaddam}, M.~{Dohler}, and S.~J. {Johnson}, ``Massive
  non-orthogonal multiple access for cellular iot: Potentials and
  limitations,'' {\em IEEE Communications Magazine}, vol.~55, pp.~55--61, Sep.
  2017.

\bibitem{Survey_comp3}
N.~{Ye}, H.~{Han}, L.~{Zhao}, and A.~hua {Wang}, ``Uplink nonorthogonal
  multiple access technologies toward 5g: A survey,'' {\em Wireless
  Communications and Mobile Computing}, vol.~2018, pp.~1--26, June 2018.

\bibitem{mMIMO12}
E.~d.~{Carvalho}, E.~{Bjornson}, J.~H. {Sorensen}, P.~{Popovski}, and E.~G.
  {Larsson}, ``Random access protocols for massive mimo,'' {\em IEEE
  Communications Magazine}, vol.~55, pp.~216--222, May 2017.

\bibitem{ML14}
S.~K. {Sharma} and X.~{Wang}, ``Towards massive machine type communications in
  ultra-dense cellular iot networks: Current issues and machine
  learning-assisted solutions,'' {\em IEEE Communications Surveys Tutorials},
  pp.~1--1, 2019.

\bibitem{Add_wifi}
L.~Oliveira, J.~Rodrigues, S.~Kozlov, R.~Rabêlo, and V.~Albuquerque, ``Mac
  layer protocols for internet of things: A survey,'' {\em Future Internet},
  vol.~11, p.~16, 01 2019.

\bibitem{Mora18}
H.~Mora, M.~T. Signes-Pont, D.~Gil, and M.~Johnsson, ``Collaborative working
  architecture for iot-based applications,'' {\em Sensors}, vol.~18, no.~6,
  2018.

\bibitem{HTCvsMTC}
Z.~{Dawy}, W.~{Saad}, A.~{Ghosh}, J.~G. {Andrews}, and E.~{Yaacoub}, ``Toward
  massive machine type cellular communications,'' {\em IEEE Wireless
  Communications}, vol.~24, pp.~120--128, February 2017.

\bibitem{bluetooth1}
``Ieee standard for telecommunications and information exchange between systems
  - lan/man - specific requirements - part 15: Wireless medium access control
  (mac) and physical layer (phy) specifications for wireless personal area
  networks (wpans),'' {\em IEEE Std 802.15.1-2002}, pp.~1--473, June 2002.

\bibitem{bluetooth2}
N.~V.~R. {Kumar}, C.~{Bhuvana}, and S.~{Anushya}, ``Comparison of zigbee and
  bluetooth wireless technologies-survey,'' in {\em 2017 International
  Conference on Information Communication and Embedded Systems (ICICES)},
  pp.~1--4, Feb 2017.

\bibitem{bluetooth8}
A.~F. {Harris III}, V.~{Khanna}, G.~{Tuncay}, R.~{Want}, and R.~{Kravets},
  ``Bluetooth low energy in dense iot environments,'' {\em IEEE Communications
  Magazine}, vol.~54, pp.~30--36, December 2016.

\bibitem{Add_BLE}
SIG, ``Bluetooth core specification, version 5.0,'' {\em
  https://www.bluetooth.com/ specifications/bluetooth-core-specification
  (Accessed 24/08/2019)}.

\bibitem{bluetooth4}
M.~{Ionascu} and M.~{Marcu}, ``Energy profiling for different bluetooth low
  energy designs,'' in {\em 2017 9th IEEE International Conference on
  Intelligent Data Acquisition and Advanced Computing Systems: Technology and
  Applications (IDAACS)}, vol.~2, pp.~1032--1036, Sep. 2017.

\bibitem{bluetooth5}
F.~S. {De Sousa}, C.~E. {Capovilla}, and I.~R.~S. {Casella}, ``Analysis of
  bluetooth low energy technology in indoor environments,'' in {\em 2016 IEEE
  International Symposium on Consumer Electronics (ISCE)}, pp.~55--56, Sep.
  2016.

\bibitem{Add_BLE1}
SIG, ``Bluetooth core specification v5.1 feature overview,'' {\em
  https://www.bluetooth.com/bluetooth-resources/
  bluetooth-core-specification-v5-1-feature-overview/ (Accessed 24/08/2019)}.

\bibitem{Add_BLE2}
SIG, ``Mesh profile bluetooth specification,'' {\em https://www.bluetooth.com
  (Accessed 10/10/2019)}.

\bibitem{Add_BLE3}
S.~Darroudi and C.~Gomez, ``Bluetooth low energy mesh networks: A survey,''
  {\em Sensors}, vol.~17, 06 2017.

\bibitem{Add_BLE5}
M.~{Baert}, J.~{Rossey}, A.~{Shahid}, and J.~{Hoebeke}, ``The bluetooth mesh
  standard: An overview and experimental evaluation,'' {\em Sensors}, vol.~18,
  2018.

\bibitem{Add_BLE4}
SIG, ``Mesh model bluetooth specification,'' {\em https://www.bluetooth.com
  (Accessed 10/10/2019)}.

\bibitem{bluetooth3}
``Bluetooth market update,'' may 2019.
\newblock https://www. bluetooth.com/.

\bibitem{bluetooth6}
B.~G.~A. {Kumar}, K.~C. {Bhagyalakshmi}, K.~{Lavanya}, and K.~H. {Gowranga},
  ``A bluetooth low energy based beacon system for smart short range
  surveillance,'' in {\em 2016 IEEE International Conference on Recent Trends
  in Electronics, Information Communication Technology (RTEICT)},
  pp.~1181--1184, May 2016.

\bibitem{bluetooth7}
Z.~{Lin}, C.~{Chang}, N.~{Chou}, and Y.~{Lin}, ``Bluetooth low energy (ble)
  based blood pressure monitoring system,'' in {\em 2014 International
  Conference on Intelligent Green Building and Smart Grid (IGBSG)}, pp.~1--4,
  April 2014.

\bibitem{zigbee1}
``Ieee approved draft standard for low-rate wireless personal area networks
  (wpans),'' {\em IEEE P802.15.4-REVc/D01, October 2015}, pp.~1--702, Jan 2015.

\bibitem{zigbee2}
C.~{Gomez} and J.~{Paradells}, ``Wireless home automation networks: A survey of
  architectures and technologies,'' {\em IEEE Communications Magazine},
  vol.~48, pp.~92--101, June 2010.

\bibitem{zigbee4}
X.~{Li } and X.~{Lu}, ``Design of a zigbee wireless sensor network node for
  aquaculture monitoring,'' in {\em 2016 2nd IEEE International Conference on
  Computer and Communications (ICCC)}, pp.~2179--2182, Oct 2016.

\bibitem{zigbee5}
S.~{Hebbar}, P.~{Pattar}, and V.~{Golla}, ``A mobile zigbee module in a traffic
  control system,'' {\em IEEE Potentials}, vol.~35, pp.~19--23, Jan 2016.

\bibitem{zigbee3}
Y.~{Kim}, S.~{Lee}, and S.~{Lee}, ``Coexistence of zigbee-based wban and wifi
  for health telemonitoring systems,'' {\em IEEE Journal of Biomedical and
  Health Informatics}, vol.~20, pp.~222--230, Jan 2016.

\bibitem{zigbee7}
I.~{Kuzminykh}, A.~{Snihurov}, and A.~{Carlsson}, ``Testing of communication
  range in zigbee technology,'' in {\em 2017 14th International Conference The
  Experience of Designing and Application of CAD Systems in Microelectronics
  (CADSM)}, pp.~133--136, Feb 2017.

\bibitem{zigbee6}
C.~{Park} and T.~S. {Rappaport}, ``Short-range wireless communications for
  next-generation networks: Uwb, 60 ghz millimeter-wave wpan, and zigbee,''
  {\em IEEE Wireless Communications}, vol.~14, pp.~70--78, August 2007.

\bibitem{Zigbee_add1}
P.~{Trelsmo}, P.~{Di Marco}, P.~{Skillermark}, R.~{Chirikov}, and J.~{Ostman},
  ``Evaluating ipv6 connectivity for ieee 802.15.4 and bluetooth low energy,''
  in {\em 2017 IEEE Wireless Communications and Networking Conference Workshops
  (WCNCW)}, pp.~1--6, March 2017.

\bibitem{Zigbee_add2}
L.~F. Del~Carpio, P.~Di~Marco, P.~Skillermark, R.~Chirikov, and K.~Lagergren,
  ``Comparison of 802.11ah, ble and 802.15.4 for a home automation use case,''
  {\em International Journal of Wireless Information Networks}, no.~3, p.~243,
  2017.

\bibitem{wifi1}
``Ieee standard for information technology - telecommunications and information
  exchange between systems - local and metropolitan area networks - specific
  requirements - part 11: Wireless lan medium access control (mac) and physical
  layer (phy) specifications - redline,'' {\em IEEE Std 802.11-2007 (Revision
  of IEEE Std 802.11-1999) - Redline}, pp.~1--1238, June 2007.

\bibitem{wifi2}
``Ieee draft standard for information technology--telecommunications and
  information exchange between system-- local and metropolitan area
  network--specific requirements part 11: Wireless lan medium access control
  (mac) and physical layer (phy) specifications amendment 5: Enhancements for
  higher throughput,'' {\em IEEE Unapproved Draft Std P802.11n/D7.0, Sep 2008},
  2008.

\bibitem{wifi3}
``Ieee standard for information technology-- telecommunications and information
  exchange between systemslocal and metropolitan area networks-- specific
  requirements--part 11: Wireless lan medium access control (mac) and physical
  layer (phy) specifications--amendment 4: Enhancements for very high
  throughput for operation in bands below 6 ghz,'' {\em IEEE P802.11ac/D5.0,
  January 2013}, pp.~1--440, Dec 2013.

\bibitem{Hal-1}
W.~Sun, M.~Choi, and S.~Choi, ``Ieee 802.11 ah: A long range 802.11 wlan at sub
  1 ghz,'' {\em Journal of ICT Standardization}, vol.~1, no.~1, pp.~83--108,
  2013.

\bibitem{Hal-2}
V.~Ba{\~n}os-Gonzalez, M.~S. Afaqui, E.~Lopez-Aguilera, and E.~Garcia-Villegas,
  ``Ieee 802.11 ah: A technology to face the iot challenge,'' {\em Sensors},
  vol.~16, no.~11, p.~1960, 2016.

\bibitem{OWC2}
M.~Z. {Chowdhury}, M.~T. {Hossan}, A.~{Islam}, and Y.~M. {Jang}, ``A
  comparative survey of optical wireless technologies: Architectures and
  applications,'' {\em IEEE Access}, vol.~6, pp.~9819--9840, 2018.

\bibitem{OWC4}
C.~W.~J. {Oh}, Z.~{Cao}, E.~{Tangdiongga}, and T.~{Koonen}, ``10 gbps
  all-optical full-duplex indoor optical wireless communication with wavelength
  reuse,'' in {\em 2016 Optical Fiber Communications Conference and Exhibition
  (OFC)}, pp.~1--3, March 2016.

\bibitem{OWC5}
C.~{Lim}, K.~{Wang}, and A.~{Nirmalathas}, ``Optical wireless communications
  for high-speed in-building personal area networks,'' in {\em 2016 18th
  International Conference on Transparent Optical Networks (ICTON)}, pp.~1--4,
  July 2016.

\bibitem{OWC6}
H.~{Le Minh}, D.~{O'Brien}, G.~{Faulkner}, O.~{Bouchet}, M.~{Wolf}, L.~{Grobe},
  and J.~{Li}, ``A 1.25-gb/s indoor cellular optical wireless communications
  demonstrator,'' {\em IEEE Photonics Technology Letters}, vol.~22,
  pp.~1598--1600, Nov 2010.

\bibitem{OWC7}
S.~R. {Teli}, S.~{Zvanovec}, and Z.~{Ghassemlooy}, ``Optical internet of things
  within 5g: Applications and challenges,'' in {\em 2018 IEEE International
  Conference on Internet of Things and Intelligence System (IOTAIS)},
  pp.~40--45, Nov 2018.

\bibitem{OWC8}
D.~R. {Dhatchayeny}, W.~A. {Cahyadi}, S.~R. {Teli}, and Y.~{Chung}, ``A novel
  optical body area network for transmission of multiple patient vital signs,''
  in {\em 2017 Ninth International Conference on Ubiquitous and Future Networks
  (ICUFN)}, pp.~542--544, July 2017.

\bibitem{OWC9}
T.~{Koonen}, ``Indoor optical wireless systems: Technology, trends, and
  applications,'' {\em Journal of Lightwave Technology}, vol.~36,
  pp.~1459--1467, April 2018.

\bibitem{OWC10}
R.~{Bian}, I.~{Tavakkolnia}, and H.~{Haas}, ``10.2 gb/s visible light
  communication with off-the-shelf leds,'' in {\em 2018 European Conference on
  Optical Communication (ECOC)}, pp.~1--3, Sep. 2018.

\bibitem{OWC11}
Y.-C. {Chi}, D.-H. {Hsieh}, C.-T. {Tsai}, H.-Y. {Chen}, H.-C. {Kuo}, and G.-R.
  {Lin}, ``450-nm gan laser diode enables high-speed visible light
  communication with 9-gbps qam-ofdm,'' {\em Opt. Express}, vol.~23,
  pp.~13051--13059, May 2015.

\bibitem{OWC12}
``Ieee standard for local and metropolitan area networks--part 15.7:
  Short-range wireless optical communication using visible light,'' {\em IEEE
  Std 802.15.7-2011}, pp.~1--309, Sep. 2011.

\bibitem{OWC13}
``Ieee draft standard for local and metropolitan area networks - part 15.7:
  Short-range optical wireless communications,'' {\em IEEE P802.15.7/D2a, June
  2018}, pp.~1--428, July 2018.

\bibitem{OWC14}
``http://www.ieee802.org/15/pub/tg13.html.''

\bibitem{OWC15}
D.~{Tsonev}, S.~{Videv}, and H.~{Haas}, ``Towards a 100 gb/s visible light
  wireless access network,'' {\em Opt. Express}, vol.~23, pp.~1627--1637, Jan
  2015.

\bibitem{OWC16}
H.~{Haas}, L.~{Yin}, Y.~{Wang}, and C.~{Chen}, ``What is lifi?,'' {\em Journal
  of Lightwave Technology}, vol.~34, pp.~1533--1544, March 2016.

\bibitem{OWC17}
T.~{Koonen}, J.~{Oh}, K.~{Mekonnen}, Z.~{Cao}, and E.~{Tangdiongga},
  ``Ultra-high capacity indoor optical wireless communication using 2d-steered
  pencil beams,'' {\em J. Lightwave Technol.}, vol.~34, pp.~4802--4809, Oct
  2016.

\bibitem{OWC18}
S.~Edirisinghege, C.~Lim, A.~Nirmalathas, E.~Wong, C.~Ranaweera, K.~Wang, and
  K.~Alameh, ``A novel mac protocol for indoor optical wireless networks,''
  {\em IET Communications}, 07 2019.

\bibitem{OWC19}
{\em A Novel Network Architecture for Indoor Optical Wireless Communication},
  2019.

\bibitem{OWC20}
A.~{Gomez}, K.~{Shi}, C.~{Quintana}, M.~{Sato}, G.~{Faulkner}, B.~C. {Thomsen},
  and D.~{O’Brien}, ``Beyond 100-gb/s indoor wide field-of-view optical
  wireless communications,'' {\em IEEE Photonics Technology Letters}, vol.~27,
  pp.~367--370, Feb 2015.

\bibitem{OWC21}
A.~{Gomez}, C.~{Quintana}, G.~{Faulkner}, and D.~{O’Brien},
  ``Point-to-multipoint holographic beamsteering techniques for indoor optical
  wireless communications,'' in {\em SPIE}, 2016.

\bibitem{OWC23}
C.~W. {Oh}, E.~{Tangdiongga}, and A.~M.~J. {Koonen}, ``42.8 gbit/s indoor
  optical wireless communication with 2-dimensional optical beam-steering,'' in
  {\em 2015 Optical Fiber Communications Conference and Exhibition (OFC)},
  pp.~1--3, March 2015.

\bibitem{ITU1}
ITU-R M.2083, {\em IMT 2020 Vision - Framework and overall objectives of the
  future development of IMT for 2020 and beyond}, December 2017.

\bibitem{ITU2}
ITU-R M.2412-0, {\em Guidelines for evaluation of radio interface technologies
  for IMT-2020}, December 2017.

\bibitem{ITU3}
ITU-R M.2410-0, {\em Minimum requirements related to technical performance for
  IMT-2020 radio interface(s)}, December 2017.

\bibitem{3G-5G}
3GPP TR 38.913, {\em Study on Scenarios and Requirements for Next Generation
  Access Technologies}, August 2016.

\bibitem{Polar1}
S.~B. {Korada}, E.~{Sasoglu}, and R.~{Urbanke}, ``Polar codes: Characterization
  of exponent, bounds, and constructions,'' {\em IEEE Transactions on
  Information Theory}, vol.~56, pp.~6253--6264, Dec 2010.

\bibitem{instruments_2018}
N.~Instruments, ``3gpp release 15 overview,'' {\em IEEE Spectrum: Technology,
  Engineering, and Science News}, Sep 2018.

\bibitem{qorvo_2017}
Qorvo, ``Getting to 5g: Comparing 4g and 5g system requirements,'' {\em
  https://www.qorvo.com/design-hub/blog/getting-to-5g-comparing-4g-and-5g-system-requirements},
  Sep 2017.

\bibitem{LowL0}
P.~{Popovski}, ``Ultra-reliable communication in 5g wireless systems,'' in {\em
  1st International Conference on 5G for Ubiquitous Connectivity},
  pp.~146--151, Nov 2014.

\bibitem{LowL6}
B.~{Soret}, P.~{Mogensen}, K.~I. {Pedersen}, and M.~C. {Aguayo-Torres},
  ``Fundamental tradeoffs among reliability, latency and throughput in cellular
  networks,'' in {\em 2014 IEEE Globecom Workshops (GC Wkshps)},
  pp.~1391--1396, Dec 2014.

\bibitem{LowL7}
P.~{Schulz}, M.~{Matthe}, H.~{Klessig}, M.~{Simsek}, G.~{Fettweis},
  J.~{Ansari}, S.~A. {Ashraf}, B.~{Almeroth}, J.~{Voigt}, I.~{Riedel},
  A.~{Puschmann}, A.~{Mitschele-Thiel}, M.~{Muller}, T.~{Elste}, and
  M.~{Windisch}, ``Latency critical iot applications in 5g: Perspective on the
  design of radio interface and network architecture,'' {\em IEEE
  Communications Magazine}, vol.~55, pp.~70--78, Feb 2017.

\bibitem{LowL1}
H.~{Chen}, R.~{Abbas}, P.~{Cheng}, M.~{Shirvanimoghaddam}, W.~{Hardjawana},
  W.~{Bao}, Y.~{Li}, and B.~{Vucetic}, ``Ultra-reliable low latency cellular
  networks: Use cases, challenges and approaches,'' {\em IEEE Communications
  Magazine}, vol.~56, pp.~119--125, December 2018.

\bibitem{LowL2}
J.~{Choi}, ``Low-latency multichannel aloha with fast retrial for machine-type
  communications,'' {\em IEEE Internet of Things Journal}, vol.~6,
  pp.~3175--3185, April 2019.

\bibitem{LowL3}
J.~{Choi}, ``An effective capacity-based approach to multi-channel low-latency
  wireless communications,'' {\em IEEE Transactions on Communications},
  vol.~67, pp.~2476--2486, March 2019.

\bibitem{LowL4}
P.~{Popovski}, J.~J. {Nielsen}, C.~{Stefanovic}, E.~d.~{Carvalho}, E.~{Strom},
  K.~F. {Trillingsgaard}, A.~{Bana}, D.~M. {Kim}, R.~{Kotaba}, J.~{Park}, and
  R.~B. {Sorensen}, ``Wireless access for ultra-reliable low-latency
  communication: Principles and building blocks,'' {\em IEEE Network}, vol.~32,
  pp.~16--23, March 2018.

\bibitem{LowL5}
H.~{Ji}, S.~{Park}, and B.~{Shim}, ``Sparse vector coding for ultra reliable
  and low latency communications,'' {\em IEEE Transactions on Wireless
  Communications}, vol.~17, pp.~6693--6706, Oct 2018.

\bibitem{Add_5g}
X.~{Jiang}, H.~{Shokri-Ghadikolaei}, G.~{Fodor}, E.~{Modiano}, Z.~{Pang},
  M.~{Zorzi}, and C.~{Fischione}, ``Low-latency networking: Where latency lurks
  and how to tame it,'' {\em Proceedings of the IEEE}, vol.~107, pp.~280--306,
  Feb 2019.

\bibitem{5G_add}
A.~{Aijaz}, M.~{Dohler}, A.~H. {Aghvami}, V.~{Friderikos}, and M.~{Frodigh},
  ``Realizing the tactile internet: Haptic communications over next generation
  5g cellular networks,'' {\em IEEE Wireless Communications}, vol.~24,
  pp.~82--89, April 2017.

\bibitem{5G_add1}
S.~{Djahel}, R.~{Doolan}, G.~{Muntean}, and J.~{Murphy}, ``A
  communications-oriented perspective on traffic management systems for smart
  cities: Challenges and innovative approaches,'' {\em IEEE Communications
  Surveys Tutorials}, vol.~17, pp.~125--151, Firstquarter 2015.

\bibitem{5G_add2}
H.~F. {Azgomi} and M.~{Jamshidi}, ``A brief survey on smart community and smart
  transportation,'' in {\em 2018 IEEE 30th International Conference on Tools
  with Artificial Intelligence (ICTAI)}, pp.~932--939, Nov 2018.

\bibitem{LA-1}
J.~G. Andrews, S.~Buzzi, W.~Choi, S.~V. Hanly, A.~Lozano, A.~C. Soong, and
  J.~C. Zhang, ``What will 5g be?,'' {\em IEEE Journal on selected areas in
  communications}, vol.~32, no.~6, pp.~1065--1082, 2014.

\bibitem{LPWAN5}
U.~{Raza}, P.~{Kulkarni}, and M.~{Sooriyabandara}, ``Low power wide area
  networks: An overview,'' {\em IEEE Communications Surveys Tutorials},
  vol.~19, pp.~855--873, Secondquarter 2017.

\bibitem{LPWAN1}
W.~{Ayoub}, A.~E. {Samhat}, F.~{Nouvel}, M.~{Mroue}, and J.~{Prévotet},
  ``Internet of mobile things: Overview of lorawan, dash7, and nb-iot in lpwans
  standards and supported mobility,'' {\em IEEE Communications Surveys
  Tutorials}, vol.~21, pp.~1561--1581, Secondquarter 2019.

\bibitem{LPWAN4}
W.~{Guibene}, K.~E. {Nolan}, and M.~Y. {Kelly}, ``Survey on clean slate
  cellular-iot standard proposals,'' in {\em 2015 IEEE International Conference
  on Computer and Information Technology; Ubiquitous Computing and
  Communications; Dependable, Autonomic and Secure Computing; Pervasive
  Intelligence and Computing}, pp.~1596--1599, Oct 2015.

\bibitem{Lora2DJ}
F.~{Adelantado}, X.~{Vilajosana}, P.~{Tuset-Peiro}, B.~{Martinez},
  J.~{Melia-Segui}, and T.~{Watteyne}, ``Understanding the limits of lorawan,''
  {\em IEEE Communications Magazine}, vol.~55, pp.~34--40, Sep. 2017.

\bibitem{LoraWAN1}
J.~{Navarro-Ortiz}, S.~{Sendra}, P.~{Ameigeiras}, and J.~M. {Lopez-Soler},
  ``Integration of lorawan and 4g/5g for the industrial internet of things,''
  {\em IEEE Communications Magazine}, vol.~56, pp.~60--67, Feb 2018.

\bibitem{Lora3}
N.~{Sornin}, M.~{Luis}, T.~{Eirich}, T.~{Kramp}, and O.~{Hersent}, ``Lora
  specification 1.0,'' {\em Lora Alliance Standard specification}, 2015.

\bibitem{Mobility_Lora}
{\'O}.~Alvear, J.~Herrera-Tapia, C.~T. Calafate, E.~Hern{\'a}ndez-Orallo, J.-C.
  Cano, and P.~Manzoni, ``Assessing the impact of mobility on lora
  communications,'' in {\em Interoperability, Safety and Security in IoT},
  (Cham), pp.~75--81, Springer International Publishing, 2018.

\bibitem{Sigfox2}
N.~I. {Osman} and E.~B. {Abbas}, ``Simulation and modelling of lora and sigfox
  low power wide area network technologies,'' in {\em 2018 International
  Conference on Computer, Control, Electrical, and Electronics Engineering
  (ICCCEEE)}, pp.~1--5, Aug 2018.

\bibitem{Sigfox3}
K.~{Mekki}, E.~{Bajic}, F.~{Chaxel}, and F.~{Meyer}, ``Overview of cellular
  lpwan technologies for iot deployment: Sigfox, lorawan, and nb-iot,'' in {\em
  2018 IEEE International Conference on Pervasive Computing and Communications
  Workshops (PerCom Workshops)}, pp.~197--202, March 2018.

\bibitem{PRACH1}
A.~{Laya}, L.~{Alonso}, and J.~{Alonso-Zarate}, ``Is the random access channel
  of lte and lte-a suitable for m2m communications? a survey of alternatives,''
  {\em IEEE Communications Surveys Tutorials}, vol.~16, pp.~4--16, First 2014.

\bibitem{PRACH2}
A.~{Zanella}, M.~{Zorzi}, A.~F. {dos Santos}, P.~{Popovski}, N.~{Pratas},
  C.~{Stefanovic}, A.~{Dekorsy}, C.~{Bockelmann}, B.~{Busropan}, and T.~A.
  H.~J. {Norp}, ``M2m massive wireless access: Challenges, research issues, and
  ways forward,'' in {\em 2013 IEEE Globecom Workshops (GC Wkshps)},
  pp.~151--156, Dec. 2013.

\bibitem{LTEM}
A.~{Hoglund}, J.~{Bergman}, X.~{Lin}, O.~{Liberg}, A.~{Ratilainen}, H.~S.
  {Razaghi}, T.~{Tirronen}, and E.~A. {Yavuz}, ``Overview of 3gpp release 14
  further enhanced mtc,'' {\em IEEE Communications Standards Magazine}, vol.~2,
  pp.~84--89, JUNE 2018.

\bibitem{LTEM1}
R.~{Ratasuk}, N.~{Mangalvedhe}, D.~{Bhatoolaul}, and A.~{Ghosh}, ``Lte-m
  evolution towards 5g massive mtc,'' in {\em 2017 IEEE Globecom Workshops (GC
  Wkshps)}, pp.~1--6, Dec 2017.

\bibitem{LTEM6}
S.~{Dawaliby}, A.~{Bradai}, and Y.~{Pousset}, ``In depth performance evaluation
  of lte-m for m2m communications,'' in {\em 2016 IEEE 12th International
  Conference on Wireless and Mobile Computing, Networking and Communications
  (WiMob)}, pp.~1--8, Oct 2016.

\bibitem{LTEM4}
{3GPP RP-170532}, ``Revised wid for further enhanced mtc for lte,'' 2017.

\bibitem{LTEM5}
{3GPP RP-170732}, ``Even further enhanced mtc for lte,'' 2017.

\bibitem{NB-IOT}
M.~{Chen}, Y.~{Miao}, Y.~{Hao}, and K.~{Hwang}, ``Narrow band internet of
  things,'' {\em IEEE Access}, vol.~5, pp.~20557--20577, 2017.

\bibitem{NBIOT3}
R.~{Ratasuk}, N.~{Mangalvedhe}, Z.~{Xiong}, M.~{Robert}, and D.~{Bhatoolaul},
  ``Enhancements of narrowband iot in 3gpp rel-14 and rel-15,'' in {\em 2017
  IEEE Conference on Standards for Communications and Networking (CSCN)},
  pp.~60--65, Sep. 2017.

\bibitem{NBIOT1}
{TR 45.820}, ``Cellular system support for ultra low complexity and low
  throughput internet of things,'' {\em V2.1.0}, 2015.

\bibitem{rico2016overview}
A.~Rico-Alvarino, M.~Vajapeyam, H.~Xu, X.~Wang, Y.~Blankenship, J.~Bergman,
  T.~Tirronen, and E.~Yavuz, ``An overview of 3gpp enhancements on machine to
  machine communications,'' {\em IEEE Communications Magazine}, vol.~54, no.~6,
  pp.~14--21, 2016.

\bibitem{hoglund2017overview}
A.~Hoglund, X.~Lin, O.~Liberg, A.~Behravan, E.~A. Yavuz, M.~Van Der~Zee,
  Y.~Sui, T.~Tirronen, A.~Ratilainen, and D.~Eriksson, ``Overview of 3gpp
  release 14 enhanced nb-iot,'' {\em IEEE Network}, vol.~31, no.~6, pp.~16--22,
  2017.

\bibitem{lee2017prediction}
J.~Lee and J.~Lee, ``Prediction-based energy saving mechanism in 3gpp nb-iot
  networks,'' {\em Sensors}, vol.~17, no.~9, p.~2008, 2017.

\bibitem{NB-IoT_Penetrate}
3GPP TR 36.888, {\em Study on provision of low-cost Machine-Type Communications
  (MTC) User Equipments (UEs) based on LTE (v12.0.0)}, August 2013.

\bibitem{NBIOT2}
R.~{Ratasuk}, N.~{Mangalvedhe}, Y.~{Zhang}, M.~{Robert}, and J.~{Koskinen},
  ``Overview of narrowband iot in lte rel-13,'' in {\em 2016 IEEE Conference on
  Standards for Communications and Networking (CSCN)}, pp.~1--7, Oct 2016.

\bibitem{Chap4_1}
M.~{Hasan}, E.~{Hossain}, and D.~{Niyato}, ``Random access for
  machine-to-machine communication in lte-advanced networks: issues and
  approaches,'' {\em IEEE Communications Magazine}, vol.~51, pp.~86--93, June
  2013.

\bibitem{mMIMO11}
M.~Sesia, M.~Toufik, and M.~Baker, {\em LTE, the UMTS long term evolution: from
  theory to practice}.
\newblock 01 2009.

\bibitem{mMIMO17}
C.~{Schlegel}, R.~{Kempter}, and P.~{Kota}, ``A novel random wireless packet
  multiple access method using cdma,'' {\em IEEE Transactions on Wireless
  Communications}, vol.~5, pp.~1362--1370, June 2006.

\bibitem{CS1}
H.~Schepker, C.~Bockelmann, and A.~Dekorsy, ``Exploiting sparsity in channel
  and data estimation for sporadic multi-user communication,'' 08 2013.

\bibitem{CS4}
G.~{Wunder}, C.~{Stefanović}, P.~{Popovski}, and L.~{Thiele}, ``Compressive
  coded random access for massive mtc traffic in 5g systems,'' in {\em 2015
  49th Asilomar Conference on Signals, Systems and Computers}, pp.~13--17, Nov
  2015.

\bibitem{CS2}
J.~{Choi}, ``Two-stage multiple access for many devices of unique
  identifications over frequency-selective fading channels,'' {\em IEEE
  Internet of Things Journal}, vol.~4, pp.~162--171, Feb 2017.

\bibitem{CS7}
J.~{Choi}, K.~{Lee}, and N.~Y. {Yu}, ``Compressive random access using multiple
  resource blocks for mtc,'' in {\em 2016 IEEE Globecom Workshops (GC Wkshps)},
  pp.~1--5, Dec 2016.

\bibitem{CS3}
J.~{Choi} and N.~Y. {Yu}, ``Compressive channel division multiple access for
  mtc under frequency-selective fading,'' {\em IEEE Transactions on
  Communications}, vol.~65, pp.~2715--2725, June 2017.

\bibitem{CS5}
D.~L. {Donoho}, ``Compressed sensing,'' {\em IEEE Transactions on Information
  Theory}, vol.~52, pp.~1289--1306, April 2006.

\bibitem{CS6}
Y.~C. {Eldar} and G.~{Kutyniok}, {\em Compressed Sensing: Theory and
  Applications}.
\newblock Cambridge, U.K.: Cambridge Univ. Press, 2012.

\bibitem{NOMA4}
Z.~{Wei}, J.~{Yuan}, D.~W.~K. {Ng}, M.~{Elkashlan}, and Z.~{Ding}, ``A survey
  of downlink non-orthogonal multiple access for 5g wireless communication
  networks,'' {\em ZTE Communications}, 09 2016.

\bibitem{NOMA5}
Y.~{Saito}, A.~{Benjebbour}, Y.~{Kishiyama}, and T.~{Nakamura}, ``System-level
  performance evaluation of downlink non-orthogonal multiple access (noma),''
  in {\em 2013 IEEE 24th Annual International Symposium on Personal, Indoor,
  and Mobile Radio Communications (PIMRC)}, pp.~611--615, Sep. 2013.

\bibitem{NOMA6}
N.~{Zhang}, J.~{Wang}, G.~{Kang}, and Y.~{Liu}, ``Uplink non-orthogonal
  multiple access in 5g systems,'' {\em IEEE Communications Letters}, vol.~20,
  pp.~458--461, March 2016.

\bibitem{NOMA7}
Z.~{Ding}, P.~{Fan}, and H.~V. {Poor}, ``Impact of user pairing on 5g
  non-orthogonal multiple-access downlink transmissions,'' {\em IEEE
  Transactions on Vehicular Technology}, vol.~65, pp.~6010--6023, Aug 2016.

\bibitem{choi2008h}
J.~Choi, ``H-arq based non-orthogonal multiple access with successive
  interference cancellation,'' in {\em IEEE GLOBECOM 2008-2008 IEEE Global
  Telecommunications Conference}, pp.~1--5, IEEE, 2008.

\bibitem{mm-NOMA}
T.~{Lv}, Y.~{Ma}, J.~{Zeng}, and P.~T. {Mathiopoulos}, ``Millimeter-wave noma
  transmission in cellular m2m communications for internet of things,'' {\em
  IEEE Internet of Things Journal}, vol.~5, pp.~1989--2000, June 2018.

\bibitem{NOMA3}
Z.~{Ding}, Y.~{Liu}, J.~{Choi}, Q.~{Sun}, M.~{Elkashlan}, C.~{I}, and H.~V.
  {Poor}, ``Application of non-orthogonal multiple access in lte and 5g
  networks,'' {\em IEEE Communications Magazine}, vol.~55, pp.~185--191,
  February 2017.

\bibitem{NOMA11}
J.~{Choi}, ``Noma-based random access with multichannel aloha,'' {\em IEEE
  Journal on Selected Areas in Communications}, vol.~35, pp.~2736--2743, Dec
  2017.

\bibitem{NOMA12}
J.~{Choi}, ``Multichannel noma-aloha game with fading,'' {\em IEEE Transactions
  on Communications}, vol.~66, pp.~4997--5007, Oct 2018.

\bibitem{NOMA8}
Z.~{Yang}, W.~{Xu}, H.~{Xu}, J.~{Shi}, and M.~{Chen}, ``Energy efficient
  non-orthogonal multiple access for machine-to-machine communications,'' {\em
  IEEE Communications Letters}, vol.~21, pp.~817--820, April 2017.

\bibitem{NOMA13}
K.~{Kang}, Z.~{Pan}, J.~{Liu}, and S.~{Shimamoto}, ``A game theory based power
  control algorithm for future mtc noma networks,'' in {\em 2017 14th IEEE
  Annual Consumer Communications Networking Conference (CCNC)}, pp.~203--208,
  Jan 2017.

\bibitem{NOMA9}
Z.~{Ding}, L.~{Dai}, and H.~V. {Poor}, ``Mimo-noma design for small packet
  transmission in the internet of things,'' {\em IEEE Access}, vol.~4,
  pp.~1393--1405, 2016.

\bibitem{NOMA10}
A.~E. {Mostafa}, Y.~{Zhou}, and V.~W.~S. {Wong}, ``Connectivity maximization
  for narrowband iot systems with noma,'' in {\em 2017 IEEE International
  Conference on Communications (ICC)}, pp.~1--6, May 2017.

\bibitem{NOMA14}
J.~{Choi}, ``Noma-based compressive random access using gaussian spreading,''
  {\em IEEE Transactions on Communications}, pp.~1--1, 2019.

\bibitem{NOMA16}
B.~{Wang}, L.~{Dai}, Y.~{Zhang}, T.~{Mir}, and J.~{Li}, ``Dynamic compressive
  sensing-based multi-user detection for uplink grant-free noma,'' {\em IEEE
  Communications Letters}, vol.~20, pp.~2320--2323, Nov 2016.

\bibitem{mMIMO1}
T.~L. {Marzetta}, ``Noncooperative cellular wireless with unlimited numbers of
  base station antennas,'' {\em IEEE Transactions on Wireless Communications},
  vol.~9, pp.~3590--3600, November 2010.

\bibitem{mMIMO2}
E.~G. {Larsson}, O.~{Edfors}, F.~{Tufvesson}, and T.~L. {Marzetta}, ``Massive
  mimo for next generation wireless systems,'' {\em IEEE Communications
  Magazine}, vol.~52, pp.~186--195, February 2014.

\bibitem{mMIMO3}
L.~{Lu}, G.~Y. {Li}, A.~L. {Swindlehurst}, A.~{Ashikhmin}, and R.~{Zhang}, ``An
  overview of massive mimo: Benefits and challenges,'' {\em IEEE Journal of
  Selected Topics in Signal Processing}, vol.~8, pp.~742--758, Oct 2014.

\bibitem{mMIMO4}
F.~{Boccardi}, R.~W. {Heath}, A.~{Lozano}, T.~L. {Marzetta}, and P.~{Popovski},
  ``Five disruptive technology directions for 5g,'' {\em IEEE Communications
  Magazine}, vol.~52, pp.~74--80, February 2014.

\bibitem{mMIMO5}
E.~{Björnson}, J.~{Hoydis}, and L.~{Sanguinetti}, ``Massive mimo has unlimited
  capacity,'' {\em IEEE Transactions on Wireless Communications}, vol.~17,
  pp.~574--590, Jan 2018.

\bibitem{mMIMO6}
H.~Q. {Ngo}, E.~G. {Larsson}, and T.~L. {Marzetta}, ``Energy and spectral
  efficiency of very large multiuser mimo systems,'' {\em IEEE Transactions on
  Communications}, vol.~61, pp.~1436--1449, April 2013.

\bibitem{mMIMO7}
E.~{Björnson}, E.~G. {Larsson}, and M.~{Debbah}, ``Massive mimo for maximal
  spectral efficiency: How many users and pilots should be allocated?,'' {\em
  IEEE Transactions on Wireless Communications}, vol.~15, pp.~1293--1308, Feb
  2016.

\bibitem{mMIMO8}
H.~{Yang} and T.~L. {Marzetta}, ``Total energy efficiency of cellular large
  scale antenna system multiple access mobile networks,'' in {\em 2013 IEEE
  Online Conference on Green Communications (OnlineGreenComm)}, pp.~27--32, Oct
  2013.

\bibitem{mMIMO9}
E.~{Björnson}, L.~{Sanguinetti}, J.~{Hoydis}, and M.~{Debbah}, ``Optimal
  design of energy-efficient multi-user mimo systems: Is massive mimo the
  answer?,'' {\em IEEE Transactions on Wireless Communications}, vol.~14,
  pp.~3059--3075, June 2015.

\bibitem{mMIMO10}
S.~{Jin}, J.~{Wang}, Q.~{Sun}, M.~{Matthaiou}, and X.~{Gao}, ``Cell coverage
  optimization for the multicell massive mimo uplink,'' {\em IEEE Transactions
  on Vehicular Technology}, vol.~64, pp.~5713--5727, Dec 2015.

\bibitem{mMIMO13}
H.~{Han}, X.~{Guo}, and Y.~{Li}, ``A high throughput pilot allocation for m2m
  communication in crowded massive mimo systems,'' {\em IEEE Transactions on
  Vehicular Technology}, vol.~66, pp.~9572--9576, Oct 2017.

\bibitem{mMIMO14}
E.~{Björnson}, E.~{de Carvalho}, E.~G. {Larsson}, and P.~{Popovski}, ``Random
  access protocol for massive mimo: Strongest-user collision resolution
  (sucr),'' in {\em 2016 IEEE International Conference on Communications
  (ICC)}, pp.~1--6, May 2016.

\bibitem{mMIMO15}
F.~{Ahsan} and A.~{Sabharwal}, ``Leveraging massive mimo spatial degrees of
  freedom to reduce random access delay,'' in {\em 2017 51st Asilomar
  Conference on Signals, Systems, and Computers}, pp.~2007--2011, Oct 2017.

\bibitem{mMIMO16}
L.~{Sanguinetti}, A.~A. {D’Amico}, M.~{Morelli}, and M.~{Debbah}, ``Random
  access in massive mimo by exploiting timing offsets and excess antennas,''
  {\em IEEE Transactions on Communications}, vol.~66, pp.~6081--6095, Dec 2018.

\bibitem{mMIMO18}
J.~{Ding}, D.~{Qu}, H.~{Jiang}, and T.~{Jiang}, ``Success probability of
  grant-free random access with massive mimo,'' {\em IEEE Internet of Things
  Journal}, vol.~6, pp.~506--516, Feb 2019.

\bibitem{mMIMO19}
J.~{Ding}, D.~{Qu}, and H.~{Jiang}, ``Optimal preamble length for spectral
  efficiency in grant-free ra with massive mimo,'' in {\em 2019 International
  Conference on Electronics, Information, and Communication (ICEIC)}, pp.~1--5,
  Jan 2019.

\bibitem{mMIMO20}
J.~{Ding}, D.~{Qu}, H.~{Jiang}, and T.~{Jiang}, ``Virtual carrier sensing-based
  random access in massive mimo systems,'' {\em IEEE Transactions on Wireless
  Communications}, vol.~17, pp.~6590--6600, Oct 2018.

\bibitem{mMIMO21}
E.~{de Carvalho}, E.~{Björnson}, J.~H. {Sørensen}, E.~G. {Larsson}, and
  P.~{Popovski}, ``Random pilot and data access in massive mimo for
  machine-type communications,'' {\em IEEE Transactions on Wireless
  Communications}, vol.~16, pp.~7703--7717, Dec 2017.

\bibitem{mMIMO22}
L.~{Liu} and W.~{Yu}, ``Massive connectivity with massive mimo—part i: Device
  activity detection and channel estimation,'' {\em IEEE Transactions on Signal
  Processing}, vol.~66, pp.~2933--2946, June 2018.

\bibitem{mMIMO23}
K.~{Senel} and E.~G. {Larsson}, ``Grant-free massive mtc-enabled massive mimo:
  A compressive sensing approach,'' {\em IEEE Transactions on Communications},
  vol.~66, pp.~6164--6175, Dec 2018.

\bibitem{mMIMO24}
U.~{Madhow}, D.~R. {Brown}, S.~{Dasgupta}, and R.~{Mudumbai}, ``Distributed
  massive mimo: Algorithms, architectures and concept systems,'' in {\em 2014
  Information Theory and Applications Workshop (ITA)}, pp.~1--7, Feb 2014.

\bibitem{mMIMO25}
A.~{Yang}, Y.~{Jing}, C.~{Xing}, Z.~{Fei}, and J.~{Kuang}, ``Performance
  analysis and location optimization for massive mimo systems with circularly
  distributed antennas,'' {\em IEEE Transactions on Wireless Communications},
  vol.~14, pp.~5659--5671, Oct 2015.

\bibitem{mMIMO26}
J.~{Joung}, Y.~K. {Chia}, and S.~{Sun}, ``Energy-efficient, large-scale
  distributed-antenna system (l-das) for multiple users,'' {\em IEEE Journal of
  Selected Topics in Signal Processing}, vol.~8, pp.~954--965, Oct 2014.

\bibitem{mMIMO27}
E.~{Björnson}, M.~{Matthaiou}, and M.~{Debbah}, ``Massive mimo with non-ideal
  arbitrary arrays: Hardware scaling laws and circuit-aware design,'' {\em IEEE
  Transactions on Wireless Communications}, vol.~14, pp.~4353--4368, Aug 2015.

\bibitem{mMIMO28}
Q.~{He}, Z.~{Chen}, T.~Q.~S. {Quek}, J.~{Choi}, and S.~{Li}, ``Compressive
  channel estimation and user activity detection in distributed-input
  distributed-output systems,'' {\em IEEE Communications Letters}, vol.~22,
  pp.~1850--1853, Sep. 2018.

\bibitem{mMIMO29}
J.~{Choi}, ``Compressive random access for mtc in distributed input distributed
  output systems,'' in {\em 2017 IEEE 85th Vehicular Technology Conference (VTC
  Spring)}, pp.~1--5, June 2017.

\bibitem{ML0}
Y.~{Sun}, M.~{Peng}, Y.~{Zhou}, Y.~{Huang}, and S.~{Mao}, ``Application of
  machine learning in wireless networks: Key techniques and open issues,'' {\em
  IEEE Communications Surveys Tutorials}, pp.~1--1, 2019.

\bibitem{ML3}
L.~{Busoniu}, R.~{Babuska}, and B.~{De Schutter}, ``A comprehensive survey of
  multiagent reinforcement learning,'' {\em IEEE Transactions on Systems, Man,
  and Cybernetics, Part C (Applications and Reviews)}, vol.~38, pp.~156--172,
  March 2008.

\bibitem{ML4}
K.~{Arulkumaran}, M.~P. {Deisenroth}, M.~{Brundage}, and A.~A. {Bharath},
  ``Deep reinforcement learning: A brief survey,'' {\em IEEE Signal Processing
  Magazine}, vol.~34, pp.~26--38, Nov 2017.

\bibitem{ML2}
Z.~M. {Fadlullah}, F.~{Tang}, B.~{Mao}, N.~{Kato}, O.~{Akashi}, T.~{Inoue}, and
  K.~{Mizutani}, ``State-of-the-art deep learning: Evolving machine
  intelligence toward tomorrow’s intelligent network traffic control
  systems,'' {\em IEEE Communications Surveys Tutorials}, vol.~19,
  pp.~2432--2455, Fourthquarter 2017.

\bibitem{ML5}
R.~C. {Daniels} and R.~W. {Heath}, ``An online learning framework for link
  adaptation in wireless networks,'' in {\em 2009 Information Theory and
  Applications Workshop}, pp.~138--140, Feb 2009.

\bibitem{ML6}
S.~{Yun} and C.~{Caramanis}, ``Reinforcement learning for link adaptation in
  mimo-ofdm wireless systems,'' in {\em 2010 IEEE Global Telecommunications
  Conference GLOBECOM 2010}, pp.~1--5, Dec 2010.

\bibitem{ML7}
X.~{Chen}, J.~{Wu}, Y.~{Cai}, H.~{Zhang}, and T.~{Chen}, ``Energy-efficiency
  oriented traffic offloading in wireless networks: A brief survey and a
  learning approach for heterogeneous cellular networks,'' {\em IEEE Journal on
  Selected Areas in Communications}, vol.~33, pp.~627--640, April 2015.

\bibitem{ML1}
N.~{Kato}, Z.~M. {Fadlullah}, B.~{Mao}, F.~{Tang}, O.~{Akashi}, T.~{Inoue}, and
  K.~{Mizutani}, ``The deep learning vision for heterogeneous network traffic
  control: Proposal, challenges, and future perspective,'' {\em IEEE Wireless
  Communications}, vol.~24, pp.~146--153, June 2017.

\bibitem{ML8}
T.~{Park} and W.~{Saad}, ``Resource allocation and coordination for critical
  messages using finite memory learning,'' in {\em 2016 IEEE Globecom Workshops
  (GC Wkshps)}, pp.~1--6, Dec 2016.

\bibitem{ML9}
V.~{Rakovic} and L.~{Gavrilovska}, ``Novel rat selection mechanism based on
  hopfield neural networks,'' in {\em International Congress on Ultra Modern
  Telecommunications and Control Systems}, pp.~210--217, Oct 2010.

\bibitem{ML10}
A.~H. {Mohammed}, A.~S. {Khwaja}, A.~{Anpalagan}, and I.~{Woungang}, ``Base
  station selection in m2m communication using q-learning algorithm in lte-a
  networks,'' in {\em 2015 IEEE 29th International Conference on Advanced
  Information Networking and Applications}, pp.~17--22, March 2015.

\bibitem{ML11}
L.~M. {Bello}, P.~{Mitchell}, and D.~{Grace}, ``Application of q-learning for
  rach access to support m2m traffic over a cellular network,'' in {\em
  European Wireless 2014; 20th European Wireless Conference}, pp.~1--6, May
  2014.

\bibitem{ML13}
Y.~{Ruan}, W.~{Wang}, Z.~{Zhang}, and V.~K.~N. {Lau}, ``Delay-aware massive
  random access for machine-type communications via hierarchical stochastic
  learning,'' in {\em 2017 IEEE International Conference on Communications
  (ICC)}, pp.~1--6, May 2017.

\bibitem{ML12}
{Jihun Moon} and {Yujin Lim}, ``Access control of mtc devices using
  reinforcement learning approach,'' in {\em 2017 International Conference on
  Information Networking (ICOIN)}, pp.~641--643, Jan 2017.

\bibitem{APP-1}
J.~Kim, D.~Kim, and S.~Choi, ``3gpp sa2 architecture and functions for 5g
  mobile communication system,'' {\em ICT Express}, vol.~3, no.~1, pp.~1--8,
  2017.

\bibitem{APP-3}
A.~{Al-Fuqaha}, M.~{Guizani}, M.~{Mohammadi}, M.~{Aledhari}, and M.~{Ayyash},
  ``Internet of things: A survey on enabling technologies, protocols, and
  applications,'' {\em IEEE Communications Surveys Tutorials}, vol.~17,
  pp.~2347--2376, Fourthquarter 2015.

\bibitem{APP-4}
N.~{Shahid} and S.~{Aneja}, ``Internet of things: Vision, application areas and
  research challenges,'' in {\em 2017 International Conference on I-SMAC (IoT
  in Social, Mobile, Analytics and Cloud) (I-SMAC)}, pp.~583--587, Feb 2017.

\bibitem{APP-5}
N.~{Dlodlo}, O.~{Gcaba}, and A.~{Smith}, ``Internet of things technologies in
  smart cities,'' in {\em 2016 IST-Africa Week Conference}, pp.~1--7, May 2016.

\bibitem{APP-6}
H.~Sundmaeker, P.~Guillemin, P.~Friess, and S.~Woelffl{\'e}, ``Vision and
  challenges for realising the internet of things,'' {\em Cluster of European
  Research Projects on the Internet of Things, European Commision}, vol.~3,
  no.~3, pp.~34--36, 2010.

\bibitem{App-7}
D.~Singh, G.~Tripathi, and A.~J. Jara, ``A survey of internet-of-things: Future
  vision, architecture, challenges and services,'' in {\em 2014 IEEE World
  Forum on Internet of Things (WF-IoT)}, pp.~287--292, IEEE, 2014.

\bibitem{APP-8}
L.~Atzori, A.~Iera, and G.~Morabito, ``The internet of things: A survey,'' {\em
  Computer networks}, vol.~54, no.~15, pp.~2787--2805, 2010.

\bibitem{Tr-1}
S.~H. Sutar, R.~Koul, and R.~Suryavanshi, ``Integration of smart phone and iot
  for development of smart public transportation system,'' in {\em 2016
  International Conference on Internet of Things and Applications (IOTA)},
  pp.~73--78, IEEE, 2016.

\bibitem{Tr-2}
A.~Zanella, N.~Bui, A.~Castellani, L.~Vangelista, and M.~Zorzi, ``Internet of
  things for smart cities,'' {\em IEEE Internet of Things journal}, vol.~1,
  no.~1, pp.~22--32, 2014.

\bibitem{Tr-3}
S.~R. Islam, D.~Kwak, M.~H. Kabir, M.~Hossain, and K.-S. Kwak, ``The internet
  of things for health care: a comprehensive survey,'' {\em IEEE Access},
  vol.~3, pp.~678--708, 2015.

\bibitem{Tr-5}
M.~Ryu, J.~Yun, T.~Miao, I.-Y. Ahn, S.-C. Choi, and J.~Kim, ``Design and
  implementation of a connected farm for smart farming system,'' in {\em 2015
  IEEE SENSORS}, pp.~1--4, IEEE, 2015.

\bibitem{App-2}
L.~Da~Xu, W.~He, and S.~Li, ``Internet of things in industries: A survey,''
  {\em IEEE Transactions on industrial informatics}, vol.~10, no.~4,
  pp.~2233--2243, 2014.

\bibitem{Tr-6}
B.~L.~R. Stojkoska and K.~V. Trivodaliev, ``A review of internet of things for
  smart home: Challenges and solutions,'' {\em Journal of Cleaner Production},
  vol.~140, pp.~1454--1464, 2017.

\bibitem{OR-1}
A.~Floris and L.~Atzori, ``Managing the quality of experience in the multimedia
  internet of things: A layered-based approach,'' {\em Sensors}, vol.~16,
  no.~12, p.~2057, 2016.

\bibitem{PS}
A.~Alabdulatif, I.~Khalil, X.~Yi, and M.~Guizani, ``Secure edge of things for
  smart healthcare surveillance framework,'' {\em IEEE Access}, vol.~7,
  pp.~31010--31021, 2019.

\bibitem{SC}
L.~D. Nadel, M.~T. Laamanen, M.~D. Garris, and C.~S. Russell, ``Recommendation:
  Closed circuit television (cctv) digital video export profile--level 0
  (revision 1),'' tech. rep., 2019.

\bibitem{MR2}
E.~Rubio-Drosdov, D.~D{\'\i}az-S{\'a}nchez, F.~Almen{\'a}rez,
  P.~Arias-Cabarcos, and A.~Mar{\'\i}n, ``Seamless human-device interaction in
  the internet of things,'' {\em IEEE Transactions on Consumer Electronics},
  vol.~63, no.~4, pp.~490--498, 2017.

\bibitem{MR1}
M.~Turunen, D.~Sonntag, K.-P. Engelbrecht, T.~Olsson, D.~Schnelle-Walka, and
  A.~Lucero, ``Interaction and humans in internet of things,'' in {\em
  Human-Computer Interaction -- INTERACT 2015} (J.~Abascal, S.~Barbosa,
  M.~Fetter, T.~Gross, P.~Palanque, and M.~Winckler, eds.), (Cham),
  pp.~633--636, Springer International Publishing, 2015.

\bibitem{lee15}
I.~Lee and K.~Lee, ``The internet of things (iot): Applications, investments,
  and challenges for enterprises,'' {\em Business Horizons}, vol.~58, no.~4,
  pp.~431--440, 2015.

\bibitem{MH}
H.~Shariatmadari, R.~Ratasuk, S.~Iraji, A.~Laya, T.~Taleb, R.~J{\"a}ntti, and
  A.~Ghosh, ``Machine-type communications: current status and future
  perspectives toward 5g systems,'' {\em IEEE Communications Magazine},
  vol.~53, no.~9, pp.~10--17, 2015.

\bibitem{GS}
R.~Li, T.~Song, N.~Capurso, J.~Yu, J.~Couture, and X.~Cheng, ``Iot applications
  on secure smart shopping system,'' {\em IEEE Internet of Things Journal},
  vol.~4, no.~6, pp.~1945--1954, 2017.

\bibitem{rev3}
J.~Qi, P.~Yang, L.~Newcombe, X.~Peng, Y.~Yang, and Z.~Zhao, ``An overview of
  data fusion techniques for internet of things enabled physical activity
  recognition and measure,'' {\em Information Fusion}, 2019.

\bibitem{rev2}
J.~Qi, P.~Yang, G.~Min, O.~Amft, F.~Dong, and L.~Xu, ``Advanced internet of
  things for personalised healthcare systems: A survey,'' {\em Pervasive and
  Mobile Computing}, vol.~41, pp.~132--149, 2017.

\bibitem{HO-1}
K.~A. Ogudo, D.~Muwawa Jean~Nestor, O.~Ibrahim~Khalaf, and H.~Daei~Kasmaei, ``A
  device performance and data analytics concept for smartphones’ iot services
  and machine-type communication in cellular networks,'' {\em Symmetry},
  vol.~11, no.~4, p.~593, 2019.

\bibitem{MO-1}
R.~Ratasuk, S.~Iraji, K.~Hugl, L.~Wang, and A.~Ghosh, ``Performance of low-cost
  lte devices for advanced metering infrastructure,'' in {\em 2013 IEEE 77th
  Vehicular Technology Conference (VTC Spring)}, pp.~1--5, IEEE, 2013.

\bibitem{NE-1}
M.~Nemati, H.~Takshi, and V.~Shah-Mansouri, ``Tag estimation in rfid systems
  with capture effect,'' in {\em 2015 23rd Iranian Conference on Electrical
  Engineering}, pp.~368--373, IEEE, 2015.

\bibitem{MO-3}
S.~Mukherjee and G.~Biswas, ``Networking for iot and applications using
  existing communication technology,'' {\em Egyptian Informatics Journal},
  vol.~19, no.~2, pp.~107--127, 2018.

\bibitem{MR3}
A.~Meddeb, ``Internet of things standards: who stands out from the crowd?,''
  {\em IEEE Communications Magazine}, vol.~54, no.~7, pp.~40--47, 2016.

\bibitem{MO-01}
R.~Fukatsu and K.~Sakaguchi, ``Millimeter-wave v2v communications with
  cooperative perception for automated driving,'' in {\em 2019 IEEE 89th
  Vehicular Technology Conference (VTC2019-Spring)}, pp.~1--5, IEEE, 2019.

\bibitem{H-01}
A.~J. Pathak and E.~Demiralp, ``Disease and fall risk assessment using depth
  mapping systems,'' April 2019.
\newblock US Patent App. 10/262,423.

\bibitem{H-02}
K.~K{\"o}lle, A.~L. Fougner, M.~A. Lundteigen, S.~M. Carlsen, R.~Ellingsen, and
  {\O}.~Stavdahl, ``Risk analysis for the design of a safe artificial pancreas
  control system,'' {\em Health and Technology}, vol.~9, no.~3, pp.~311--328,
  2019.

\bibitem{Eric1-1}
A.~Ericsson, ``Ericsson mobility report june 2019,'' {\em
  https://www.ericsson.com/media-cdn/49d1d9/siteassets/mobility
  -report/documents/2019/ericsson-mobility-report-june-2019.pdf (Accessed
  09/07/2019)}, pp.~9--12, 2019.

\bibitem{Eric}
A.~Ericsson, ``Ericsson mobility report june 2018,'' {\em
  https://www.ericsson.com/assets/local/mobility-report
  /documents/2018/ericsson-mobility-report-june-2018.pdf (Accessed
  09/07/2019)}, pp.~3--13, 2018.

\bibitem{noura2019}
M.~Noura, M.~Atiquzzaman, and M.~Gaedke, ``Interoperability in internet of
  things: Taxonomies and open challenges,'' {\em Mobile Networks and
  Applications}, vol.~24, no.~3, pp.~796--809, 2019.

\bibitem{3G-2016}
3GPP TS 22.891 V14.2.0, {\em Technical Specification Group Services and System
  Aspects; Feasibility Study on New Services and Markets Technology Enablers},
  September 2016.

\bibitem{IoTALL}
IoT For All, {\em An Introduction To Connectivity In IoT}, January 2019.

\bibitem{3G-B}
3GPP R1-162204, {\em Numerology requirements}, April 2016.

\bibitem{HD-1}
``Ieee standard for high data rate wireless multi-media networks,'' {\em IEEE
  Std 802.15.3-2016 (Revision of IEEE Std 802.15.3-2003}, pp.~1--510, July
  2016.

\bibitem{HD-3}
L.~Rakotondrainibe, Y.~Kokar, G.~Zaharia, and G.~El~Zein, ``Millimeter-wave
  system for high data rate indoor communications,'' in {\em 2009 International
  Symposium on Signals, Circuits and Systems}, pp.~1--4, IEEE, 2009.

\bibitem{HD-4}
L.~Kong, M.~K. Khan, F.~Wu, G.~Chen, and P.~Zeng, ``Millimeter-wave wireless
  communications for iot-cloud supported autonomous vehicles: Overview, design,
  and challenges,'' {\em IEEE Communications Magazine}, vol.~55, no.~1,
  pp.~62--68, 2017.

\bibitem{B1}
D.~Chandramouli, R.~Liebhart, and J.~Pirskanen, {\em 5G for the Connected
  World}.
\newblock Wiley Online Library, 2019.

\bibitem{KC}
Z.~He, ``Automatic cooking system,'' Feb 2019.
\newblock US Patent App. 16/155,895.

\bibitem{L1}
C.~Yi, J.~Cai, and Z.~Su, ``A multi-user mobile computation offloading and
  transmission scheduling mechanism for delay-sensitive applications,'' {\em
  IEEE Transactions on Mobile Computing}, 2019.

\bibitem{LA-2}
S.~Pandit, F.~H. Fitzek, and S.~Redana, ``Demonstration of 5g connected cars,''
  in {\em 2017 14th IEEE Annual Consumer Communications \& Networking
  Conference (CCNC)}, pp.~605--606, IEEE, 2017.

\bibitem{C001}
A.~{Pal} and K.~{Kant}, ``Nfmi: Connectivity for short-range iot
  applications,'' {\em Computer}, vol.~52, pp.~63--67, Feb 2019.

\bibitem{MR4}
P.~KUCHHAL and U.~C. SHARMA, ``Battery waste management,''

\bibitem{MR5}
M.~Merry, ``Environmental problems that batteries cause,'' {\em Sciencing}, Mar
  2019.

\bibitem{P01}
A.~Froytlog, T.~Foss, O.~Bakker, G.~Jevne, M.~A. Haglund, F.~Y. Li, J.~Oller,
  and G.~Y. Li, ``Ultra-low power wake-up radio for 5g iot,'' {\em IEEE
  Communications Magazine}, vol.~57, no.~3, pp.~111--117, 2019.

\bibitem{P02}
Z.~Qin, F.~Y. Li, G.~Y. Li, J.~A. McCann, and Q.~Ni, ``Low-power wide-area
  networks for sustainable iot,'' {\em IEEE Wireless Communications}, 2019.

\bibitem{PR-1}
Newark-An AVNET company, {\em Calculating Battery Life in IoT Applications},
  May 2017.

\bibitem{P003}
J.~W. Diachina, N.~Johansson, and P.~Schliwa-Bertling, ``Extended discontinuous
  receive (edrx) cycles,'' Mar.~5 2019.
\newblock US Patent App. 10/225,101.

\bibitem{BC-1}
N.~Van~Huynh, D.~T. Hoang, X.~Lu, D.~Niyato, P.~Wang, and D.~I. Kim, ``Ambient
  backscatter communications: A contemporary survey,'' {\em IEEE Communications
  Surveys \& Tutorials}, vol.~20, no.~4, pp.~2889--2922, 2018.

\bibitem{RE-1}
B.~Safaei, A.~M.~H. Monazzah, M.~B. Bafroei, and A.~Ejlali, ``Reliability
  side-effects in internet of things application layer protocols,'' in {\em
  2017 2nd International Conference on System Reliability and Safety (ICSRS)},
  pp.~207--212, IEEE, 2017.

\bibitem{M-01}
N.~A. Mohammed, A.~M. Mansoor, and R.~B. Ahmad, ``Mission-critical machine-type
  communications: An overview and perspectives towards 5g,'' {\em IEEE Access},
  2019.

\bibitem{B010}
M.~B. Mollah, S.~Zeadally, and M.~A.~K. Azad, ``Emerging wireless technologies
  for internet of things applications: Opportunities and challenges,'' 2019.

\bibitem{ICI}
M.~Nemati and H.~Arslan, ``Low ici symbol boundary alignment for 5g numerology
  design,'' {\em IEEE Access}, vol.~6, pp.~2356--2366, 2017.

\bibitem{Mob-1}
J.~{Wu} and P.~{Fan}, ``A survey on high mobility wireless communications:
  Challenges, opportunities and solutions,'' {\em IEEE Access}, vol.~4,
  pp.~450--476, 2016.

\bibitem{Abramson}
N.~Abramson, ``The aloha system: another alternative for computer
  communications,'' in {\em Proceedings of the November 17-19, 1970, fall joint
  computer conference}, pp.~281--285, ACM, 1970.

\bibitem{ARS-1}
A.~Matoba, M.~Hanada, H.~Kanemitsu, and M.~W. Kim, ``Asymmetric rts/cts for
  exposed node reduction in ieee 802.11 ad hoc networks,'' {\em Journal of
  Computing Science and Engineering}, vol.~8, no.~2, pp.~107--118, 2014.

\bibitem{Ars}
W.~Diepstraten, G.~Ennis, and P.~Belanger, ``Distributed foundation wireless
  medium access control,'' {\em IEEE P802 11-93}, vol.~190, 1993.

\bibitem{Candes}
E.~J. Cand{\`e}s {\em et~al.}, ``Compressive sampling,'' in {\em Proceedings of
  the international congress of mathematicians}, vol.~3, pp.~1433--1452,
  Madrid, Spain, 2006.

\bibitem{Donoho}
D.~L. Donoho {\em et~al.}, ``Compressed sensing,'' {\em IEEE Transactions on
  information theory}, vol.~52, no.~4, pp.~1289--1306, 2006.

\bibitem{Choi_ISWCS17}
J.~{Choi}, ``Noma: Principles and recent results,'' in {\em 2017 International
  Symposium on Wireless Communication Systems (ISWCS)}, pp.~349--354, Aug 2017.

\bibitem{CoverBook}
T.~M. {Cover} and J.~A. {Thomas}, {\em Elements of Information Theory}.
\newblock Wiley-Interscience, 2006.

\bibitem{mMIMO00}
C.~Shepard, H.~Yu, N.~Anand, E.~Li, T.~Marzetta, R.~Yang, and L.~Zhong,
  ``Argos: Practical many-antenna base stations,'' in {\em Proceedings of the
  18th Annual International Conference on Mobile Computing and Networking},
  Mobicom '12, (New York, NY, USA), pp.~53--64, ACM, 2012.

\end{thebibliography}

 \begin{IEEEbiography}[{\includegraphics[width=1in,height=1.25in,clip,keepaspectratio]{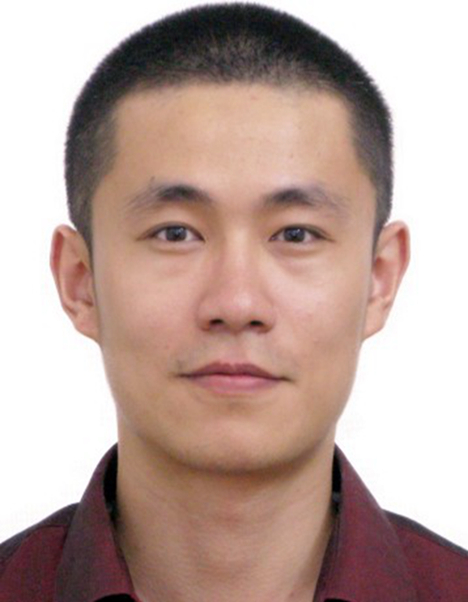}}]{Jie Ding} received the Ph.D. degree in communication engineering from Macquarie University, Sydney, Australia, in 2016. He is now with the School of Information Technology, Burwood, Deakin University, Australia. His research interests include machine-type communication, massive MIMO, and random access.
\end{IEEEbiography}

 \begin{IEEEbiography}[{\includegraphics[width=1in,height=1.25in,clip,keepaspectratio]{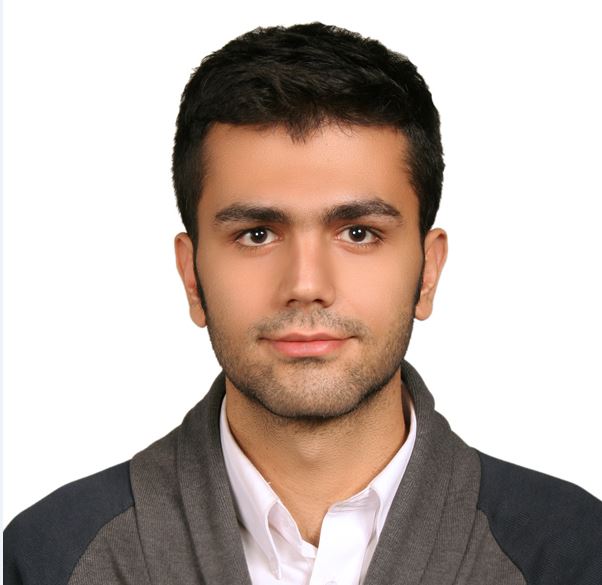}}]{Mahyar Nemati} received the B.S. degree in Electrical Engineering and Telecommunications from the University of Tehran, Iran in 2015; and the M.S. degree in electrical, electronics, and cyber systems with Istanbul Medipol University, Turkey in 2017. He is currently a research assistant with the school of IT at Deakin University where he is involved in the field of wireless communication. His research interests include digital communications, signal processing techniques at the physical and medium access layer, multi- carrier schemes, waveform design in wireless networks, and IoT, MTC, and URLL communications.
\end{IEEEbiography}

 \begin{IEEEbiography}[{\includegraphics[width=1in,height=1.25in,clip,keepaspectratio]{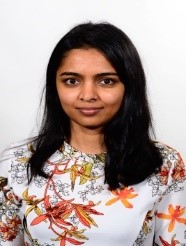}}]{Chathurika Ranaweera} received the B.Sc. and PhD degrees in from The University of Peradeniya, Sri Lanka and The University of Melbourne, Australia, respectively. She is currently a Senior Lecturer at the School of Information Technology, Deakin University, Australia. Her research interests include IoT connectivity, optical transport \& wireless networks design, network optimisation, quality of service management, network energy efficiency, and Smart grid communication.
\end{IEEEbiography}

\begin{IEEEbiography}[{\includegraphics[width=1in,height=1.25in,keepaspectratio]{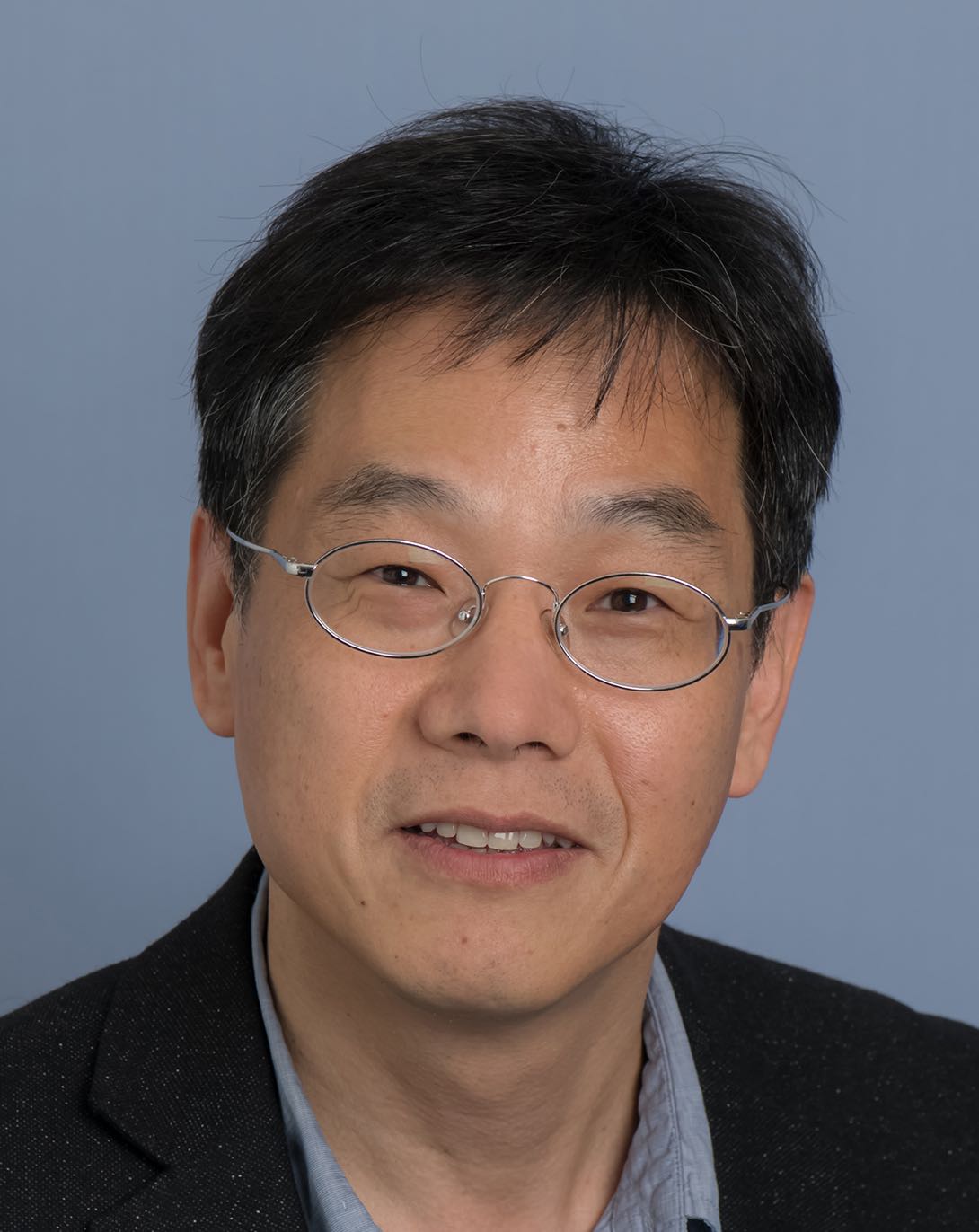}}]
{Jinho~Choi}
(SM'02) was born in Seoul, Korea. He received
B.E. (magna cum laude) degree
in electronics engineering in 1989 from Sogang University, Seoul,
and M.S.E. and Ph.D. degrees in electrical engineering from
Korea Advanced Institute of Science and Technology (KAIST) in 1991 and 1994,
respectively.
He is with the School of Information
Technology, Burwood, Deakin University, Australia, as a Professor.
Prior to joining Deakin in 2018, he was with
Swansea University, United Kingdom, as a Professor/Chair in Wireless,
and Gwangju Institute of Science
and Technology (GIST), Korea, as a Professor.
His research interests include
the Internet of Things (IoT),
wireless communications, and statistical
signal processing.
He authored two books published by
Cambridge University Press in 2006 and 2010. Prof. Choi received
the 1999 Best Paper Award for Signal Processing from EURASIP, 2009
Best Paper Award from WPMC (Conference), and is Senior Member of IEEE.
Currently, he is an Editor of IEEE Trans. Communications and
IEEE Wireless Communications Letters and a Division Editor of
Journal of Communications and Networks (JCN).
We also had served as an Associate Editor or Editor of other journals
including IEEE Communications Letters, JCN, IEEE Trans.
Vehicular Technology, and ETRI journal.

\end{IEEEbiography}




\end{document}